\documentclass[twocolumn, apj, numberedappendix, appendixfloats, showkeys]{openjournal_dhayaa}

\usepackage{xcolor}
\definecolor{linkcolor}{rgb}{0.0,0.3,0.5}
\usepackage[utf8]{inputenc}
\usepackage[english]{babel}

\usepackage{hyperref}
\hypersetup{
    unicode,
    colorlinks=true,
    linkcolor=linkcolor,
    citecolor=linkcolor,
    filecolor=linkcolor,
    urlcolor=linkcolor,
}
\usepackage{xurl}
\usepackage{color,colortbl}
\usepackage{tensind}
\tensordelimiter{?}
\DeclareGraphicsExtensions{.bmp,.png,.jpg,.pdf}
\usepackage{verbatim}
\usepackage[normalem]{ulem}
\usepackage{orcidlink}
\usepackage{soul}
\usepackage{lineno}
\renewcommand{\AA}{%
  \leavevmode\setbox0=\hbox{A}\copy0\kern-\wd0
  \raise1.3ex\hbox to\wd0{\hss{\fontsize{6}{6}\selectfont$\circ$}\hss\kern.01em\kern.04em}%
}
\usepackage{xspace}
\usepackage{enumitem}
\usepackage{array}
\usepackage{amsmath}

\newcommand{\SSSSS}{${S}^5$\xspace}
\newcommand{\gaia}{\textit{Gaia}\xspace}
\newcommand{\vlosUnits}{$\mathrm{km~s}^{-1}$\xspace}
\newcommand{\pmUnits}{mas yr$^{-1}$\xspace}

\begin{document}
\title{Bo\"otes III is a Tidally Disrupting Ultra-Faint Dwarf Galaxy on an Eccentric Polar Orbit\altaffilmark{*}}

\shorttitle{Bo\"otes III tidal disruption}
\shortauthors{Li et al. (\SSSSS Collaboration)}

\altaffiltext{*}{We are deeply grateful to Carl~J.\ Grillmair, who passed away in February~2026. Inter alia, he discovered Bo\"otes~III and the Styx stream \citep{Grillmair2009} and kindly shared the Styx stream trajectory shown in Figure~\ref{fig:stream_comparison}. This work builds directly on his extraordinary legacy. Many of us fondly remember his encouraging words over the years, and we dedicate this paper to him.}
\makeatletter
\def\frontmatter@collaboration@above{\centering}
\makeatother
\altaffiltext{$\dagger$}{Corresponding author: \texttt{ting.li@astro.utoronto.ca}. Author affiliations are listed at the end of this paper.}

\author{Ting~S.~Li~\orcidlink{0000-0002-9110-6163}$^{\dagger,1,2,3}$}
\noaffiliation
\author{Denis~Erkal~\orcidlink{0000-0002-8448-5505}$^{4}$}
\noaffiliation
\author{Andrew~B.~Pace~\orcidlink{0000-0002-6021-8760}$^{5}$}
\noaffiliation
\author{Jiaxun~Yang~\orcidlink{0009-0005-6976-1734}$^{1}$}
\noaffiliation
\author{Sergey~E.~Koposov~\orcidlink{0000-0003-2644-135X}$^{6,7}$}
\noaffiliation
\author{Jo~Bovy~\orcidlink{0000-0001-6855-442X}$^{1,2}$}
\noaffiliation
\author{Nathan~R.~Sandford~\orcidlink{0000-0002-7393-3595}$^{1}$}
\noaffiliation
\author{Andrew~P.~Li~\orcidlink{0009-0005-5355-5899}$^{1,8}$}
\noaffiliation
\author{Gustavo~E.~Medina~\orcidlink{0000-0003-0105-9576}$^{1,2}$}
\noaffiliation
\author{Lara~R.~Cullinane~\orcidlink{0000-0001-8536-0547}$^{9}$}
\noaffiliation
\author{Gary~S.~Da~Costa~\orcidlink{0000-0001-7019-649X}$^{10}$}
\noaffiliation
\author{Alexander~P.~Ji~\orcidlink{0000-0002-4863-8842}$^{11,12,13}$}
\noaffiliation
\author{Kyler~Kuehn~\orcidlink{0000-0003-0120-0808}$^{14}$}
\noaffiliation
\author{Geraint~F.~Lewis~\orcidlink{0000-0003-3081-9319}$^{15}$}
\noaffiliation
\author{Guilherme~Limberg~\orcidlink{0000-0002-9269-8287}$^{11,13}$}
\noaffiliation
\author{Sarah~L.~Martell~\orcidlink{0000-0002-3430-4163}$^{16}$}
\noaffiliation
\author{Aldo~Mura-Guzm{\'a}n~\orcidlink{0000-0003-1711-1981}$^{17,18}$}
\noaffiliation
\author{Nora~Shipp~\orcidlink{0000-0003-2497-091X}$^{19}$}
\noaffiliation
\author{Yong~Yang~\orcidlink{0000-0001-7609-1947}$^{15}$}
\noaffiliation
\author{Daniel~B.~Zucker~\orcidlink{0000-0003-1124-8477}$^{17,18}$}
\noaffiliation
\author{Kaia~R.~Atzberger~\orcidlink{0000-0001-9649-8103}$^{5}$}
\noaffiliation
\author{Joss~Bland-Hawthorn~\orcidlink{0000-0001-7516-4016}$^{15}$}
\noaffiliation
\author{John~D.~Dixon~\orcidlink{0000-0001-6168-3130}$^{20,21}$}
\noaffiliation
\collaboration{The \SSSSS\ Collaboration}

\begin{abstract}
We present updated systemic properties of the ultra-faint dwarf galaxy Bo\"otes~III from the Southern Stellar Stream Spectroscopic Survey (\SSSSS).
We identify 21 high-probability members and measure a velocity dispersion of $\sigma_{v} = 1.69^{+1.03}_{-0.85}$ \vlosUnits, about six times smaller than the previously reported $10.7 \pm 3.5$ \vlosUnits, and a mean metallicity of $[\mathrm{Fe/H}] = -2.34 \pm 0.11$. The revised dispersion brings Bo\"otes~III in line with other tidally disrupting dwarfs such as Antlia~II and Crater~II.
Orbit integrations in a Milky Way (MW) + Large Magellanic Cloud (LMC) potential confirm a highly eccentric ($e \approx 0.8$), polar ($i \approx 89.5^\circ$) orbit with a recent pericentric passage $\sim 0.14$~Gyr ago at $r_{\mathrm{peri}} \approx 9.5$~kpc. 
Bo\"otes~III is thus likely actively tidally disrupting, as its tidal radius at pericenter, $r_t \approx 164$~pc, is only $\sim 0.35$ of its half-light radius.
The unusually low dispersion also implies that Bo\"otes~III has either lost most of its dark matter to tides or hosts a cored inner density profile, making it a probe of the nature of dark matter.
Simulated tidal streams are broadly consistent with the Styx stellar stream, though the predicted track and kinematics are sensitive to the MW halo mass, LMC mass, and solar velocity.
Bo\"otes~III overlaps the Typhon stream in integrals-of-motion space but has a much lower mean metallicity, suggesting the two are not the same system but may have had a common group infall origin.
Sagittarius-stream contamination prevents a direct tidal-tail detection, so deep spectroscopic follow-up remains essential, both to confirm Styx as a genuine stream and to establish it as Bo\"otes~III's tidal tail.
\end{abstract}


\section{Introduction} \label{section:intro}

Dwarf spheroidal galaxies (dSphs) are among the most dark matter-dominated systems known, making them powerful probes of the nature of dark matter \citep[see, e.g.,][]{Bullock2017ARAA}.
The number of satellite galaxies in a host galaxy can constrain warm and fuzzy dark matter particle masses \citep{Nadler2021}, while internal kinematics of a dwarf galaxy can constrain the self-interaction cross section of self-interacting dark matter \citep[SIDM;][]{Zavala2013, Correa2021, Ando2025}.
However, inferring the dark matter density profile from kinematics alone is challenging due to the degeneracy between the density profile and the velocity anisotropy \citep{Lokas2002, Guerra2023}.

The tidal disruption of dSphs offers an alternative approach to constraining their dark matter density profiles.
The dense, cuspy profiles predicted by cold dark matter (CDM) are far more resilient to tidal stripping than the cored profiles arising in SIDM models \citep{Penarrubia2010}. \citet{Errani2023} demonstrated that cuspy Navarro--Frenk--White (NFW) subhalos never fully disrupt, whereas cored subhalos undergo runaway mass loss once tides penetrate the core, leading to complete destruction. The survival or disruption of a dwarf galaxy on a close orbit therefore places direct constraints on the maximum core size permitted by observations, and thus provides insights into the nature of dark matter. Tidal disruption also enters comparisons between simulated and observed satellite populations: cosmological hydrodynamic simulations of Milky Way (MW)-mass hosts predict that a large fraction of surviving satellite galaxies are heavily tidally disrupted \citep{Shipp:2023, Riley:2025, Shipp:2025}, and it is not yet clear whether the observed MW satellite population is consistent with these predicted disruption rates. Identifying and characterizing individual disrupting systems is therefore an important step toward testing this prediction.

Wide-field photometric surveys, including the Sloan Digital Sky Survey (SDSS; \citealt{York2000}), the Dark Energy Survey (DES; \citealt{DES2016, Abbott2018_DESDR1}), and the Panoramic Survey Telescope and Rapid Response System (Pan-STARRS; \citealt{Chambers2016}), have driven the rapid discovery of the MW's satellite galaxies over the past two decades \citep[see][and references therein]{Simon2019}. Interpreting the tidal features of these systems requires careful modeling of the gravitational potential, including the influence of the Large Magellanic Cloud (LMC). The LMC induces a barycentric acceleration of the MW that significantly perturbs satellite orbits and the morphology of tidal streams \citep{Vera-Ciro+2013,Gomez+2015,Erkal+2018,Erkal2019, Erkal2020, Nadler2020,Patel+2020}. \citet{Pace2022} computed orbits for 46~dSphs in a combined MW+LMC potential and found that 40\% of their sample experienced a $>25\%$ change in pericenter and/or apocenter when the LMC was included. They further identified several systems likely undergoing tidal disruption based on comparing each satellite's average density to the MW density at its orbital pericenter. Besides the well-known Sagittarius (Sgr) dwarf and its extensive tidal stream \citep{Vasiliev2021}, several other MW satellites have been observationally identified as likely undergoing tidal disruption; these include Tucana~III \citep{DrlicaWagner2015, Simon2017}, Antlia~II \citep{Torrealba2019}, and Crater~II \citep{Torrealba2016, Caldwell2017, Ji2021, Limberg2025}.

Spectroscopic follow-up with the Two-degree Field (2dF; \citealt{Lewis2002}) fiber positioner and the AAOmega spectrograph \citep{Sharp2006} on the 3.9\,m Anglo-Australian Telescope (AAT) has provided key evidence in each case. In particular, spectroscopic characterization of these faint systems has been greatly advanced by the Southern Stellar Stream Spectroscopic Survey (\SSSSS; \citealt{Li2019,Li2022}). Originally designed to map stellar streams in the Southern Hemisphere discovered by the DES \citep{Shipp2018}, \SSSSS has also enabled detailed spectroscopic studies of dwarf galaxies and their tidal features. In a pathfinder program of \SSSSS\ that demonstrated the feasibility of spectroscopic stream studies with AAT/AAOmega and motivated the design of \SSSSS, \citet{Li2018tucIII} identified members in the tidal tails of Tucana~III and measured a velocity gradient along the stream, confirming the tails as genuine tidal extensions and making Tucana~III the first ultra-faint dwarf (UFD) with a spectroscopically confirmed tidal stream. Subsequently within \SSSSS, \citet{Ji2021} detected a velocity gradient aligned with the orbital direction in Antlia~II and a tentative gradient in Crater~II; \citet{Limberg2025} then spectroscopically confirmed Crater~II's stellar stream and used the ratio of stream-to-galaxy velocity dispersion to constrain its dark matter density profile. In a related study, \citet{Sandford2026} reported a tentative velocity gradient in Bo\"{o}tes~I aligned with its orbital direction, which, if confirmed, could indicate incipient tidal disruption in that system as well.

Bo\"otes~III (hereafter Boo~III) is a UFD discovered by \citet{Grillmair2009} as a stellar overdensity in SDSS photometry. With a semi-major half-light radius of $a_h \sim 0.55$\,kpc, corresponding to an azimuthally-averaged $r_h \sim 0.45$\,kpc \citep{Moskowitz2020}, Boo~III is unusually extended for its luminosity ($M_V \sim -5.8$), giving it a low surface brightness which led to early speculation that the system is being tidally stripped \citep{Carlin2009, Correnti2009}. \citet{Carlin2009} conducted the first spectroscopic study of Boo~III, measuring a systemic line-of-sight velocity of $v_{\rm los} = 197.5 \pm 3.8$ \vlosUnits and a velocity dispersion of $\sigma_v = 14.0 \pm 3.2$ \vlosUnits from 20 probable members observed with the Multi-Mirror Telescope (MMT)/Hectospec. \citet{Carlin2018} subsequently refined the velocity dispersion to $\sigma_v = 10.7 \pm 3.5$ \vlosUnits through an improved membership analysis and also measured proper motions from \gaia DR2 \citep{GaiaDR2}, revealing a retrograde, highly eccentric orbit. 
The Styx stream is a diffuse, low-surface-brightness stellar stream identified by \citet{Grillmair2009} in the same SDSS data that revealed Boo~III: it traces a long arc across the sky that passes through the Boo~III field at a comparable heliocentric distance ($\sim 45$~kpc), with Boo~III lying near its midpoint and debris extending to either side. Since its discovery, Styx has gone unconfirmed by any subsequent deep imaging, \gaia, or spectroscopic survey, leaving both its reality as a genuine stream and any physical link to Boo~III as open questions. However, the \gaia\ DR2 proper motion of Boo~III \citep{Carlin2018} aligns with the orbit that \citet{Grillmair2009} predicted for the Styx stream from its on-sky track -- circumstantial evidence that Styx is a genuine stream physically associated with Boo~III. This alignment led \citet{Carlin2018} to propose that Boo~III is a dwarf galaxy actively disrupting to form the Styx stream, which would make it a rare example of a UFD caught in the act of tidal disruption.

The large velocity dispersion reported by \citet{Carlin2009} and \citet{Carlin2018} is, however, puzzling when compared to other tidally disrupting dwarfs. Both Antlia~II and Crater~II have much smaller velocity dispersions -- $\sigma_v = 5.98$ \vlosUnits \citep{Ji2021} and $\sigma_v = 2.51$ \vlosUnits \citep{Limberg2025}, respectively -- despite having significantly larger half-light radii ($\sim 2.5$\,kpc and $\sim 1.1$\,kpc; \citealt{Ji2021, Torrealba2016}, respectively) and being much more luminous ($M_V = -9.86$ and $-8.2$; \citealt{Ji2021, Torrealba2016}, respectively). This raises the question of whether the previously measured velocity dispersion of Boo~III was inflated by contamination or small sample size, or whether Boo~III has a fundamentally different tidal evolution history.

In this work, we address this question using new spectroscopic data from \SSSSS DR2 combined with astrometry from \gaia DR3. We re-measure Boo~III's systemic properties, and use these updated parameters to integrate its orbit in a MW+LMC potential and generate mock tidal streams to assess the Boo~III--Styx connection. This paper is organized as follows. Section~\ref{section:data} introduces the \SSSSS\ DR2 spectroscopic data and the cleaning cuts applied to build the input catalog of candidate Boo~III stars. Section~\ref{section:membership} presents the Gaussian-mixture-model (GMM) membership analysis that simultaneously re-derives Boo~III's systemic heliocentric velocity, velocity dispersion, mean metallicity, intrinsic metallicity dispersion, and \gaia\ DR3 proper motion, together with an updated spatial centroid, and compares them with previous spectroscopic studies. Section~\ref{section:distance} derives a new heliocentric distance from five \gaia\ DR3 RR~Lyrae (RRL) stars in the field. In Section~\ref{section:sim} we integrate Boo~III's orbit and simulate tidal streams using \texttt{galpy} \citep{Bovy2015}, comparing results with and without the LMC. Section~\ref{section:discussion} discusses the implications: evidence for active tidal disruption from the orbit, tidal-radius, and density arguments; restricted $N$-body models of the disrupting remnant, tested against the observed velocity dispersion and velocity gradient; the near-progenitor stream kinematics as a probe of the MW halo mass; the association with the Typhon stream \citep{Tenachi2022} in integrals-of-motion and action space; and the observational search for an extended Boo~III tidal tail. We conclude in Section~\ref{section:conclusion}.

\section{Data} \label{section:data}

\subsection{\SSSSS\ Spectroscopy}
This paper uses data from the second public data release (DR2) of \SSSSS\ (T.\,S.~Li et al., in preparation).
\SSSSS\ is a spectroscopic follow-up of stellar streams and dwarf galaxies discovered in photometric surveys in the southern hemisphere, with a particular focus on dwarf galaxies potentially undergoing tidal stripping, carried out with the dual-arm AAOmega spectrograph \citep{Sharp2006} fed by the 2dF \citep{Lewis2002} fiber positioner on the AAT.
Each AAT configuration observes up to 392 science fibers simultaneously over a 2\ensuremath{^{\circ}}-diameter field of view. With a 580V grating in the blue-arm and a 1700D grating in the red-arm, it yields a low-resolution spectrum ($3800$--$5800$\,\AA, $R \sim 1300$) used for stellar parameters and metallicity, and a high-resolution spectrum ($8420$--$8820$\,\AA, $R \sim 10{,}000$) covering the calcium triplet (CaT) region for precise radial velocities.
The survey design, target selection strategy, and data reduction pipeline are described in detail in \citet{Li2019} and \citet{Li2022}.
Boo~III was observed over multiple AAT nights and covered with several 2dF pointings centered on \citet{Grillmair2009}'s reported centroid, with targets drawn from \gaia\ DR3 \citep{GaiaMission, GaiaDR3} astrometry and photometry from the Legacy Survey DR9 catalog \citep{DECaLS}, following the standard \SSSSS\ targeting strategy: candidate members were required to be consistent with the color--magnitude diagram (CMD) of an old, metal-poor stellar population at the distance of Boo~III and to have proper motions broadly consistent with the system. Throughout this work the photometry quoted as $g_0, r_0$ refers to the dereddened magnitudes with the \citet{Schlegel1998} $E(B-V)$ map and the DES extinction coefficients ($R_g = 3.185$, $R_r = 2.140$; \citealt{Abbott2018_DESDR1}); the same dereddening is used for both target selection and the membership / CMD analysis below.

Stellar parameters ($T_{\rm eff}$, $\log g$, $[\mathrm{Fe/H}]$) and heliocentric line-of-sight velocities are derived by jointly fitting the blue- and red-arm spectra against the \texttt{rvspecfit} template grid \citep{Koposov2019_rvspecfit} using the \SSSSS\ spectral-fitting pipeline. DR2 includes a number of improvements to this pipeline relative to DR1, summarized in \citet{Ji2021}, \citet{Sandford2026}, \citet{Limberg2025}, and T.\,S.~Li et al., in preparation, that yield more robust line-of-sight velocity and $[\mathrm{Fe/H}]$ measurements, particularly for faint low signal-to-noise ratio (S/N) stars.
We adopt the DR2 catalog throughout this paper; all line-of-sight velocities $v_{\rm los}$ quoted below are heliocentric.

\subsection{Input Sample Selection} \label{section:input_sample} 
Sky positions are measured as angular separations from the Boo~III centroid $(\alpha_0,\delta_0)$, taken initially from \citet{Grillmair2009} for the target selection.  This centroid is re-derived self-consistently in Section~\ref{section:membership} from the 21 high-probability GMM members and the updated value, $(\alpha_0,\delta_0)_{\rm new} = (209.557^\circ, 26.553^\circ)$, is used in all subsequent analysis.  The tangent-plane coordinates are computed as
\begin{align}
\xi  &= \cos\delta\,\sin(\alpha-\alpha_0), \\
\eta &= \sin\delta\cos\delta_0 - \cos\delta\sin\delta_0\cos(\alpha-\alpha_0),
\end{align}
i.e.\ the orthographic (SIN) projection of the celestial sphere onto the plane tangent at $(\alpha_0,\delta_0)$, defining the tangent-plane coordinates $(\xi, \eta)$ used throughout this paper. In the small-angle limit relevant here, $\xi \approx \Delta\alpha\cos\delta$ and $\eta \approx \Delta\delta$. 
We then apply the following quality and sanity cuts to all \SSSSS\ DR2 sources that were targeted as part of the Boo~III field:
\begin{itemize}[itemsep=0pt]
    \item signal-to-noise ratio \texttt{best\_sn\_1700d} $> 2$;\footnote{\texttt{best\_sn\_1700d} is the highest average signal-to-noise ratio (averaged across pixels) achieved in the 1700D red-arm grating setting among the multiple observations of a star; \SSSSS\ DR2 fits all repeat observations of a star simultaneously to derive its stellar parameters, a new feature relative to DR1 (T.\,S.~Li et al., in preparation).}
    \item heliocentric line-of-sight velocity uncertainty \texttt{vel\_calib\_std} $< 10$~\vlosUnits;
    \item good-star probability flag \texttt{good\_star\_pb} $> 0.5$\footnote{The \texttt{good\_star\_pb} quantity is the probability that a source is a good-quality stellar spectrum with a reliable line-of-sight velocity, returned by a random-forest classifier trained on features of the \texttt{rvspecfit} template fit -- the fit $\chi^2$, the $v_{\rm los}$ uncertainty and its posterior skewness and kurtosis, the effective temperature, median S/N, and the relative template residual; see \citet[][their Section~4.5]{Li2019} for details.};
    \item finite \gaia\ DR3 proper motions in both components;
    \item $|v_{\rm los}| < 600$~\vlosUnits\ (to exclude pipeline failures);
    \item $[\mathrm{Fe/H}] < 0$~dex (excluding metal-rich disk stars).
\end{itemize}
All structural parameters of Boo~III used in this work are taken from the literature; the adopted values, including the ellipticity, position angle (PA), and the circularized half-light radius $r_h$ (related to the semi-major-axis half-light radius by $a_h = r_h / \sqrt{1-\epsilon}$), are listed in Table~\ref{tab:params}. Throughout this paper we use $r_h = 33.03'$ \citep{Moskowitz2020} as the canonical half-light radius unless otherwise noted.

\begin{table*}[ht]
\centering
\caption{Comparison of literature and updated Boötes III parameters. \label{tab:params}}
\footnotesize
\begin{tabular}{c l l l l l}
\hline\hline
Row & Quantity & Units & Literature & This work & References \\
\hline
(1)  & RA                            & deg           & $209.300$                  & $209.557 \pm 0.3$              & \citet{Grillmair2009}; this work \\
(2)  & Dec                           & deg           & $26.800$                   & $26.553 \pm 0.3$               & \citet{Grillmair2009}; this work \\
(3)  & $D_\odot$                     & kpc           & $46.5 \pm 2.0$             & $48.5 \pm 1.9$                 & \citet{Carlin2018}; this work \\
(4)  & $\bar v_{\rm los}$            & \vlosUnits    & $197.5 \pm 3.8$            & $191.22^{+0.64}_{-0.80}$       & \citet{Carlin2009}; this work \\
(5)  & $\bar\mu_\alpha\cos\delta$    & \pmUnits      & $-1.168 \pm 0.018$         & $-1.163^{+0.017}_{-0.019}$     & \citet{Pace2022}; this work \\
(6)  & $\bar\mu_\delta$              & \pmUnits      & $-0.890 \pm 0.014$         & $-0.883^{+0.013}_{-0.014}$     & \citet{Pace2022}; this work \\
(7)  & $\sigma_v$                    & \vlosUnits    & $10.7 \pm 3.5$             & $1.69^{+1.03}_{-0.85}$         & \citet{Carlin2018}; this work \\
(8)  & $\overline{[\mathrm{Fe/H}]}$  & dex           & $-2.1 \pm 0.2$             & $-2.34^{+0.11}_{-0.11}$        & \citet{Carlin2018}; this work \\
(9)  & $\sigma_{[\mathrm{Fe/H}]}$    & dex           & $0.55 \pm 0.19$            & $0.36^{+0.10}_{-0.08}$         & \citet{Carlin2018}; this work \\
(10) & $M_V$                         & mag           & $-5.8 \pm 0.5$             & ---                            & \citet{Correnti2009} \\
(11) & $\epsilon$                    & ---           & $0.33 \pm 0.09$            & ---                            & \citet{Moskowitz2020} \\
(12) & PA                            & deg           & $278.91 \pm 7.5$           & ---                            & \citet{Moskowitz2020} \\
(13) & $r_h$ (az.\ avg.)             & arcmin        & $33.03 \pm 2.50$           & ---                            & \citet{Moskowitz2020} \\
(14) & $a_h$ (semi-major)            & arcmin        & ---                        & $40.35 \pm 4.08$               & derived from (11), (13) \\
(15) & $r_h$ (az.\ avg.)             & pc            & ---                        & $466 \pm 39$                   & derived from (3), (13) \\
(16) & $a_h$ (semi-major)            & pc            & ---                        & $569 \pm 62$                   & derived from (3), (14) \\
(17) & $M_{1/2}$                & ${\rm M}_\odot$ & ---                   & $1.24^{+1.96}_{-0.93}\times 10^6$ & this work \\
(18) & $r_{\rm peri}$ (w/o LMC)      & kpc           & ---                        & $9.04^{+2.16}_{-1.89}$  & this work \\
(19) & $r_{\rm peri}$ (w/ LMC)       & kpc           & ---                        & $9.48^{+2.20}_{-1.93}$  & this work \\
(20) & $r_{\rm apo}$ (w/o LMC)       & kpc           & ---                        & $107.67^{+9.36}_{-7.89}$ & this work \\
(21) & $r_{\rm apo}$ (w/ LMC)        & kpc           & ---                        & $97.60^{+8.16}_{-6.81}$ & this work \\
(22) & $e$ (w/o LMC)                 & ---           & ---                        & $0.85^{+0.02}_{-0.02}$  & this work \\
(23) & $e$ (w/ LMC)                  & ---           & ---                        & $0.82^{+0.02}_{-0.02}$  & this work \\
(24) & $i$ (orbital inclination) & deg & --- & $89.5^{+0.5}_{-0.5}$ & this work \\
(25) & $|\nabla v_{\rm los}|$          & \vlosUnits~deg$^{-1}$ & ---                        & $3.28^{+2.0}_{-1.7}$    & this work \\
(26) & PA$(\nabla v_{\rm los})$      & deg           & ---                        & $170^{+15}_{-21}$       & this work \\
\hline
\end{tabular}

\vspace{1ex}
\noindent\parbox{\textwidth}{\footnotesize \textbf{Notes.} Rows are grouped by quantity type. Rows~1--9 are kinematic and chemical parameters (the 6D phase-space coordinates plus internal velocity and metallicity dispersions). Rows~10--17 are structural parameters: rows~10--13 are adopted directly from the cited literature; rows~14--17 are derived in this work from the literature inputs and our new distance. Rows~18--24 are orbital properties computed in this work, and rows~25--26 are the observed line-of-sight velocity gradient. The new RA and Dec in rows~(1, 2) are the iterative-median centroid of the 21 GMM members (Section~\ref{section:center}) within $1\,r_h$ of the running center, converged in three iterations; the $\pm 0.3^\circ$ uncertainty covers both individual-axis shifts ($\xi = +0.23^\circ$, $\eta = -0.25^\circ$ from the \citealt{Grillmair2009} value). The heliocentric distance in row~(3) is from five \gaia~DR3 RRL stars (Section~\ref{section:distance}, Table~\ref{tab:rrl_distances}) using the period-independent $M_G([\mathrm{Fe/H}])$ calibration of \citet{Garofalo2022} evaluated at the system-mean $\overline{[\mathrm{Fe/H}]} = -2.34$. The systemic velocity, proper motion, dispersion, and metallicity values in this work (rows~4--9) are from the GMM fit (Section~\ref{section:membership}, full prior table in Table~\ref{tab:priors}). Row~(14) is the semi-major half-light radius, derived as $a_h = r_h/\sqrt{1-\epsilon}$, with uncertainty propagated from (11) and (13). Rows~(15)--(16) convert (13) and (14) to physical units using the new $D_\odot$, propagating both the angular and distance uncertainties. Row~(17) is the half-light mass using $M_{1/2} = 930\,\sigma_v^2\,r_h$ \citep{Wolf2010} with the circularized half-light radius in pc (row~15) for consistency with \citet{Pace2022}. Rows~(25)--(26) are the observed line-of-sight velocity-gradient amplitude and position angle (east of north), from the four-parameter single-Gaussian fit to the 21 members (Section~\ref{section:vgrad_compare}). The pericenter, apocenter, and eccentricity (rows~18--23) are computed from the McMillan17 potential \citep{McMillan2017} with 1000 MC realizations of the 6D phase-space integrated 4~Gyr backward (See \S\ref{section:orbit} and Figures~\ref{fig:orbits}--\ref{fig:orbit_params}); the inclination in row~(24) is $i = \arccos(L_z/|L|)$, derived from the present-day angular momentum and independent of the assumed potential ($90^\circ$ corresponds to a polar orbit). Uncertainties for rows 4--9 and 17--26 are 16/84-percentile spreads.}
\end{table*}

We restrict the spatial footprint to within the 3 $r_h$ ellipse, i.e.\ an elliptical aperture with semi-major axis $3\,a_h$. An elliptical aperture, rather than a simple circle, is chosen to best reflect the observed light distribution of the system.
We also apply a $4\times 4$~\pmUnits\ proper-motion box centered on the \citet{Pace2022} Boo~III value,
\begin{align}
-3.17 &< \mu_\alpha\cos\delta < 0.83~\mathrm{mas\,yr^{-1}}, \\
-2.88 &< \mu_\delta < 1.12~\mathrm{mas\,yr^{-1}},
\end{align}
to restrict the input catalog to the region of proper-motion space where Boo~III members can plausibly lie.  
Finally, we cross-match the remaining stars to the \gaia\ DR3 RRL catalog \citep{GaiaRRL_reference} within a 2$''$ tolerance and remove all RRL cross-matches from the input catalog; their pulsation-induced line-of-sight velocity scatter would otherwise bias the kinematic fit of Section~\ref{section:membership}. We do not impose an explicit blue-horizontal-branch (BHB) filter, but the per-star line-of-sight velocity uncertainty cut (\texttt{vel\_calib\_std} $< 10$~\vlosUnits) implicitly removes most BHB candidates at Boo~III's $g_0 \approx 19$ horizontal branch, since their broader hydrogen-line absorption produces per-star line-of-sight velocity errors typically $\gtrsim 15$~\vlosUnits\ in \SSSSS\ at that magnitude.
After all of the above cuts, the input catalog consists of 127 non-RRL stars within the 3 $r_h$ ellipse centered on the \citet{Grillmair2009} discovery position; the same cuts evaluated about the updated centroid derived in Section~\ref{section:membership} below yield 120 stars.  We have re-run the GMM membership analysis on both the 127-star (original-center) and 120-star (updated-center) input catalogs and recover the same 21 high-probability ($P_{\rm mem} > 0.95$) members in both cases; results quoted in the remainder of the paper are from the 120-star, updated-center fit. This catalog is the input for the GMM membership classification described in Section~\ref{section:membership}.

\begin{figure*}
    \centering
    \includegraphics[width=\textwidth]{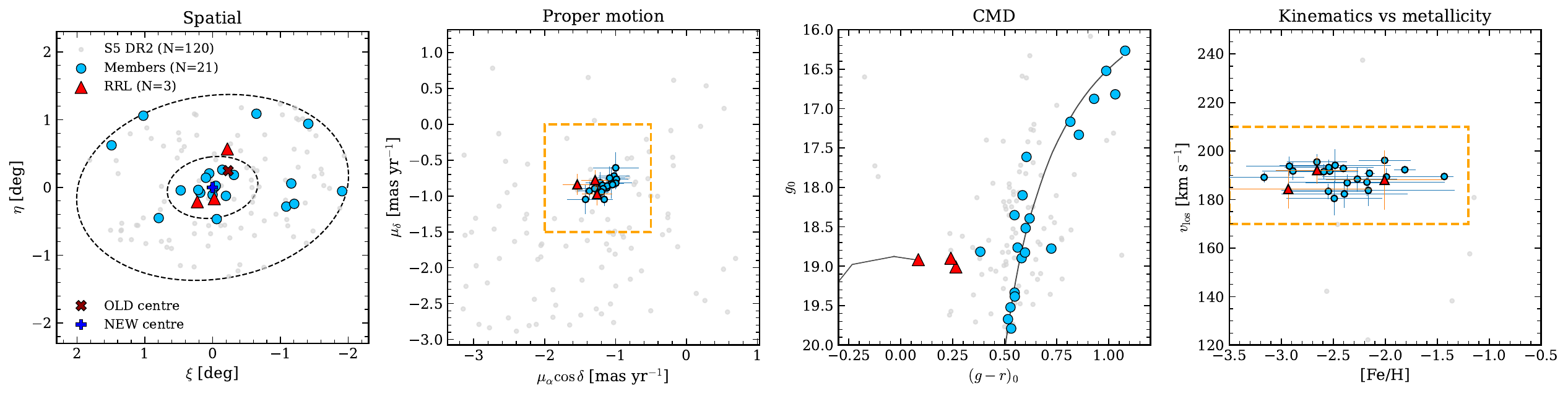}
    \caption{Membership analysis of Boo~III based on \SSSSS\ DR2, in four observational spaces. Gray dots show the 120-star updated-center input catalog (Section~\ref{section:input_sample}): non-RRL \SSSSS\ DR2 stars inside the 3 $r_h$ ellipse and the $4\times 4$~\pmUnits\ proper-motion box. Blue filled circles mark the 21 GMM members ($P_{\rm mem} > 0.95$). Red triangles mark three \gaia\ DR3 RRL kinematically and photometrically consistent with Boo~III, removed from the GMM fit but added back to the final 24-star member sample. \emph{Panel 1 (spatial):} dashed ellipses show 1 and 3 times the elliptical half-light radius; the dark-red cross and blue plus mark the \citet{Grillmair2009} and updated (this work) centroids, respectively. \emph{Panel 2 (proper motion):} the orange dashed rectangle marks the hand-tuned proper-motion box cut of Eqs.~\eqref{eq:boxcut_pmra}--\eqref{eq:boxcut_pmdec}, shown for reference. \emph{Panel 3 (CMD):} the solid curve is a 12.5~Gyr, $[\mathrm{Fe/H}] = -2.25$ Dartmouth isochrone \citep{Dotter2008} shifted to our updated distance of Boo~III ($m-M = 18.43$, $D_\odot = 48.5$~kpc), and the horizontal line is the empirical M92 horizontal-branch ridgeline. \emph{Panel 4 ($v_{\rm los}$ vs.\ $[\mathrm{Fe/H}]$):} the orange dashed rectangle marks the box cut of Eqs.~\eqref{eq:boxcut_rv} and \eqref{eq:boxcut_feh}.}
    \label{fig:main_members}
\end{figure*}

\begin{figure*}
    \centering
    \includegraphics[width=\textwidth]{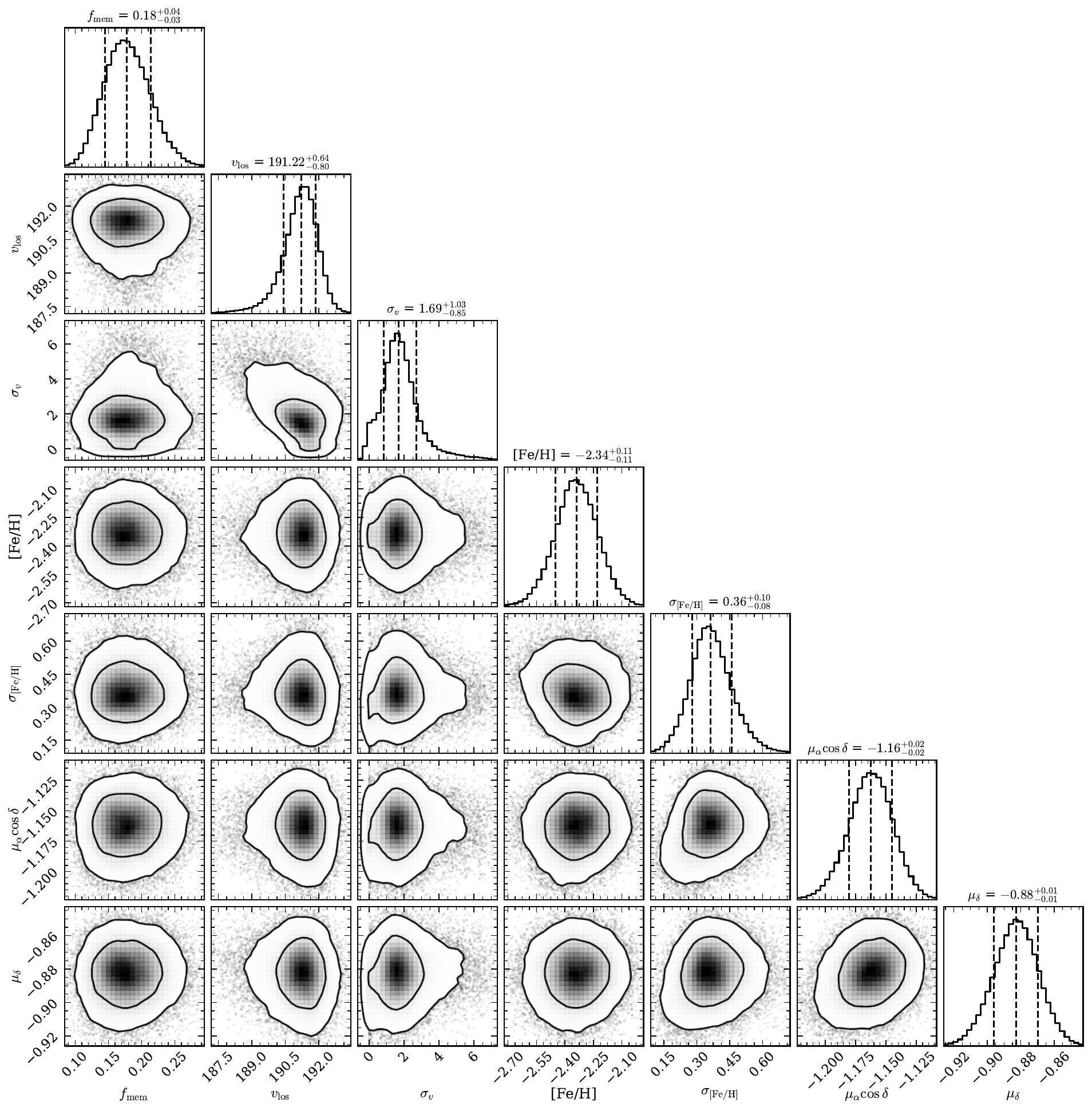}
    \caption{Posterior distributions of Boo~III's seven systemic GMM parameters (Section~\ref{section:membership}), fit jointly to the 120-star (updated-center) input catalog of Section~\ref{section:input_sample}: the membership fraction $f_{\rm mem}$, the systemic heliocentric velocity $\bar v_{\rm los}$, the line-of-sight velocity dispersion $\sigma_v$, the mean metallicity $\overline{[\mathrm{Fe/H}]}$ and its intrinsic dispersion $\sigma_{[\mathrm{Fe/H}]}$, and the two systemic proper-motion components $\bar\mu_\alpha\cos\delta$ and $\bar\mu_\delta$. The four MW field-background parameters are also fit but are omitted from this plot for clarity (full prior table in Table~\ref{tab:priors}). Diagonal panels show 1D marginals; off-diagonals show 2D contours at 0.68 and 0.95 confidence. Numerical values are quoted at the 16th, 50th and 84th percentiles.}
    \label{fig:boo3_params}
\end{figure*}

\section{Membership and Systemic Properties} \label{section:membership}

Rather than identifying Boo~III members by successive hand-tuned line-of-sight velocity and CMD box cuts, we classify the input catalog from Section~\ref{section:input_sample} with a GMM that simultaneously fits Boo~III's systemic properties and the MW field-star background. The model and sampling scheme follow the mixture-model framework of \citet{Pace2019} and \citet{Pace2022} for the astrometric component, and \citet{Sandford2026} for the joint inclusion of $v_{\rm los}$ and $[\mathrm{Fe/H}]$ in the likelihood, on which the present implementation is most closely modeled. We refer the reader to those works for the full likelihood derivation.

\subsection{Mixture Model}
Each star $i$ is assigned a likelihood that is the weighted sum of a dwarf-galaxy (``gal'') component and an MW field-background (``bg'') component:
\begin{equation} \label{eq:gmm_mix}
\mathcal{L}_i = f_{\rm mem}\,\mathcal{L}_i^{\rm gal} + (1 - f_{\rm mem})\,\mathcal{L}_i^{\rm bg},
\end{equation}
where $f_{\rm mem}$ is the galaxy membership fraction, a free parameter of the fit.

For each observable ($v_{\rm los}$, $[\mathrm{Fe/H}]$, $\mu_\alpha\cos\delta$, $\mu_\delta$), the \emph{Boo~III galaxy} component is a Gaussian centered at the systemic value with a scatter that combines the intrinsic dispersion and the per-star measurement error in quadrature; for the proper motions we use the full per-star \gaia\ covariance, including the $\mu_\alpha\cos\delta$--$\mu_\delta$ correlation. For $v_{\rm los}$ and $[\mathrm{Fe/H}]$ we fit both the mean and a constant intrinsic dispersion (four parameters: $\bar v_{\rm los}$, $\sigma_v$, $\overline{[\mathrm{Fe/H}]}$, $\sigma_{[\mathrm{Fe/H}]}$); for the proper motions we fit only the two means ($\bar\mu_\alpha\cos\delta$, $\bar\mu_\delta$), with the intrinsic proper-motion dispersion fixed to zero, because an internal dispersion comparable to the measured $\sigma_v$ ($\sim 1$~\vlosUnits) is unresolved by \gaia\ at Boo~III's heliocentric distance ($D_\odot = 48.5$~kpc, derived in Section~\ref{section:distance}).

The \emph{MW field background} is modeled as a truncated Gaussian in $v_{\rm los}$ with mean $\bar v^{\rm bg}$ and dispersion $\sigma_v^{\rm bg}$ (truncated to $|v_{\rm los}| < 600$~\vlosUnits), times a truncated Gaussian in $[\mathrm{Fe/H}]$ with mean $\overline{[\mathrm{Fe/H}]}^{\rm bg}$ and dispersion $\sigma_{[\mathrm{Fe/H}]}^{\rm bg}$ (truncated to $[\mathrm{Fe/H}] \in [-4, 0]$, the range of the \texttt{rvspecfit} template grid), times a uniform density over the $4 \times 4$~\pmUnits\ proper-motion box of Section~\ref{section:input_sample}. The truncated-Gaussian forms in $v_{\rm los}$ and $[\mathrm{Fe/H}]$ capture the actual peaked, non-uniform distribution of MW field stars in the Boo~III phase-space neighborhood; the uniform proper-motion background is appropriate because the proper-motion box was already tuned to bracket Boo~III, so the residual MW contamination is approximately flat in $(\mu_\alpha\cos\delta, \mu_\delta)$.

The GMM therefore samples a total of eleven parameters: the membership fraction $f_{\rm mem}$; the six systemic galaxy parameters ($\bar v_{\rm los}, \sigma_v, \overline{[\mathrm{Fe/H}]}, \sigma_{[\mathrm{Fe/H}]}, \bar\mu_\alpha\cos\delta, \bar\mu_\delta$); and the four MW field-background parameters ($\bar v^{\rm bg}, \sigma_v^{\rm bg}, \overline{[\mathrm{Fe/H}]}^{\rm bg}, \sigma_{[\mathrm{Fe/H}]}^{\rm bg}$). We adopt deliberately weakly informative priors that do not depend on any published Boo~III systemic value. All means are given linear-uniform priors and all dispersions log-uniform (Jeffreys') priors. The ranges of the four galaxy parameters are set to the orange box of Figure~\ref{fig:main_members} (Eqs.~\eqref{eq:boxcut_rv}--\eqref{eq:boxcut_feh}), while the background mean ranges span the full input box of Section~\ref{section:input_sample}. The full prior table is given in Table~\ref{tab:priors}.

\begin{table}[ht]
\centering
\caption{Priors used in the Boo~III GMM fit. \label{tab:priors}}
\footnotesize
\begin{tabular}{l l l}
\hline\hline
Parameter & Description & Prior \\
\hline
\multicolumn{3}{c}{\textit{Boo~III Galaxy component}} \\
\hline
$f_{\rm mem}$                          & galaxy mixing fraction                        & $\mathcal{U}(0,\,1)$                         \\
$\bar v_{\rm los}$                     & systemic $v_{\rm los}$ [\vlosUnits]           & $\mathcal{U}(170,\,210)$                     \\
$\log_{10}\sigma_v$                    & log intrinsic $v_{\rm los}$ dispersion         & $\mathcal{U}(-2.0,\,+1.3)$                   \\
$\overline{[\mathrm{Fe/H}]}$           & mean metallicity [dex]                          & $\mathcal{U}(-3.5,\,-1.2)$                   \\
$\log_{10}\sigma_{[\mathrm{Fe/H}]}$    & log intrinsic $[\mathrm{Fe/H}]$ dispersion                 & $\mathcal{U}(-2.0,\,+0.3)$                   \\
$\bar\mu_\alpha\cos\delta$             & systemic $\mu_\alpha\cos\delta$ [\pmUnits]      & $\mathcal{U}(-2.0,\,-0.5)$                   \\
$\bar\mu_\delta$                       & systemic $\mu_\delta$ [\pmUnits]                & $\mathcal{U}(-1.5,\, 0.0)$                   \\
\hline
\multicolumn{3}{c}{\textit{MW background component}} \\
\hline
$\bar v^{\rm bg}$                      & mean $v_{\rm los}$ [\vlosUnits]               & $\mathcal{U}(-600,\,+600)$                   \\
$\log_{10}\sigma_{v}^{\rm bg}$         & log $v_{\rm los}$ dispersion                     & $\mathcal{U}(+1.0,\,+2.5)$                   \\
$\overline{[\mathrm{Fe/H}]}^{\rm bg}$  & mean $[\mathrm{Fe/H}]$ [dex]                                & $\mathcal{U}(-4.0,\,\,0.0)$                  \\
$\log_{10}\sigma_{[\mathrm{Fe/H}]}^{\rm bg}$ & log $[\mathrm{Fe/H}]$ dispersion                       & $\mathcal{U}(-1.0,\,+0.3)$                   \\
\hline
\end{tabular}

\vspace{1ex}
\noindent\parbox{0.9\columnwidth}{\footnotesize \textbf{Notes.} 
Weakly informative priors. The four galaxy parameter ranges ($\bar v_{\rm los}$, $\bar\mu_\alpha\cos\delta$, $\bar\mu_\delta$, $\overline{[\mathrm{Fe/H}]}$) are the box-cut limits of Eqs.~\ref{eq:boxcut_rv}--\ref{eq:boxcut_feh} (the orange selection box drawn in Figure~\ref{fig:main_members}); the background mean ranges span the full input-box of Section~\ref{section:input_sample}. All dispersion priors are log-uniform (Jeffreys' prior for a scale parameter).}
\end{table}

\subsection{Fit Results}
We sample the posterior with \texttt{emcee} \citep{emcee}, using 64 walkers run for 8000 steps with a 2000-step burn-in.
We find a systemic heliocentric velocity of $\bar v_{\rm los} = 191.22^{+0.64}_{-0.80}$~\vlosUnits\ and a small velocity dispersion of $\sigma_v = 1.69^{+1.03}_{-0.85}$~\vlosUnits.
The systemic proper motion is $(\bar\mu_\alpha\cos\delta, \bar\mu_\delta) = (-1.163^{+0.017}_{-0.019}, -0.883^{+0.013}_{-0.014})$~\pmUnits, and the mean metallicity and intrinsic metallicity dispersion are $\overline{[\mathrm{Fe/H}]} = -2.34 \pm 0.11$~dex and $\sigma_{[\mathrm{Fe/H}]} = 0.36^{+0.10}_{-0.08}$~dex.
Our velocity dispersion is about six times smaller than the $10.7 \pm 3.5$~\vlosUnits\ reported by \citet{Carlin2018}, and our systemic velocity is offset from theirs by $\sim 6$~\vlosUnits; the metallicity is consistent with \citet{Carlin2018} within the joint uncertainties.  Our systemic proper motion is consistent with both the \gaia~DR2 measurement of \citet{Carlin2018} and the more precise \gaia~DR3 measurement of \citet{Pace2022}. A detailed comparison with prior spectroscopic measurements (\citealt{Carlin2009}, \citealt{Carlin2018}, and the recent \citealt{Geha2026} Keck/DEIMOS sample) is presented in Section~\ref{section:lit_compare}.
The posterior corner plot of the six Boo~III galaxy parameters plus the membership fraction $f_{\rm mem}$ is shown in Figure~\ref{fig:boo3_params} (the four MW field-background nuisance parameters are also fit but omitted for clarity), and all systemic parameters are summarized in Table~\ref{tab:params}.

\subsection{Membership}
The GMM yields an exceptionally clean membership classification.
Of the 120 input stars, 21 have $P_{\rm mem} > 0.95$ (20 with $P_{\rm mem} > 0.99$), while 99 have $P_{\rm mem} < 0.01$.
There are \emph{no} stars in the ambiguous $0.01 < P_{\rm mem} < 0.95$ regime.
This bimodal membership distribution reflects the fact that Boo~III is kinematically and chemically well separated from the MW field in all four observables.
We verify this classification by applying a hand-tuned box cut to the same input catalog.
The box cut selects stars satisfying all four of the conditions below (illustrated as orange dashed rectangles in the proper-motion and $v_{\rm los}$-vs-$[\mathrm{Fe/H}]$ panels of Figure~\ref{fig:main_members}):
\begin{align}
170 &< v_{\rm los} < 210~\mathrm{km~s^{-1}}, \label{eq:boxcut_rv}\\
-2.0 &< \mu_\alpha\cos\delta < -0.5~\mathrm{mas\,yr^{-1}}, \label{eq:boxcut_pmra}\\
-1.5 &< \mu_\delta <  0.0~\mathrm{mas\,yr^{-1}}, \label{eq:boxcut_pmdec}\\
[\mathrm{Fe/H}] &< -1.2. \label{eq:boxcut_feh}
\end{align}
The box cut recovers identically the same 21 stars as the GMM, confirming that the membership assignment is robust to the choice of method. Because this simple cut is a faithful surrogate for the full mixture-model classification, we adopt it as the search criterion for tidal-tail candidates beyond the 3 $r_h$ ellipse in Appendix~\ref{section:appendix_tail}.

\subsubsection{Updated centroid}\label{section:center}
With the 21 GMM members in hand we re-derive the Boo~III center by iterating the median (RA, Dec) of all members within $1\,r_h$ of the running center, using the 1\,$r_h$ ellipse in \citet{Moskowitz2020}. Starting from the \citet{Grillmair2009} discovery center, the center converges in three iterations to $(\alpha, \delta)_{\rm new} = (209.557^\circ \pm 0.3^\circ,\ 26.553^\circ \pm 0.3^\circ)$, with 11 members inside $1\,r_h$ at convergence. The new center is offset by $\xi = +13.8'$ (east) and $\eta = -14.8'$ (south), for a total shift of $20.2'$, or $0.61\,r_h$. The shift reflects an asymmetry in the spatial distribution of the spectroscopically-confirmed members within the $3\,r_h$ ellipse: a clump of high-probability members sits $\sim 20'$ south-east of the \citet{Grillmair2009} center (Figure~\ref{fig:main_members}, left panel). The $\pm 0.3^\circ$ uncertainty is a deliberately conservative choice that brackets the full shift from the literature \citet{Grillmair2009} centroid to our updated value; this $\pm 0.3^\circ$ width then propagates into the Monte Carlo (MC) orbit integrations of Section~\ref{section:orbit}. Because the \SSSSS\ target selection function is approximately uniform over the field of view, the offset is unlikely to be an artifact of selection bias and instead reflects a genuine shift of the Boo~III centroid as traced by its spectroscopically-confirmed members.  The earlier \citet{Carlin2009, Carlin2018} spectroscopic samples used a single $\sim 1\,r_h$ pointing centered near the \citet{Grillmair2009} discovery position and therefore have very limited leverage on the larger-scale centroid; their member sub-samples cluster tightly within $\lesssim 1\,r_h$ of that pointing by construction.  Boo~III's only published structural fit \citep{Moskowitz2020} adopts the \citet{Grillmair2009} centroid (itself a by-eye estimate from the SDSS data) as a fixed prior rather than fitting it.  Given these constraints, our updated centroid is best understood as a refined spatial center, anchored to the on-sky positions of the spectroscopically-confirmed members.

\subsubsection{RR Lyrae members}
We also examine the RRL within the 3 $r_h$ ellipse from \gaia\ DR3 (i.e. those removed from the input catalog in Section~\ref{section:input_sample}).
Three of them fall within the hand-tuned box-cut region (Eqs.~\ref{eq:boxcut_rv}--\ref{eq:boxcut_pmdec}) and have apparent magnitudes $g_0 \in [18.5, 19.5]$, consistent with an RRL at the approximate distance of Boo~III.  We therefore consider these three RRLs to be additional Boo~III members, shown as red triangles in Figure~\ref{fig:main_members}, and add them back to the member sample for the subsequent analysis, bringing the total to 24 stars.  The 21 red giant branch (RGB) members from the GMM fit are listed in Table~\ref{tab:boo3_S5_sample}; the three RRL members (with photometry, kinematics, and the per-star distance moduli derived from the \citealt{Garofalo2022} calibration) are listed separately in Table~\ref{tab:rrl_distances} (Section~\ref{section:distance}). As discussed in Section~\ref{section:distance}, Table~\ref{tab:rrl_distances} also includes the properties of two additional RRLs that are likely Boo~III members from the literature.
The observed properties of these 24 members (21 RGB + 3 RRL) in sky position, proper motion, CMD space, and $v_{\rm los}$ vs.\ $[\mathrm{Fe/H}]$ are summarized in Figure~\ref{fig:main_members}.

\begin{table*}[ht]
\centering
\caption{Spectroscopic RGB sample of Boo~III in \SSSSS\ DR2: 21 high-probability GMM members with $P_{\rm mem} > 0.95$. \label{tab:boo3_S5_sample}}
\footnotesize
\setlength{\tabcolsep}{4pt}
\begin{tabular}{rrrrrrrrc}
\hline\hline
Gaia Source ID & R.A. & Decl. & $g_0$ & $r_0$ & $v_{\rm los}$ & $[\mathrm{Fe/H}]$ & S/N & $P_{\rm mem}$ \\
 & (deg) & (deg) & (mag) & (mag) & (\vlosUnits) & (dex) &  &  \\
\hline
1258541764098047104 & 210.4413 & 26.0983 & 16.52 & 15.53 & 190.83$\pm$0.70 & -2.15$\pm$0.05 & 64.4 & 1.000 \\
1260240127310861184 & 211.2317 & 27.1648 & 19.52 & 18.99 & 188.41$\pm$5.05 & -2.27$\pm$0.38 & 3.5 & 1.000 \\
1450579200830093312 & 207.4225 & 26.4815 & 16.82 & 15.79 & 189.50$\pm$0.79 & -1.43$\pm$0.09 & 30.1 & 0.985 \\
1450593490185773568 & 208.3450 & 26.2653 & 16.88 & 15.95 & 192.29$\pm$0.97 & -1.81$\pm$0.10 & 16.1 & 1.000 \\
1450596243260368896 & 208.2133 & 26.3043 & 17.33 & 16.48 & 189.36$\pm$3.43 & -1.99$\pm$0.62 & 5.7 & 1.000 \\
1450639119918594560 & 208.2587 & 26.6050 & 18.78 & 18.05 & 187.16$\pm$5.75 & -2.18$\pm$0.75 & 3.7 & 1.000 \\
1450726977769346176 & 209.4887 & 26.0856 & 18.52 & 17.92 & 193.78$\pm$4.07 & -2.92$\pm$0.42 & 6.0 & 1.000 \\
1450752026018903168 & 209.5567 & 26.4448 & 18.76 & 18.20 & 191.48$\pm$3.12 & -2.59$\pm$0.33 & 5.2 & 1.000 \\
1450756767662816768 & 209.5261 & 26.5529 & 18.35 & 17.80 & 195.50$\pm$3.35 & -2.66$\pm$0.29 & 6.8 & 1.000 \\
1450757042540728960 & 209.5046 & 26.5773 & 19.33 & 18.79 & 180.45$\pm$7.01 & -2.49$\pm$0.46 & 3.4 & 1.000 \\
1450763227293876608 & 209.7604 & 26.4746 & 18.90 & 18.32 & 186.90$\pm$3.55 & -2.37$\pm$0.40 & 5.9 & 1.000 \\
1450763777049701888 & 209.7924 & 26.5161 & 18.83 & 18.23 & 193.22$\pm$3.12 & -2.54$\pm$0.43 & 6.4 & 1.000 \\
1450766839361412608 & 210.0826 & 26.5086 & 17.61 & 17.01 & 196.15$\pm$1.42 & -2.01$\pm$0.25 & 12.8 & 1.000 \\
1450782713560551808 & 209.6672 & 26.6952 & 16.27 & 15.19 & 192.83$\pm$0.72 & -2.40$\pm$0.04 & 47.5 & 1.000 \\
1450795289224753536 & 209.3370 & 26.4279 & 17.17 & 16.35 & 191.71$\pm$1.05 & -2.53$\pm$0.13 & 23.5 & 1.000 \\
1450820608056155648 & 209.2047 & 26.7368 & 19.67 & 19.16 & 182.31$\pm$5.80 & -2.39$\pm$0.61 & 2.5 & 0.999 \\
1450830267438457344 & 209.6123 & 26.7595 & 19.38 & 18.84 & 194.11$\pm$6.79 & -2.48$\pm$0.54 & 2.8 & 1.000 \\
1450833596037626752 & 209.3966 & 26.8129 & 18.39 & 17.77 & 189.15$\pm$2.08 & -3.16$\pm$0.39 & 6.7 & 1.000 \\
1451135553713792000 & 208.8271 & 27.6383 & 18.10 & 17.51 & 191.75$\pm$3.70 & -2.89$\pm$0.30 & 4.5 & 1.000 \\
1451865659498945664 & 207.9657 & 27.4836 & 19.79 & 19.26 & 183.76$\pm$6.38 & -2.16$\pm$0.83 & 4.1 & 0.998 \\
1452433527190255744 & 210.7094 & 27.6068 & 18.82 & 18.43 & 183.41$\pm$3.23 & -2.55$\pm$0.35 & 7.5 & 1.000 \\
\hline
\end{tabular}

\vspace{1ex}
\end{table*}

\begin{table*}[ht]
\centering
\caption{The five \gaia\ DR3 RRL stars associated with Boo~III used in the distance derivation (Section~\ref{section:distance}) \label{tab:rrl_distances}}
\footnotesize
\setlength{\tabcolsep}{3pt}
\begin{tabular}{rrrrrrrrrcc}
\hline\hline
Gaia Source ID & R.A. & Decl. & $g_0$ & $r_0$ & $G_{\rm int}$ & $v_{\rm los}$ & S/N & $A_G$ & $m-M$ & $D_\odot$ \\
& (deg) & (deg) & (mag) & (mag) & (mag) & (\vlosUnits) &  & (mag) & (mag) & (kpc) \\
\hline
1450735121027824000 & 209.8078 & 26.3404 & 19.01 & 18.74 & 18.704 & 192.08$\pm$5.66 & 3.5 & 0.034 & $18.39 \pm 0.14$ & $47.7 \pm 3.1$ \\
1450750170592551040 & 209.5292 & 26.3856 & 18.92 & 18.83 & 18.811 & 188.08$\pm$12.29 & 2.4 & 0.038 & $18.50 \pm 0.14$ & $50.0 \pm 3.2$ \\
1451041850411971840 & 209.3121 & 27.1200 & 18.90 & 18.66 & 18.714 & 184.33$\pm$7.99 & 3.0 & 0.048 & $18.39 \pm 0.14$ & $47.6 \pm 3.1$ \\
1450796178282259072\tablenotemark{a} & 209.2222 & 26.4655 & --- & --- & 18.761 & 191.83$\pm$4.86 & --- & 0.040 & $18.44 \pm 0.14$ & $48.8 \pm 3.1$ \\
1258556500130302080\tablenotemark{b} & 210.1439 & 25.9313 & --- & --- & 18.728 & 173.00$\pm$13.00 & --- & 0.034 & $18.42 \pm 0.14$ & $48.2 \pm 3.1$ \\
\hline
\multicolumn{8}{r}{Mean of 5 (Section~\ref{section:distance})} & --- & $18.43 \pm 0.09$ & $48.5 \pm 1.9$ \\
\hline
\end{tabular}

\vspace{1ex}
\noindent\parbox{\textwidth}{\footnotesize \textbf{Notes.} Three identified by cross-matching the \SSSSS\ DR2 footprint with the \gaia\ DR3 RRL catalog (top three rows).\\
$^{a}$ Identified as a Boo~III member in the Keck/DEIMOS catalog of \citet{Geha2026}; the listed $v_{\rm los}$ is from the Keck measurement.  \quad \\ $^{b}$ Identified as a Boo~III member in the MMT/Hectospec catalog of \citet{Carlin2018} (their identifier \texttt{BooIII\_RR1}); the listed $v_{\rm los}$ is from Hectospec.  \\
Note that $v_{\rm los}$ for RRLs is pulsation-phase-dependent \citep{2023MNRAS.519.5689M, 2025arXiv250402924M} and therefore is not used in the measurement of kinematic properties.\quad}
\end{table*}

\subsection{Comparison with Previous Spectroscopic Studies} \label{section:lit_compare}
We compare our \SSSSS\ DR2 + \gaia\ DR3 measurements with the three published spectroscopic studies of Boo~III: the discovery-era MMT/Hectospec analyses of \citet{Carlin2009} and \citet{Carlin2018}, and the more recent Keck/DEIMOS catalog of \citet{Geha2026}.

\citet{Carlin2009} observed Boo~III with MMT/Hectospec using a single pointing centered on the Boo~III centroid, and identified 20 probable members yielding $\bar v_{\rm los} = 197.5 \pm 3.8$~\vlosUnits\ and $\sigma_v = 14.0 \pm 3.2$~\vlosUnits.  \citet{Carlin2018} reanalyzed the same data with refined membership criteria and additional \gaia DR2 proper motions, lowering the dispersion to $10.7 \pm 3.5$~\vlosUnits\ for 16 retained members.  Their 16-member sample includes one \gaia\ DR3 RR Lyrae star (their \texttt{BooIII\_RR1}; the same star now used in our 5-RRL distance derivation, Section~\ref{section:distance}).  None of the 16 Carlin members have a counterpart within 2$''$ of any of our 21 GMM members: their sample is systematically fainter ($g_{\rm PanSTARRS}$ median 20.1, faintest 21.0, versus our median $g_0 = 18.5$, faintest 19.8), so most stars are below our effective S/N limit and the two datasets sample largely disjoint Boo~III populations.  Our measurements differ from \citet{Carlin2018} in two ways: $\sigma_v$ is about six times smaller, and $\bar v_{\rm los}$ is offset by $\sim 6$~\vlosUnits.  We note that overall, our sample has a higher per-star line-of-sight velocity precision ($\sigma_{v_{\rm los}} \lesssim 3$~\vlosUnits\ at $g \lesssim 19$, versus median $\sigma_{v_{\rm los}} \approx 10$~\vlosUnits\ for the Hectospec data), which may partially explain these discrepancies. 

The recent Keck/DEIMOS catalog of \citet{Geha2026} reports a Boo~III dispersion of $\sigma_v = 5.27^{+2.10}_{-1.67}$~\vlosUnits\ from 16 member stars (their $P_{\rm mem,novar} > 0.5$ sample, i.e.\ $P_{\rm mem} > 0.5$ \emph{and} not flagged as a velocity variable).\footnote{\citet{Geha2026} also identified one RRL member, but this RRL is \emph{not} in their 16-member $\sigma_v$ sample; it is present in their broader $P_{\rm mem} > 0.5$ catalog (20 members) but excluded by their velocity-variable cut.} 
A direct on-sky cross-match between our 21 GMM members and the \citet{Geha2026} sample finds no overlap, as the Keck/DEIMOS pointing covers a substantially smaller area than the \SSSSS\ footprint but reaches considerably deeper magnitudes.   Their systemic velocity, $\bar v_{\rm los} = 189.9 \pm 1.8$~\vlosUnits, agrees with our $191.22^{+0.64}_{-0.80}$~\vlosUnits\ within $\sim 0.7\sigma$.  Their velocity dispersion agrees with our $1.69^{+1.03}_{-0.85}$~\vlosUnits\ within $\sim 1.8\sigma$.

The updated $\sigma_v$ resolves a previously puzzling discrepancy with other tidally disrupting dwarfs. 
 With $\sigma_{v} = 1.69^{+1.03}_{-0.85}$ \vlosUnits\ at $M_V = -5.8$, Boo~III now fits more
naturally alongside Antlia~II ($\sigma_v = 5.98$~\vlosUnits, $M_V = -9.86$; \citealt{Ji2021}) and Crater~II
($\sigma_v = 2.51$~\vlosUnits, \citealt{Limberg2025}; $M_V = -8.2$, \citealt{Torrealba2016}) as examples of disrupting dwarf galaxies with low velocity dispersions. The implications of this low $\sigma_v$ for the system's dynamical state are discussed in Section~\ref{section:tidaldisrupt}.

\section{Heliocentric Distance to Boo~III} \label{section:distance}

The heliocentric distance to Boo~III has been measured several times since its discovery in SDSS by \citet{Grillmair2009}, who reported a BHB-based estimate of $\sim 46$~kpc. \citet{Carlin2009} adopted that distance for their initial spectroscopic study. \citet{Correnti2009} subsequently used a red-clump fit to SDSS photometry to derive an independent distance modulus $(m-M)_0 = 18.58 \pm 0.15$, corresponding to $D_\odot = 52.0 \pm 3.6$~kpc, roughly 6~kpc larger than the BHB-based estimate. \citet{Sesar2014} then identified an RRL star coincident with the Boo~III field whose distance is $D_\odot = 46 \pm 2$~kpc and agrees with the lower value of \citet{Grillmair2009}. The currently adopted value, $D_\odot = 46.5 \pm 2.0$~kpc, comes from the analysis of \citet{Carlin2018}, who performed a least-squares fit of the M15 horizontal-branch ridgeline to the Boo~III BHB stars in PanSTARRS-1 photometry. \citet{Pace2022} adopt this value, while \citet{Moskowitz2020} adopt the \citet{Correnti2009} red-clump value of $52 \pm 3.6$~kpc directly. The $\sim 6$~kpc tension between the red-clump route and the RRL/isochrone route is at the $\sim 1.5\sigma$ level and is unresolved.

We also derive an independent distance to Boo~III from five \gaia~DR3 RRL stars associated with the system (Table~\ref{tab:rrl_distances}): the three RRLs identified directly in the \SSSSS\ footprint via the box-cut and \gaia DR3 RRL catalog cross-match of Section~\ref{section:membership}, plus one additional RRL match each in the published \citet{Carlin2018} and \citet{Geha2026} spectroscopic member samples. For each RRL we use the intensity-averaged \gaia~$G$ magnitude (\texttt{int\_average\_g}). For the extinction we adopt the \citet{Schlegel1998} $E(B-V)$ rescaled by the $0.86$ factor of \citet{Schlafly2011} and convert to $A_G$ using the color-dependent extinction law of \citet{Babusiaux2018}. The absolute magnitude is computed from the period-independent $M_G$--$[\mathrm{Fe/H}]$ calibration of \citet{Garofalo2022} (their Eq.~19): $M_G = (0.33 \pm 0.02)\,[\mathrm{Fe/H}] + (1.05 \pm 0.03)$, with hierarchical-Bayesian intrinsic dispersion $\sigma_{\rm int} = 0.03$~mag. Evaluated at Boo~III's system-mean $[\mathrm{Fe/H}] = -2.34$ (the per-RRL spectroscopic $[\mathrm{Fe/H}]$ is unreliable due to pulsation-phase broadening of the equivalent widths), this gives $M_G = +0.28$~mag. 

The mean across the five stars yields $\langle m-M \rangle = 18.43 \pm 0.09$, corresponding to $\langle D_\odot \rangle = 48.5 \pm 1.9$~kpc.  The five per-star distances span 47.6--50.0~kpc with an internal scatter of $1.0$~kpc.  The 0.09 mag uncertainty in distance modulus comes from a combination of the statistical uncertainties $\sigma_{m-M}^{\rm stat} = 0.123/\sqrt{5} = 0.055$~mag and systematic uncertainties $\sigma_{m-M}^{\rm sys} = 0.066$~mag.  The dominant per-star statistical term is the propagation of the galaxy-wide $[\mathrm{Fe/H}]$ dispersion $\sigma_{[\mathrm{Fe/H}]} = 0.36$~dex through the calibration slope (i.e., ${\rm slope} \times \sigma_{[\mathrm{Fe/H}]} = 0.33 \times 0.36 = 0.119$~mag; each RRL is an independent draw from the galaxy $[\mathrm{Fe/H}]$ distribution, so the term reduces as $1/\sqrt{N}$); adding the calibration's intrinsic-dispersion contribution $\sigma_{\rm int} = 0.03$~mag in quadrature gives the total per-star statistical $0.123$~mag.  The systematic budget combines three terms that are shared by all five RRLs and do not reduce with $N$: the slope-times-$[\mathrm{Fe/H}]$ term ($\sigma_{\rm slope} \times |[\mathrm{Fe/H}]| = 0.02 \times 2.34 = 0.047$~mag), the intercept uncertainty ($\sigma_{\rm intercept} = 0.03$~mag), and the propagation of the uncertainty on the galaxy-mean $[\mathrm{Fe/H}]$ through the slope (${\rm slope} \times \sigma_{\overline{[\mathrm{Fe/H}]}} = 0.33 \times 0.11 = 0.036$~mag), which sum in quadrature to 0.066~mag. 

For comparison, repeating the same analysis with the earlier \citet{Muraveva2018} $G$-band calibration based on \gaia DR2, i.e. $M_G = (0.32 \pm 0.04)\,[\mathrm{Fe/H}] + (1.11 \pm 0.06)$ with intrinsic dispersion $\sigma_{\rm int} = 0.17$~mag, yields $\langle m-M \rangle_{\rm Mur18} = 18.34 \pm 0.15$, or $\langle D_\odot \rangle_{\rm Mur18} = 46.6 \pm 3.2$~kpc. This is consistent with the \citet{Garofalo2022} value within their joint uncertainty, and agrees almost exactly with the $46.5 \pm 2.0$~kpc value of \citet{Carlin2018}. The $\sim 1.8$~kpc offset between the two RRL calibrations reflects the move from a DR2-based calibration in \citet{Muraveva2018} to an EDR3-based calibration in \citet{Garofalo2022}. The DR2-to-EDR3 shift corresponds to a $\sim 0.08$~mag systematic offset in $M_G$ between the two calibrations and places the EDR3-based value between \citet{Carlin2018} and the red-clump value of \citet{Correnti2009} adopted by \citet{Moskowitz2020}, agreeing with both at $\lesssim 1\sigma$. We adopt the \citet{Garofalo2022}-based 5-RRL value, $D_\odot = 48.5 \pm 1.9$~kpc, throughout the present analysis as it incorporates the more recent EDR3 parallax solution and has both tighter statistical and systematic precision than \citet{Muraveva2018}.

\section{Orbit and Stream Simulations} \label{section:sim}

\subsection{Orbit} \label{section:orbit}
We integrate the orbit of Boo~III using the systemic velocity and proper motion measured in Section~\ref{section:membership}, combined with the on-sky position in Section~\ref{section:center} and heliocentric distance in Section~\ref{section:distance} listed in Table~\ref{tab:params}. The integration is performed with the \texttt{galpy} Python package \citep{Bovy2015}, adopting the \texttt{McMillan17} potential \citep{McMillan2017}. This potential assumes a solar radius of $R_0 = 8.21$~kpc and circular velocity of $V_0 = 233.1$ \vlosUnits. We adopt the solar peculiar motion $(U,V,W)_\odot = (+12.9, +12.5, +7.78)$ \vlosUnits chosen so that the Sun's total velocity in the Galactocentric frame matches the \texttt{astropy}~v4.0 default $(V_x, V_y, V_z)_\odot = (12.9, 245.6, 7.78)$ \vlosUnits \citep{Drimmel2018}, giving a total solar velocity of $V_{\phi,\odot} = 245.6$ \vlosUnits with respect to the Galactic center. The same convention is used for the LMC orbit integration described below.

We include the gravitational influence of the LMC, which induces a reflex motion of the MW barycenter that must be accounted for in the orbit integration \citep{Erkal2019, Petersen2021}. We model the LMC as a moving Hernquist potential \citep{Hernquist1990} added to the \texttt{McMillan17} base potential, and include the resulting non-inertial frame correction using \texttt{galpy}’s \texttt{NonInertialFrameForce} to account for the MW's barycentric acceleration \citep{2025ApJ...984...56A}.

We adopt the LMC parameters from \citet{Pace2022}: a total mass of $1.38 \times 10^{11}~{\rm M}_\odot$ \citep{Erkal2019} and an enclosed mass of $1.7 \times 10^{10}~{\rm M}_\odot$ at 8.7~kpc \citep{vanderMarel2014}, in good agreement with the more recent \citet{Cullinane2020} estimate of $(1.8 \pm 0.3) \times 10^{10}~{\rm M}_\odot$ at $\sim 10.5$~kpc.  Solving the \citet{Hernquist1990} $M(<r) = M_{\rm tot}\,r^2/(r+a)^2$ relation for these values yields a scale radius $a = 16.09$~kpc and a half-mass radius $r_{\rm hm} = a(1+\sqrt{2}) = 38.84$~kpc, which we adopt in the LMC potential below. For the LMC's present-day phase-space coordinates we adopt the proper motion of \citet{Kallivayalil2013}, the distance of \citet{Pietrzynski2019}, the systemic velocity of \citet{vandermarel2002}, and the LMC center of \citet{vanderMarel2014}; all of these inputs are summarized in Table~\ref{tab:streamspray_params}.

\begin{figure*}
    \centering
    \includegraphics[width=\textwidth]{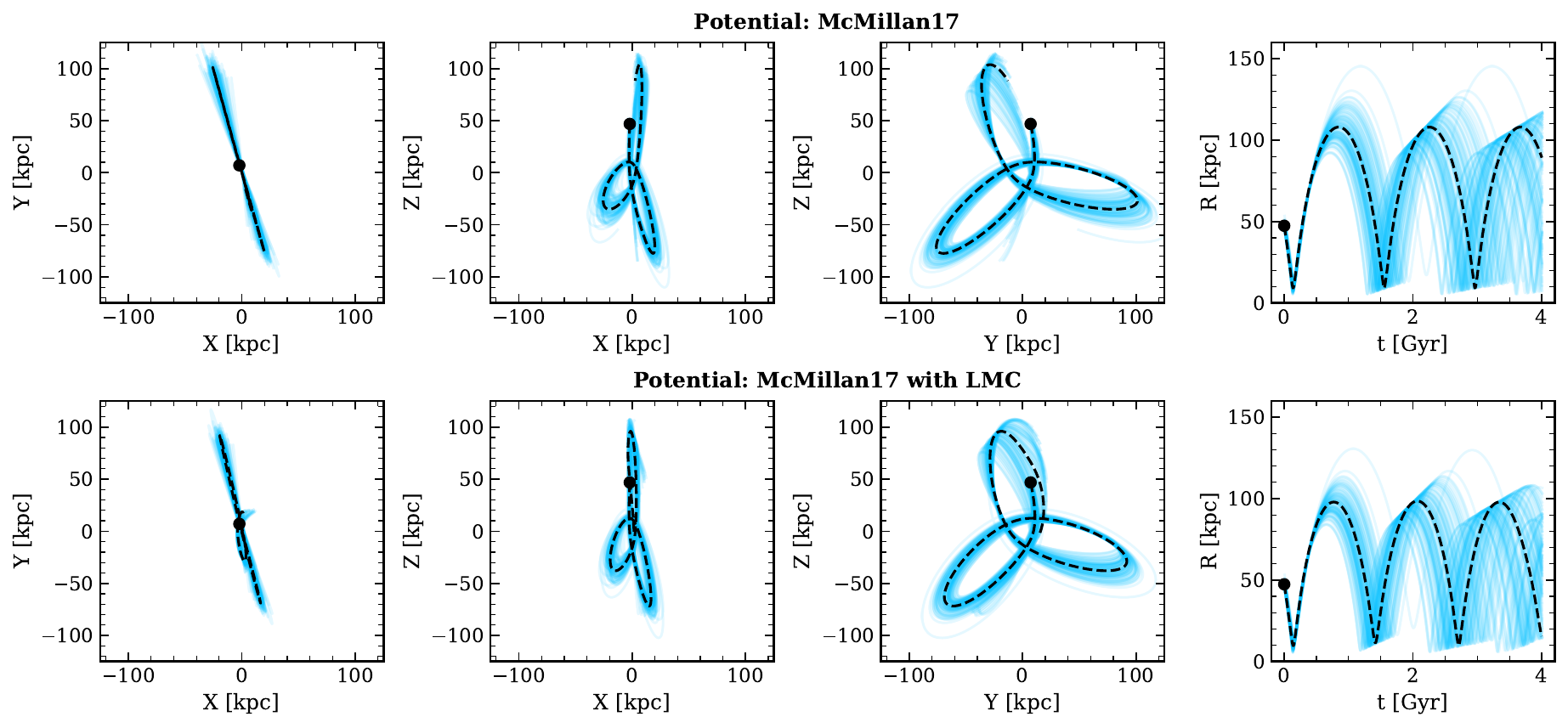}
    \caption{Boo~III orbit integrated 4~Gyr backwards from the present, with and without the LMC, using the \texttt{McMillan17} model as base potential. The orbits are plotted in Galactocentric Cartesian coordinates: the dashed black line shows the integrated orbit using the median 6D parameters, while the semi-transparent blue lines show 1000 realizations drawn from the 6D error distribution to indicate uncertainty on the orbit. The black point marks the present-day position. Top row: orbits integrated in the \texttt{McMillan17} potential (without the LMC). Bottom row: orbits integrated in the \texttt{McMillan17} potential, the LMC as a moving Hernquist potential, and the corresponding MW barycentric acceleration. The rightmost column shows the Galactocentric distance vs. lookback time, where $t = 0$~Gyr is the present.}
    \label{fig:orbits}
\end{figure*}

\begin{figure*}
    \centering
    \includegraphics[width=\textwidth]{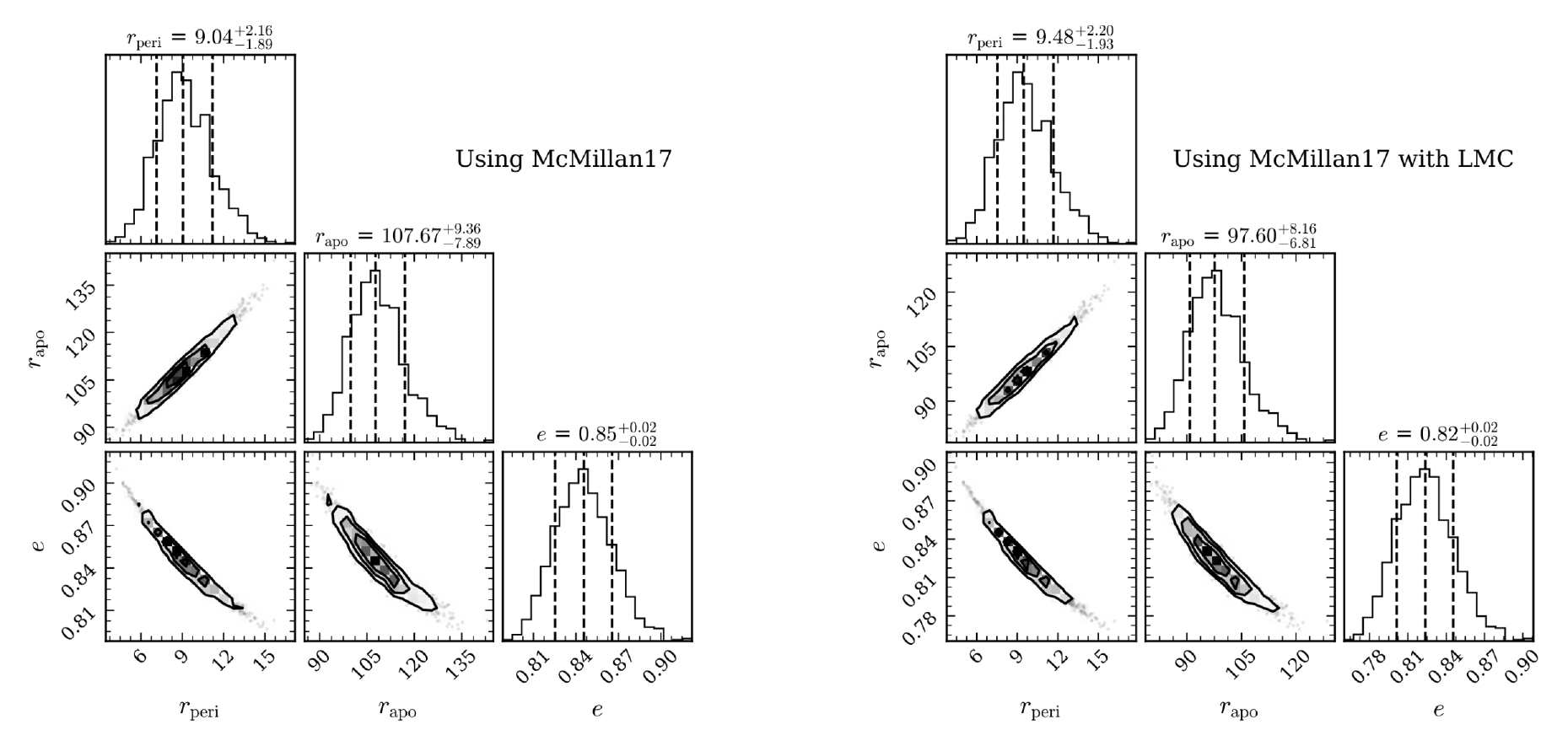}
    \caption{Posterior distributions of the Boo~III orbital parameters: pericenter and apocenter (both in kpc), and eccentricity. The nominal value and its uncertainties are obtained from the median, 16th percentile, and 84th percentile of the 1000 realizations of the orbit parameters. Left: the orbital parameters with \texttt{McMillan17} only. Right: the orbital parameters with the modified potential that accounts for the LMC's effects. The apocenter value decreases when we switch from a static, ordinary MW potential to one that includes the LMC and the MW reflex motion.}
    \label{fig:orbit_params}
\end{figure*}

Figure~\ref{fig:orbits} shows the resulting orbits integrated 4~Gyr backwards, with and without the LMC. To propagate observational uncertainties into the inferred orbital parameters, we generate 1000 MC realizations of the 6D phase-space coordinates by drawing each component independently from a Gaussian distribution centered on the median value and with width equal to the $1\sigma$ uncertainty listed in Table~\ref{tab:params}. The pericenter, apocenter, and eccentricity are taken as the median of these realizations, with the 16th and 84th percentiles as uncertainties (Figure~\ref{fig:orbit_params}). Including the LMC reduces the apocenter and slightly increases the pericenter. In addition, we run a per-component MC, perturbing each of the six phase-space coordinates in turn while holding the others fixed, to identify which observational uncertainty dominates the inferred orbit; the heliocentric distance is by far the dominant driver (Appendix~\ref{section:appendix_6dmc}).
We also present the orbital parameters in Table~\ref{tab:params}.

As seen in Figure~\ref{fig:orbits}, the motion of Boo~III is largely confined to the Galactocentric $YZ$-plane, indicating a polar orbit, with an orbital inclination of $i = (89.5 \pm 0.5)^\circ$ between its angular momentum and the Galactic plane (Table~\ref{tab:params}; $90^\circ$ corresponds to an exactly polar orbit). In the static MW potential, Boo~III passed its most recent pericenter $0.139 \pm 0.006$~Gyr ago at $r_{\rm peri} \approx 9.0^{+2.2}_{-1.9}$~kpc; including the LMC gives a similar time of $0.137 \pm 0.006$~Gyr ago at $r_{\rm peri} \approx 9.5^{+2.2}_{-1.9}$~kpc (Table~\ref{tab:params}).  Both the small pericenter distance and the recent passage are essential for the tidal-disruption interpretation discussed in Section~\ref{section:tidaldisrupt}. The primary effect of the LMC is to reduce the apocenter distance (Figure~\ref{fig:orbit_params}), confining Boo~III to a tighter orbit.

For comparison, we repeat the orbit integration in the lighter \texttt{MWPotential2014} \citep{Bovy2015}. The pericenter and apocenter depend on the assumed MW potential: the lighter halo produces more modest increases in the pericenter and eccentricity but yields a much larger apocenter, nearly doubling it (from $r_{\rm apo} \approx 98$ to $\approx 164$~kpc with the LMC) and lengthening the orbital period from $\sim 1.3$ to $\sim 3$~Gyr, while the qualitative character of the orbit (polar, highly eccentric, with a recent pericentric passage) is unchanged. The full trajectories, posteriors, and numerical comparison are given in Appendix~\ref{section:appendix_mwpot}.

\subsection{Stream}\label{section:stream}
To model the tidal stream expected from Boo~III's disruption, we use the particle spray code \texttt{fardal15spraydf} \citep{Qian2022} in \texttt{galpy} \citep{Bovy2015}, an implementation of the \citet{Fardal2015} spray model.  This method integrates the progenitor's orbit from a chosen disruption time to the present, continuously releasing particles at the L1 and L2 Lagrange points to build up a mock stream.  We note that because particles near the progenitor are dominated by this stochastic Lagrange-point release, the spray model is not a reliable predictor of the kinematics in the immediate vicinity of Boo~III; we model that region instead with the restricted $N$-body simulations of Section~\ref{section:nbody}.

The input parameters for the particle spray model are the progenitor mass, the 6D phase-space coordinates, the disruption time, and the Galactic potential. For the progenitor mass we adopt the \citet{Wolf2010} half-light dynamical mass $M_{1/2} \approx 1.24 \times 10^{6}\,{\rm M}_\odot$ derived in Section~\ref{section:tidaldisrupt} (Eq.~\eqref{eq:Mhalf}), which combines our GMM line-of-sight dispersion with the literature angular size and our new RRL distance.  We adopt a disruption time of 3~Gyr and use the same MW+LMC potential as in the orbit integration. All fiducial stream parameters are listed in Table~\ref{tab:streamspray_params}.

\begin{figure*}[ht]
    \centering
    \includegraphics[width=\textwidth]{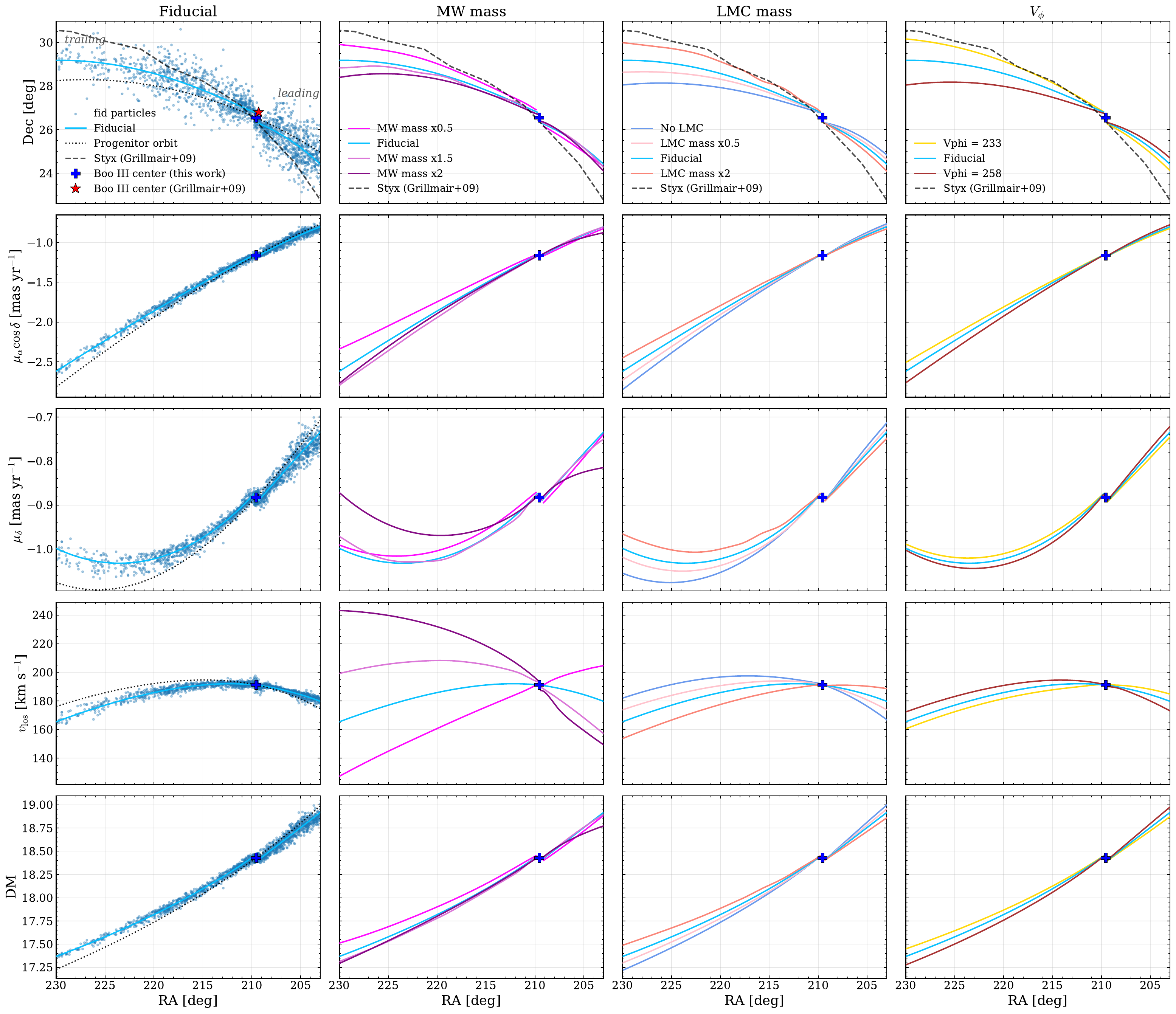}
    \caption{Boo~III predicted tidal stream from particle spray simulations using \texttt{galpy}'s \texttt{fardal15spraydf}, with progenitor mass $M_{1/2} = 1.24 \times 10^{6}\,{\rm M}_\odot$, 1000 particles per arm, $t_{\rm disrupt} = 3$~Gyr, and \texttt{galpy}'s \texttt{streamTrack} interpolation evaluated with \texttt{smoothing\_factor}~$= 4$. Rows: Dec,  $\mu_\alpha\cos\delta$, $\mu_\delta$, $v_{\rm los}$, distance modulus, all vs.~RA. The blue $+$ marker indicates the updated Boo~III center (this work) in every panel; the red star in the top-left (Fiducial) panel marks the literature \citet{Grillmair2009} center for direct comparison. The black dashed curve in the top (Dec) row is the  Styx stream track from \citet[priv.\ comm.]{Grillmair2009}, shown in all columns. The updated Boo~III centroid adopted in this work lies closer to this track than the discovery centroid (Table~\ref{tab:params}). \emph{Column 1 (Fiducial):} blue dots are the raw spray particles from the fiducial \texttt{McMillan17}~+~LMC run, and the black dotted curve overlaid in every row is the integrated Boo~III progenitor orbit with the same potential,  shown for direct comparison with the spray streamTrack curves in cyan; subsequent columns show the leading and trailing streamTrack curves only. \emph{Column 2 (MW mass scan):} MW halo amplitude scaled by $\times 0.5$, $\times 1$ (fiducial), $\times 1.5$, $\times 2$. Decreasing the MW mass brings the predicted stream closer to the Styx track. \emph{Column 3 (LMC mass scan):} no-LMC, $\times 0.5$, $\times 1$ (fiducial), $\times 2$ LMC mass. Without the LMC, the stream is farther from Styx; doubling the LMC mass shifts it closer. \emph{Column 4 ($V_\phi$ scan):} Sun's total $V_\phi = 233$ and $258$~\vlosUnits, bracketing the fiducial $V_\phi = 245.6$~\vlosUnits\ (\texttt{astropy}~v4.0). The lower $V_\phi$ brings the stream model closer to Styx.}

    \label{fig:stream_comparison}
\end{figure*}

To study the sensitivity of the predicted stream to the assumed potential, we generate several variations by changing one parameter at a time relative to the fiducial model:
\begin{itemize}[itemsep=0pt]
    \item LMC mass halved, doubled, or removed from potential entirely
    \item MW halo mass halved,  increased by a factor of 1.5, or doubled
    \item solar velocity decreased to $V_\phi = 233$ \vlosUnits, or increased to $V_\phi = 258$ \vlosUnits
\end{itemize}
Figure~\ref{fig:stream_comparison} compares the fiducial particle spray model and these one-parameter-at-a-time variations to the on-sky Styx stream track from \citet{Grillmair2009} (the track data themselves are not published in that paper and were supplied by C.~Grillmair, priv.~comm.). For each model run we summarize the spray-particle cloud with \texttt{galpy}'s \texttt{streamTrack} estimator, which fits a smooth one-dimensional centroid track to the leading and trailing arms of the cloud as a function of arc length, providing an interpolated stream centerline that can be over-plotted on the data; the leftmost (`Fiducial') column shows the raw spray particles together with this interpolated track, and subsequent columns show the \texttt{streamTrack} centerlines only.

For each stream model we plot the predicted stream tracks of \emph{five} observables (declination, $\mu_\alpha\cos\delta$, $\mu_\delta$, heliocentric $v_{\rm los}$, and distance modulus), all as functions of right ascension along the trailing tail, against the on-sky position of the Styx track in the top row. The three one-parameter scans show how the predicted stream moves with: (i) the MW halo mass (column 2: $\times 0.5$, $\times 1$, $\times 1.5$, $\times 2$ of \texttt{McMillan17}); (ii) the LMC mass (column 3: no-LMC, $\times 0.5$, $\times 1$, $\times 2$ of the fiducial value); and (iii) the Sun's circular speed $V_\phi$ (column 4: $233$~\vlosUnits, fiducial $245.6$~\vlosUnits, $258$~\vlosUnits). Of the five observables, declination and proper motion are partially degenerate between the three changes, but the $v_{\rm los}$ track, particularly within $\pm 10^\circ$ of Boo~III in RA, separates the model variations most cleanly.

The simulated streams follow tracks broadly consistent with the Styx stream, providing additional support for the Boo~III--Styx connection beyond the proper motion alignment noted by \citet{Carlin2018}. The fiducial (\texttt{McMillan17}~+~LMC) prediction lies close to, but slightly south of, the Styx track. This remaining offset is not necessarily a problem for the association: a lighter MW halo, a heavier LMC, a smaller solar $V_\phi$, or some combination of the three can bring the predicted tail closer to the observed Styx track under our current 6D measurement of the progenitor\footnote{The predicted tail location also depends on Boo~III's own 6D phase-space coordinates; the corresponding orbital sensitivities are studied separately in Appendix~\ref{section:appendix_6dmc}.}. We also caution that any such comparison should be made against the debris track rather than the progenitor's orbit: on such a highly eccentric orbit the tidal debris does not trace the orbit, and, as the fiducial column of Figure~\ref{fig:stream_comparison} shows, the trailing arm is increasingly displaced from the orbit with angular distance from the progenitor, toward the Styx track. Testing the association against Boo~III's orbit alone would therefore overstate the mismatch. Since the model variations separate most cleanly in the line-of-sight velocity, future spectroscopic observations of Styx stream members would be particularly valuable for jointly constraining the MW halo mass and the LMC mass, which are presently degenerate with the solar reflex velocity and with Boo~III's own 6D phase-space coordinates in our stream-track fits.

\begin{table*}[!t]
\centering
\caption{Fiducial input parameters for the \texttt{galpy} particle-spray stream model (Section~\ref{section:stream}). The MW potential, the LMC mass and structure, and the LMC present-day phase space are also adopted in the restricted $N$-body simulations (Section~\ref{section:nbody_setup}). \label{tab:streamspray_params}}
\footnotesize
\begin{tabular}{l l l}
\hline\hline
Parameter & Value & Reference \\
\hline
Progenitor mass                 & $1.24 \times 10^6~{\rm M}_\odot$ & this work \\
Boo~III orbit 6D                & See Table~\ref{tab:params}       & this work \\
Time of disruption              & 3~Gyr                            & \\
MW potential                    & \texttt{McMillan17}~+~LMC        & \citet{McMillan2017} \\
LMC mass                        & $1.38 \times 10^{11}~{\rm M}_\odot$ & \citet{Erkal2019} \\
LMC scale radius $a$            & 16.09~kpc                        & \citet{vanderMarel2014} \\
LMC half-mass radius $r_{\rm hm}$ & 38.84~kpc                      & derived \\
Solar $V_\phi$                  & $245.6~\mathrm{km~s}^{-1}$       & \texttt{astropy}~v4.0 default \\
\hline
\multicolumn{3}{l}{\textit{LMC present-day phase space}}\\
\quad $(\alpha,\delta)_{\rm LMC}$ & $(78.76^\circ,\ -69.19^\circ)$ & \citet{vanderMarel2014} \\
\quad $D_{\odot,\,\rm LMC}$      & $49.59$~kpc                      & \citet{Pietrzynski2019} \\
\quad $(\mu_{\alpha}\cos\delta,\ \mu_{\delta})_{\rm LMC}$ & $(1.910,\ 0.229)$~\pmUnits & \citet{Kallivayalil2013} \\
\quad $v_{\rm los,\,LMC}$        & $262.2~\mathrm{km~s}^{-1}$       & \citet{vandermarel2002} \\
\hline
\end{tabular}
\end{table*}

\section{Discussion} \label{section:discussion}
\subsection{Tidal Disruption of Boo~III} \label{section:tidaldisrupt}

We assess whether Boo~III is currently undergoing tidal disruption using three complementary diagnostics, presented in the order in which we treat them below: (i) the timing and geometry of its most recent pericentric passage; (ii) its tidal radius (also known as the Jacobi radius) at pericenter, computed from the half-light dynamical mass and the enclosed MW mass profile; and (iii) the equivalent density criterion, comparing its mean density within the half-light radius to the enclosed MW density at pericenter. All three independently favor active tidal stripping, and we discuss them in turn.

\subsubsection{Recent pericentric passage} Section~\ref{section:orbit} found that Boo~III's last pericentric passage was just $\sim 0.14$~Gyr ago, at a Galactocentric distance of $r_{\mathrm{peri}} \approx 9.5$~kpc (with LMC; Table~\ref{tab:params}).  This is a small pericenter for a satellite at the present-day Galactocentric distance of $\sim 50$~kpc, and a very recent passage compared with Boo~III's $\sim 1.3$~Gyr orbital period in the \texttt{McMillan17}~+~LMC integration. Both are necessary preconditions for an actively stripping picture: a system with a substantially larger pericenter would never experience a strong instantaneous tidal field, and a system whose last pericentric passage occurred many Gyr ago would by now have had time to relax back to a quasi-equilibrium configuration.

\begin{figure}
    \centering
    \includegraphics[width=\columnwidth]{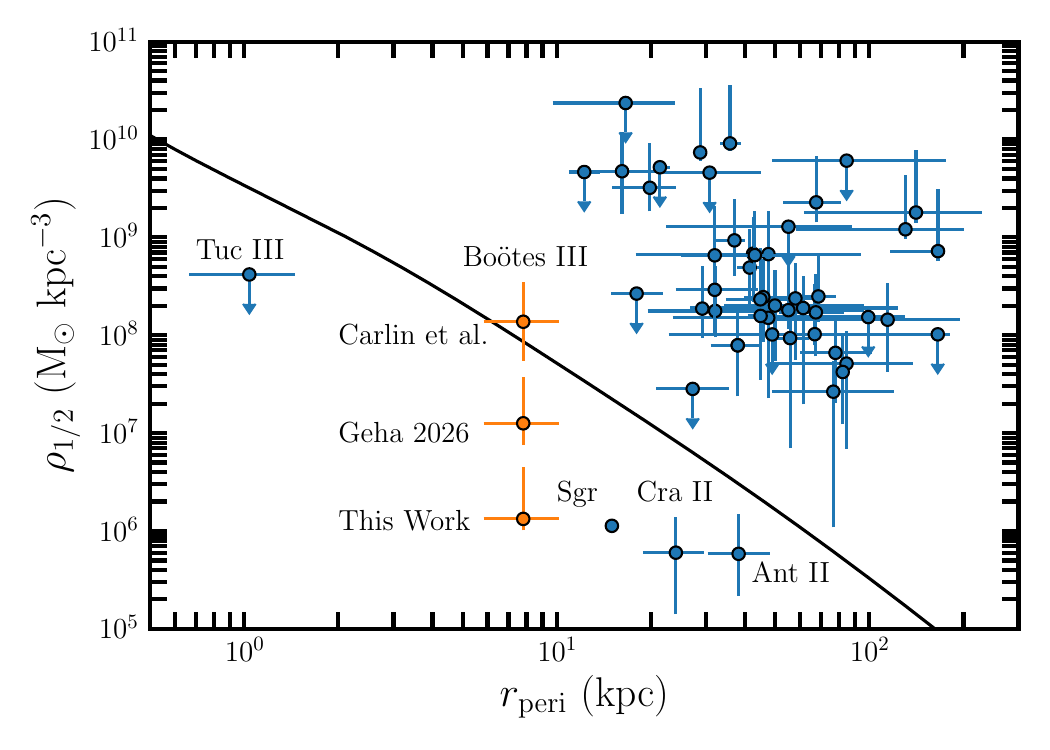}
    \caption{Comparison of the orbital pericenter and the average density within the half-light radius for MW dwarf galaxies (blue) and Boo~III (orange), adapted from Figure~5 of \citet{Pace2022}. The black line shows twice the average enclosed MW density as a function of radius; objects below this line have tidal radii smaller than their half-light radii and are expected to be tidally disrupting. We highlight the comparison systems near or below the line (from left to right: Tucana~III, Sgr, Crater~II, and Antlia~II). For Boo~III (orange) we show three half-light densities at its pericenter, each from a different velocity-dispersion measurement and labeled by source: $\sigma_v = 10.7 \pm 3.5$~\vlosUnits\ from \citet{Carlin2018}, $\sigma_v = 5.27^{+2.10}_{-1.67}$~\vlosUnits\ from the Keck/DEIMOS sample of \citet{Geha2026}, and our updated $\sigma_v = 1.69^{+1.03}_{-0.85}$~\vlosUnits\ (this work). As the measured dispersion has decreased, the implied half-light density drops by nearly two orders of magnitude, moving Boo~III from above the line (Carlin~et~al.) to well below it (Geha~et~al.\ and this work), increasingly consistent with active tidal disruption.}
    \label{fig:pericenter_density}
\end{figure}

\subsubsection{Tidal radius far smaller than the half-light radius}
The tidal radius for a satellite on a circular orbit in a host with a flat rotation curve \citep[][\S19.4.2]{Bovy2026} is
\begin{equation}\label{eq:rt}
    r_t = r \left( \frac{m}{2 M(<r)} \right)^{1/3},
\end{equation}
where $r$ is the satellite's distance from the MW, $m$ is the satellite mass, and $M(<r)$ is the enclosed MW mass.  We evaluate Equation~\eqref{eq:rt} for Boo~III at its most recent pericenter. 
For $m$ we adopt the satellite mass within its half-light radius, and for $M(<r)$ the MW mass enclosed at the pericentric distance.

The half-light dynamical mass is computed via the \citet{Wolf2010} estimator
\begin{equation}\label{eq:Mhalf}
    M_{1/2} \;=\; 930 \, \sigma_v^2 \, r_h
    \;\approx\; 1.24 \times 10^{6}\,{\rm M}_\odot,
\end{equation}
where the constant $930$ has units of ${\rm M}_\odot \, \rm pc^{-1} \, (km \, s^{-1})^{-2}$, and we use our mixture-model line-of-sight dispersion $\sigma_v = 1.69$~\vlosUnits\ (Section~\ref{section:membership}) together with the (circularized) half-light radius $r_h = 466$~pc (from the \citet{Moskowitz2020} angular size combined with our updated distance of $48.5$~kpc; see Table~\ref{tab:params}), matching the convention of \citet{Pace2022}. This $M_{1/2}$ value is the canonical half-light mass used throughout the paper, including as the progenitor mass for the particle spray stream model of Section~\ref{section:stream}. The implied dynamical mass-to-light ratio within the half-light radius is $(M/L_V)_{1/2} \sim 140\,{\rm M}_\odot/{\rm L}_\odot$ albeit with a large uncertainty. The 16-84 percent credible interval is $\sim 30-390{\rm M}_\odot/{\rm L}_\odot$, primarily due to uncertainty on the velocity dispersion. The large mass-to-light ratio confirms that Boo~III remains strongly dark-matter dominated even in its tidally-stripped state. We caution that this estimator assumes the system is in dynamical equilibrium and supported by an isotropic velocity dispersion; for an actively disrupting galaxy like Boo~III these assumptions are not strictly satisfied, so $M_{1/2}$ (and quantities derived from it) should be regarded as an indicative dynamical mass rather than a rigorous equilibrium measurement, though it remains the best estimate available from the current data.

For the MW enclosed mass we use the spherically equivalent value
\begin{equation}
    M(<r) \;=\; v_c^2(r)\,r/G,
\end{equation}
evaluated in the static \texttt{McMillan17} potential. Although the orbit integration in Section~\ref{section:orbit} also includes the LMC's reflex acceleration on the MW, the LMC perturbation is small near the inner Galaxy and the local circular velocity at $r \sim 9$~kpc is well described by \texttt{McMillan17} alone. Adopting the with-LMC pericentric distance from Section~\ref{section:orbit}, $r_{\mathrm{peri}} = 9.48$~kpc, we obtain $v_c(r_{\mathrm{peri}}) = 233$~\vlosUnits\ and $M(<r_{\mathrm{peri}}) \approx 1.20 \times 10^{11}\,{\rm M}_\odot$.

Substituting into Equation~\eqref{eq:rt} yields $r_t \approx 164$~pc at pericenter, or $\approx 12'$ on the sky at Boo~III's distance, which is $\sim 0.35\,r_h$, about a factor of three smaller than the half-light radius. The dominant uncertainty is the velocity dispersion: propagating the $1\sigma$ range $\sigma_v = 1.69^{+1.03}_{-0.85}$~\vlosUnits\ alone gives $r_t \approx 100$--$225$~pc ($\approx 0.2$--$0.5\,r_h$), all substantially smaller than $r_h$. For comparison, at Boo~III's present-day Galactocentric distance ($r \approx 47$~kpc), where the tidal field is far weaker, Equation~\eqref{eq:rt} still gives $r_t \approx 510$~pc $\approx 1.1\,r_h$, barely larger than the half-light radius. Even at this weaker-field location, then, Boo~III is only marginally able to retain material within $r_h$, so it remains strongly affected by Galactic tides and is unlikely to be in dynamical equilibrium. We caution that Equation~\eqref{eq:rt} assumes a circular orbit and dynamical equilibrium, and we evaluate it using the instantaneous tidal field at pericenter; these assumptions are not strictly valid for an actively disrupting system on an eccentric orbit, so $r_t$ should be regarded as an order-of-magnitude estimate of the bound-region size at closest approach. Equation~\eqref{eq:rt} also treats the satellite mass as a single value $m$ (the half-light dynamical mass) rather than modeling Boo~III's extended internal mass profile; the precise tidal radius depends on that internal mass distribution, which we do not attempt to model here.

\subsubsection{Density below the MW envelope}
The result $r_t < r_h$ for Boo~III can also be phrased as a density condition.  When the half-light radius equals the tidal radius, Equation~\eqref{eq:rt} reduces to $\bar\rho_{1/2} = 2\,\bar\rho_{\mathrm{MW}}(r_{\mathrm{peri}})$, i.e.\ a satellite is expected to be significantly affected by tidal stripping when its average internal density falls below twice the mean enclosed MW density at its orbital pericenter \citep{Pace2022}. Figure~\ref{fig:pericenter_density}, adapted from fig.~5 (left panel, with LMC) of \citet{Pace2022} with Boo~III updated to our new measurements, shows this diagnostic (the black line marks the threshold) and makes the key impact of our revised velocity dispersion immediately apparent. With the previous measurement of $\sigma_v = 10.7 \pm 3.5$~\vlosUnits\ \citep{Carlin2018}, Boo~III sits above the line, suggesting that its internal density is sufficient to resist tidal disruption despite its small pericenter. With our updated value of $\sigma_v = 1.69^{+1.03}_{-0.85}$~\vlosUnits, the implied dynamical mass, and hence the average density within the half-light radius, decreases dramatically, placing Boo~III well below the line and firmly in the tidally-affected regime. Independently, \citet{Geha2026b} recreates this pericenter--density diagnostic for a uniform DEIMOS sample of MW satellites and likewise finds Boo~III below the tidal-disruption threshold (their Figure~3), consistent with the picture presented here.

This result places Boo~III among the clearly tidally disrupting satellites in the MW system, alongside Antlia~II, Crater~II, Sgr, and Tucana~III. The combination of a small pericenter ($r_{\mathrm{peri}} \sim 9.5$\,kpc), a very recent pericentric passage ($\sim 0.14$\,Gyr ago), and a low velocity dispersion is fully consistent with Boo~III being in a state of active tidal disruption. Additionally, the low density and small pericenter of Boo~III are consistent with its puffy, extended structure relative to other dSphs of similar luminosity: its half-light radius ($r_h \approx 466$~pc) is roughly twice the $\sim 200$~pc typical of $M_V \approx -5.8$ systems such as Hercules ($r_h \approx 213$~pc) and Bo\"{o}tes~I ($r_h \approx 192$~pc), as tides can alter, and in some regimes inflate, a satellite's half-light radius during close pericentric passages \citep{Penarrubia2009, Errani2015}.

\subsubsection{Implications for Dark Matter Annihilation}

The MW dSphs are among the best targets for dark matter indirect detection searches due to their proximity, high dark matter fractions, and low astrophysical backgrounds \citep[e.g.,][]{2004PhRvD..69l3501E, 2024PhRvD.109f3024M, 2026JCAP...03..035A}. 
For dark matter annihilation, the predicted signal is generally decomposed into two parts: a particle-physics component (e.g., dark matter particle mass, cross section, and branching fractions into standard model particles) and an astrophysical component that depends on the dark matter distribution. The latter is referred to as the J-factor and scales as the density profile squared, $J=\int \rho_{\rm DM}^2~{\rm d}\ell{\rm d}\Omega$. 

A recent search has found a $3.1\sigma$ excess near Boo~III in IceCube neutrino data that is consistent with dark matter annihilation \citep{2026arXiv260601853J}.
However, no detailed modeling exists for the expected dark matter distribution of Boo~III, and we therefore estimate the expected J-factor of Boo~III using the J-factor scaling relations of \citet{Pace2019MNRAS.482.3480P},

{\small
\begin{equation}\label{eq:jfactor}
J(0.5^{\circ})/({\rm GeV^2~cm^{-5}}) = 10^{17.87}
\left( \frac{\sigma_v}{5~{\rm km~s^{-1}}} \right)^{4}
\left( \frac{D_\odot}{100~{\rm kpc}} \right)^{-2}
\left( \frac{r_h}{100~{\rm pc}} \right)^{-1}
\end{equation}
}

Fixing the distance and half-light radius to the values in Table~\ref{tab:params}, we estimate the J-factor for several velocity dispersion measurements in the literature and computed here. 
With $\sigma_v=10.7, 5.27, 1.69$~\vlosUnits we find $\log_{10}{\left(J(0.5^\circ)/{\rm GeV^2~cm^{-5}}\right)}=19.15,~17.92,~15.95$, respectively.  These velocity dispersion measurements correspond to the \gaia~DR2-cleaned Hectospec sample of \citet{Carlin2018}, recent results with Keck/DEIMOS \citep{Geha2026}, and our results here. Notably, the velocity dispersion and corresponding J-factor have continually decreased with better stellar-kinematic datasets. 
The largest MW dSph J-factor values are $\log_{10}{\left(J(0.5^\circ)/{\rm GeV^2~cm^{-5}}\right)}\sim 19$, whereas the bulk of the population has values in the $17 \lesssim\log_{10}{\left(J(0.5^\circ)/{\rm GeV^2~cm^{-5}}\right)}\lesssim 19$ range \citep{2015MNRAS.453..849B, 2015ApJ...801...74G, Pace2019MNRAS.482.3480P}. 
While the initial spectroscopic sample suggested a large J-factor, the more recent velocity dispersion measurements imply that Boo~III has a low density with a low J-factor. 
The Boo~III stellar kinematics therefore disfavor the dark matter annihilation solution of the neutrino flux \citep{2026arXiv260601853J}. 

\subsection{$N$-body Modeling of the Boo~III-like Progenitor and Its Vicinity} \label{section:nbody}

The particle spray model of Section~\ref{section:stream} predicts where the tidal debris should lie, but it treats the stripped stars as massless test particles and therefore carries no information about the internal kinematics of the surviving bound remnant, which is precisely the region probed by our \SSSSS\ members. To model the remnant itself, we turn to restricted $N$-body simulations and compare them with the \SSSSS\ data in two ways. First, we ask which dark-matter halo mass reproduces the observed velocity dispersion (Section~\ref{section:tidaldisrupt_vdisp}). Second, to probe whether the observed kinematics deviate from a bound, dispersion-supported equilibrium, we measure the on-sky velocity gradient of the 21 GMM members and compare it to three independent model predictions: a solid-body / perspective-rotation expectation, an orbit-based prediction, and the $N$-body simulation.

\subsubsection{$N$-body simulation setup} \label{section:nbody_setup}

To compare with the observed structure of the Boo~III dwarf, we run restricted $N$-body simulations of Boo~III-like progenitors. This restricted $N$-body technique has been shown to closely reproduce the results of $N$-body simulations \citep[][Senkevich et al. in prep., Vasiliev et al. in prep.]{Vasiliev2021} and has been recently used to produce models of Crater~II \citep{Limberg2025}. As in an $N$-body simulation, the dwarf is modeled with both star and dark matter particles. However, instead of accounting for the force of each particle, a low-order multipole expansion is fit to the particles at a series of timesteps. Between these timesteps, the particles are integrated as tracers in the combined potential of the progenitor, the MW, and the LMC. For the potential of the progenitor, we use a predictor-corrector approach to interpolate the progenitor's potential during these timesteps between the initial and final potentials. The main advantage of this restricted $N$-body approach is that, by construction, the dwarf ends up at the observed present-day phase-space coordinates.

We use a simulation setup almost identical to that of Section~\ref{section:sim}. We evolve Boo~III in two potentials for the MW: the \texttt{McMillan17} potential from \citet{McMillan2017} and \texttt{MWPotential2014} from \citet{Bovy2015}. We model the LMC as a Hernquist profile with the same mass, scale radius, and present-day phase-space coordinates as the particle spray model of Section~\ref{section:sim} (Table~\ref{tab:streamspray_params}), and include dynamical friction following \citet{Vasiliev2021}. We use the same solar radius, solar height above the Galactic plane, and solar motion as in Section~\ref{section:sim}. Finally, we also account for the reflex motion of the MW in response to the LMC, which is included as a non-inertial force. 

We model Boo~III with stars represented by a Plummer sphere embedded in a spherical NFW dark matter halo. For the Plummer sphere, we use a scale radius of 0.466 kpc, based on its half-light radius from Table~\ref{tab:params}, and a stellar mass of \mbox{$3.573\times10^4~{\rm M}_\odot$}, based on its observed luminosity and assuming a mass-to-light ratio of 2. For each NFW dark matter mass, we obtain the concentration using the mass-concentration relation from \citet{Dutton+2014} assuming a Hubble constant of $67.4~\mathrm{km\,s^{-1}\,Mpc^{-1}}$ \citep{Planck+2020}. We initialize this system using \textsc{agama} \citep{Vasiliev2019}. To avoid initializing particles very far from the progenitor, we use a truncation radius of 10 scale radii for the Plummer sphere and 2 virial radii for the NFW profile.  

Since we are interested in the properties of the progenitor, and since we are initializing the progenitor using its present-day stellar properties, we choose to evolve our models of Boo~III starting from their second-most-recent apocenter. This allows them to experience two pericentric passages, which reshape the progenitor. For the \texttt{McMillan17} potential and \texttt{MWPotential2014} potential, this corresponds to a duration of 2.04 Gyr and 4.13 Gyr, respectively. 

For each of the two MW potential choices, we perform three simulations that differ only in the initial NFW dark-matter halo mass, set at the second-most-recent apocenter where the integration begins: $M_{\rm DM} = 10^6$, $10^7$, and $10^8\,{\rm M}_\odot$. The $10^8\,{\rm M}_\odot$ value is the canonical halo mass for UFDs in $\Lambda$CDM \citep[e.g.,][]{Simon2019} and is roughly the minimum mass at which a galaxy can form under standard models of galaxy formation \citep{BenitezLlambay2020, Bullock2017ARAA}; the $10^7$ and $10^6\,{\rm M}_\odot$ runs represent progressively stripped progenitors. As we show in Section~\ref{section:tidaldisrupt_vdisp} below, only the $10^6\,{\rm M}_\odot$ run reproduces our observed line-of-sight velocity dispersion. We therefore adopt the $10^6\,{\rm M}_\odot$ simulation as the fiducial run for all subsequent figures and gradient comparisons in this section, with the $10^7$ and $10^8\,{\rm M}_\odot$ runs retained only for the dispersion comparison of Section~\ref{section:tidaldisrupt_vdisp}.

\subsubsection{Velocity-dispersion comparison} \label{section:tidaldisrupt_vdisp}
We measure the line-of-sight velocity dispersion of the star particles within a projected aperture of radius $r_h$ for each of the three \texttt{McMillan17} simulations, and find a clear monotonic trend with progenitor halo mass: $\sigma_v = 2.12$~\vlosUnits\ at ${\rm M}_{\rm DM} = 10^6\,{\rm M}_\odot$, $\sigma_v = 3.52$~\vlosUnits\ at $10^7\,{\rm M}_\odot$, and $\sigma_v = 5.64$~\vlosUnits\ at $10^8\,{\rm M}_\odot$. As an example, the spatial $\sigma_v$ map for the fiducial $10^6\,{\rm M}_\odot$ run is shown in the bottom-right panel of Figure~\ref{fig:fig7_gradient}.
Comparing against our observed $\sigma_v = 1.69^{+1.03}_{-0.85}$~\vlosUnits\ (Section~\ref{section:membership}), only the $10^6\,{\rm M}_\odot$ run is consistent with the data; the $10^7\,{\rm M}_\odot$ run is marginally inconsistent at $\sim 1.8\sigma$, and the $10^8\,{\rm M}_\odot$ run is incompatible at the $\sim 4\sigma$ level.  The corresponding values in the \texttt{MWPotential2014} potential are systematically lower by $\sim 0.1$--$0.4$~\vlosUnits\ at each progenitor mass,  indicating that
  $\sigma_v$ is governed primarily by the progenitor halo mass rather than by the choice of MW potential.

However, $10^6\,{\rm M}_\odot$ is too low for a halo mass to host an in-situ galaxy under standard $\Lambda$CDM galaxy-formation models, so this cannot represent the progenitor's birth halo.  The natural interpretation is that Boo~III formed in a halo of order $10^8\,{\rm M}_\odot$ (consistent with
CDM expectations for a UFD of its luminosity), but has since lost a substantial fraction of its dark-matter mass to tidal stripping, leaving the present-day surviving bound mass much smaller than the
natal halo mass.  The observed low $\sigma_v$ thus tracks the much-reduced bound mass that remains, in line with the actively-disrupting picture of Section~\ref{section:tidaldisrupt}.

A complementary interpretation is that the dark-matter density profile of Boo~III differs from the cuspy NFW profile adopted in our simulations.  A cored profile, such as the \citet{Burkert1995} profile or the cores produced by SIDM \citep{Spergel2000, TulinYu2018}, yields a shallower central
potential and therefore a lower $\sigma_v$ at fixed total halo mass than a cuspy NFW halo, which could naturally reconcile the observed low dispersion with the canonical $10^8\,{\rm M}_\odot$ halo mass.  A detailed comparison between the observed dispersion of disrupting UFDs like Boo~III and simulations spanning a range of inner-profile shapes could therefore serve as a powerful probe of the nature of dark matter (CDM cusp vs.\ SIDM core, etc.), but we defer such a study to future work.

This low-dispersion, underdense-halo puzzle is not unique to Boo~III: it is central to the interpretation of two clearly disrupting, ultra-cold dSphs, Antlia~II and Crater~II. Both are unusually cold and diffuse, with Crater~II in particular described as the most extreme low-density outlier among the MW satellites \citep{Caldwell2017, Ji2021, Limberg2025}. Detailed modeling of Crater~II finds that even with tidal stripping a cuspy NFW progenitor requires an implausibly low birth mass ($\sim 10^{7.6}\,{\rm M}_\odot$, well below the $\sim 10^{8.6}\,{\rm M}_\odot$ expected from abundance matching), whereas a cored or self-interacting halo, or, more radically, modified Newtonian dynamics (MOND), more naturally reproduces its size and coldness \citep{Limberg2025}. The same phenomenon extends to M31: its satellites Andromeda~XIX and Andromeda~XXI are comparably extended and dynamically cold ($\sigma_v \approx 7.8$ and $6.1$~\vlosUnits, with half-light radii up to $\sim 3$~kpc; \citealt{Collins2020, Collins2021}) and are likewise attributed to tidal reshaping in a low-density halo, with And~XIX explicitly identified as Antlia~II's M31 counterpart. More broadly, a systematic trend toward low dark-matter densities has recently been reported across the M31 dwarf-galaxy system \citep{Pickett2026}. With an even smaller $\sigma_v$ and the direct disruption signatures established in Section~\ref{section:tidaldisrupt}, Boo~III is a natural addition to this small class of cold, disrupting systems whose kinematics jointly constrain the efficiency of tidal stripping and the inner dark-matter profiles of dwarf galaxies.
  
\subsubsection{Velocity-gradient comparison} \label{section:vgrad_compare}
The line-of-sight velocity gradient observed across a dwarf galaxy can be decomposed into two contributions: a purely geometric component arising from the projection of the system's bulk transverse motion onto the line of sight (perspective rotation; see e.g., \citealt{Kaplinghat2008, Walker2008, Sandford2026}), and an intrinsic component reflecting the dynamics of the system itself. For a system at a distance $D_\odot$ moving with bulk proper motion $\boldsymbol{\mu}$, the perspective gradient amplitude scales as $|\nabla v_{\rm persp}| \approx 4.74\,|\boldsymbol{\mu}|\,D_\odot$ (with $\boldsymbol{\mu}$ in mas yr$^{-1}$, $D_\odot$ in kpc, and the resulting gradient in \vlosUnits~rad$^{-1}$), and it is aligned with the \emph{apparent} proper-motion vector.  For a fully \emph{bound}, dispersion-supported dwarf with no internal rotation, the observed gradient is purely perspective,  i.e., the system acts as a solid body, so the intrinsic gradient is recovered as $\nabla v_{\rm int} = \nabla v_{\rm obs} - \nabla v_{\rm persp}$ (as applied to Boo~I by \citealt{Sandford2026}); perspective rotation also inflates the apparent velocity dispersion across the system if not subtracted, although for most compact UFDs, this contribution is small compared with the intrinsic $\sigma_v$.  For a fully \emph{disrupting} stream-like system, in contrast, the bulk of the stars are no longer bound and move along the local orbit, so the observed gradient instead approaches the orbital $v_{\rm los}$ slope at the present-day position; the ``observed $-$ perspective'' decomposition above no longer cleanly corresponds to the system's internal dynamics in this limit.

\begin{figure*}
    \centering
    \includegraphics[width=\textwidth]{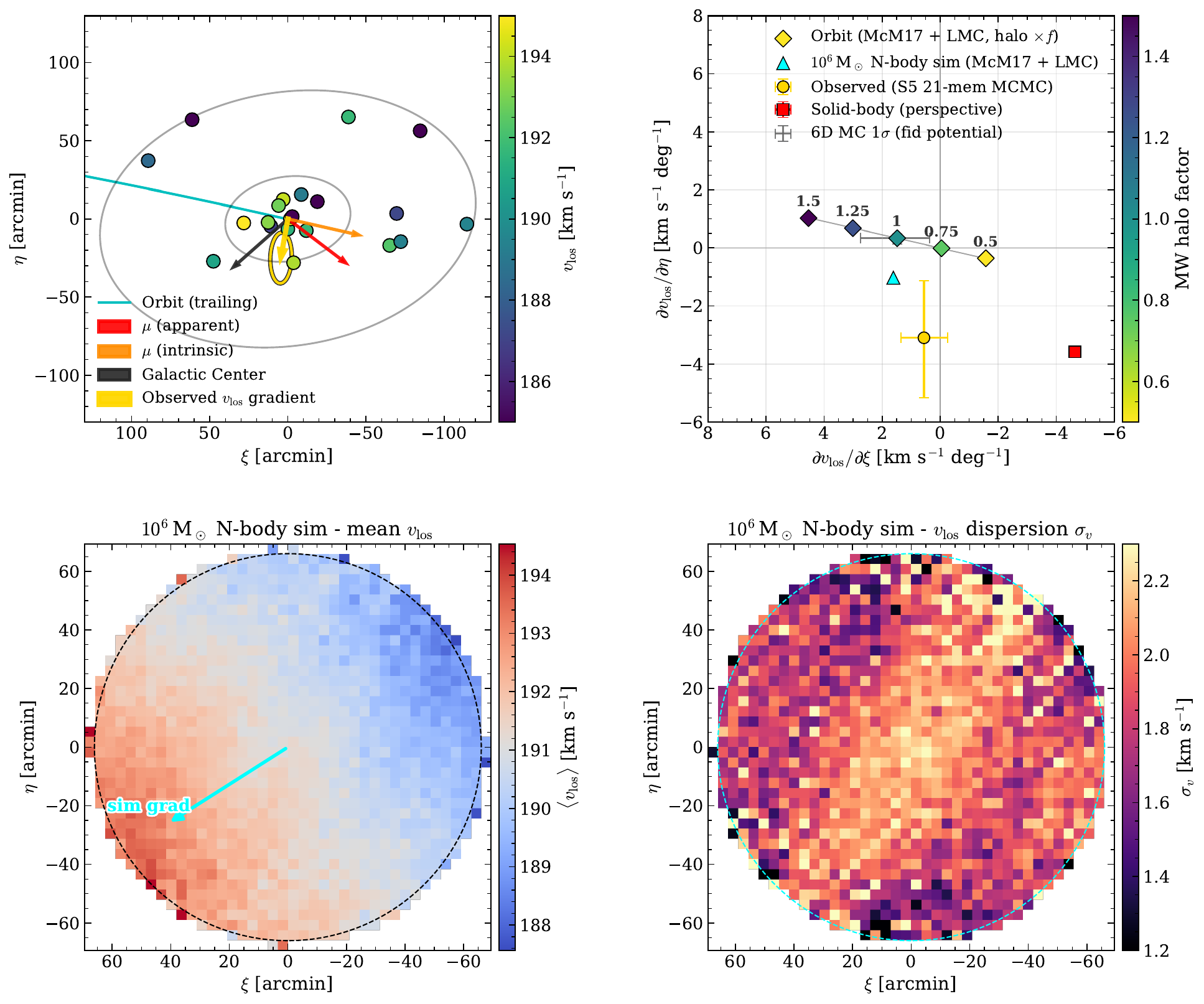}
    \caption{Velocity-gradient diagnostic for Boo~III. \emph{Top-left:} the 21 \SSSSS\ members on the sky, color-coded by $v_{\rm los}$, with arrows indicating the apparent proper-motion direction (red), the solar-reflex-corrected intrinsic proper-motion direction (orange), the Galactic-Center direction (black), and the observed velocity gradient (yellow); the $1\sigma$ uncertainty ellipse on the observed gradient is shown in yellow. The cyan curve shows the integrated Boo~III orbit in the trailing-tail direction. Gray ellipses mark 1 and 3 times the elliptical half-light radius. \emph{Top-right:} comparison of the observed gradient (yellow circle) with three model predictions in $(\partial v_{\rm los}/\partial\xi,\,\partial v_{\rm los}/\partial\eta)$ space: the solid-body / perspective-rotation prediction (red square), the orbit prediction across halo-mass scalings $f$ (colored diamonds, labeled by $f$), and the $10^6\,{\rm M}_\odot$ $N$-body simulation within $2\,r_h$ (cyan triangle); the gray error bars on the $f = 1$ diamond show the $1\sigma$ spread from the 6D MC of Appendix~\ref{section:appendix_6dmc}. \emph{Bottom-left:} mean $v_{\rm los}$ map of the $N$-body simulation particles within $2\,r_h$, with the linear-fit gradient direction shown by the cyan arrow. \emph{Bottom-right:} line-of-sight velocity-dispersion ($\sigma_v$) map of the same $N$-body simulation particles within $2\,r_h$.}
    \label{fig:fig7_gradient}
\end{figure*}

To measure the observed gradient, we refit the 21 high-probability members of Section~\ref{section:membership} with a four-parameter single-Gaussian velocity model ($v_{\rm los,0}$, $\partial v_{\rm los}/ \partial\xi$, $\partial v_{\rm los}/\partial\eta$, $\sigma_v$; membership is fixed so no MW background component is needed).  We obtain $|\nabla v_{\rm obs}| = 3.28^{+2.0}_{-1.7}$~\vlosUnits~deg$^{-1}$
at $\mathrm{PA} = (170^{+15}_{-21})^\circ$ (both also listed in Table~\ref{tab:params}; yellow point in Figure~\ref{fig:fig7_gradient}, top-right panel; the corresponding sky-panel arrow is shown in yellow in the top-left panel). We caution, however, that this is only a marginal ($\sim 1.8\sigma$) detection (the gradient amplitude exceeds zero by less than $2\sigma$), so both the amplitude and, especially, its PA are weakly constrained, and the measurement should be interpreted with corresponding caution.

We compare this measurement against the two limiting predictions introduced above (the fully-bound perspective rotation expectation and the fully-disrupting orbital prediction) together with the intermediate prediction from our $N$-body simulation, all shown in the top-right panel of Figure~\ref{fig:fig7_gradient}.
The \emph{perspective prediction} (red square; the fully-bound limit) is computed by assuming the 21 members move rigidly at the GMM-fit progenitor proper motion and projecting their predicted heliocentric $v_{\rm los}$ onto the sky; for Boo~III's $D_\odot = 48.5$~kpc and $|\boldsymbol{\mu}| = 1.46$~\pmUnits\ this yields $|\nabla v_{\rm persp}| = 5.85$~\vlosUnits~deg$^{-1}$ at $\mathrm{PA} \approx 232^\circ$ (i.e., along the direction of Boo~III's apparent proper motion).
The \emph{orbit prediction} (diamond markers, color-coded by halo-mass scaling factor; the fully-disrupting limit) is the local linear $v_{\rm los}$ slope along the integrated \texttt{McMillan17}~+~LMC orbit at the present-day position, evaluated for halo masses scaled by $f \in \{0.5, 0.75, 1.0, 1.25, 1.5\}$ relative to the fiducial \texttt{McMillan17} value.  We use the smooth progenitor orbit's local slope here, rather than a direct 2-D least-squares fit to the spray particles within a few degrees of the progenitor, because spray particles near the center are dominated by the stochastic Lagrange-point release.  The gray error bar on the $f = 1$ diamond shows the $1\sigma$ spread induced by propagating Boo~III's 6D phase-space measurement uncertainties (Table~\ref{tab:params}) through the orbit integration; Appendix~\ref{section:appendix_6dmc} breaks this down per component.
The \emph{$N$-body simulation prediction} (cyan triangle) is the linear gradient measured from the stellar particles within $2\,r_h$ in our fiducial $10^6\,{\rm M}_\odot$ Boo~III simulation (integrated in the same \texttt{McMillan17}~+~LMC potential; Section~\ref{section:nbody_setup}): $|\nabla v_{\rm sim}| = 1.91$~\vlosUnits~deg$^{-1}$ at $\mathrm{PA} = 122^\circ$. The mean $v_{\rm los}$ map from which this gradient is measured is shown in the lower-left panel of Figure~\ref{fig:fig7_gradient}, with the gradient direction indicated by the cyan arrow. Interestingly, the gradient direction lies roughly along the line from Boo~III to the Galactic Center, shown as the black arrow in the upper-left panel of the same Figure.

The observed gradient sits between the fully-bound (perspective)
expectation and the fully-unbound (orbital) prediction shown in the
top-right panel of Figure~\ref{fig:fig7_gradient}, broadly consistent
with Boo~III being a partially-disrupted system in which a small bound
core is surrounded by tidally stripped stars whose motions deviate from
rigid-body translation.  The $N$-body simulation prediction (cyan
triangle) is similarly displaced from both endpoints, supporting this
intermediate-disruption picture for our fiducial $10^6\,{\rm M}_\odot$
progenitor halo.

The diamond track in the same panel additionally shows that the
orbit-predicted gradient depends on the assumed MW halo mass:
scaling the \texttt{McMillan17} halo by $f \in [0.5, ..., 1.5]$ produces a
clear, monotonic shift in the orbit-predicted $(\partial v_{\rm los}/
\partial\xi,\,\partial v_{\rm los}/\partial\eta)$.  A sufficiently
precise measurement of the observed gradient could therefore in
principle constrain the MW halo mass, a program we explore
further in the next section using the predicted
Boo~III tidal stream (if detected) as a complementary probe.  With
only 21 high-purity members in our current sample, however, the
gradient uncertainty is too large to provide a useful
halo-mass constraint or to cleanly separate the bound, unbound, and
intermediate scenarios; a larger member sample from future deep
spectroscopic surveys will be required.

\subsection{Stream Kinematics along $\phi_1$ as a Probe of the MW Halo Mass} \label{section:phi1_stream}
The comparison with the Styx track in Section~\ref{section:stream} tested the predicted debris on large scales; we now return to the stream prediction in the immediate vicinity of the progenitor, where our \SSSSS\ data can test it directly. We work in the great-circle stream coordinates $(\phi_1,\phi_2)$ defined in Appendix~\ref{section:appendix_phi}: the centroid lies at $\phi_1=0$, the $\phi_1$ axis is aligned with Boo~III's reflex-corrected (intrinsic) proper motion so that positive $\phi_1$ runs along the leading tail and negative $\phi_1$ along the trailing tail, and the orbit stays close to $\phi_2=0$. As already evident from Figure~\ref{fig:stream_comparison}, $v_{\rm los}$ is the most halo-mass-sensitive feature of the predicted stream, so we display the $v_{\rm los}(\phi_1)$ track for a range of MW halo-mass scalings and compare these predictions with the \SSSSS\ members, including the two tidal-tail candidates identified beyond the $3\,r_h$ ellipse in Appendix~\ref{section:appendix_tail}.

Figure~\ref{fig:phi1_stream} compares the predicted heliocentric $v_{\rm los}$ and heliocentric distance as functions of $\phi_1$ to our observations. The top panel shows
$v_{\rm los}(\phi_1)$ for the fiducial particle spray model of Section~\ref{section:stream} together with three MW-mass scalings ($f=0.5$, $1.5$, and $2$, all with the LMC retained), and for the $10^6\,{\rm M}_\odot$ $N$-body simulation (the fiducial mass) evolved in the two MW potentials of Section~\ref{section:nbody_setup}: \texttt{McMillan17}~+~LMC, the same potential as the spray model (orange), and the lighter \texttt{MWPotential2014}~+~LMC (green). Notably, the two potentials predict $v_{\rm los}(\phi_1)$ slopes of opposite sign near the progenitor, so the stream kinematics in the immediate vicinity of Boo~III by themselves discriminate between MW potentials. In both panels we overlay, in
matching colors, the corresponding \emph{progenitor orbits} as dashed lines, i.e.\ the trajectory that each MW-mass potential predicts for a test particle launched from
Boo~III's present-day 6D phase-space position. Within the $\pm 3\,a_h$ near-progenitor footprint, the spray \texttt{streamTrack} curves (Section~\ref{section:stream})
closely follow the underlying progenitor orbit in all four MW-mass cases, justifying our use of the smooth orbit gradient as the model reference in the velocity-gradient
comparison of Figure~\ref{fig:fig7_gradient} (top-right). Open red circles show the 21 GMM RGB members with their $v_{\rm los}$ measurements, and green stars mark the two
candidate Boo~III members lying \emph{outside} the $3\,r_h$ ellipse that we identify in Appendix~\ref{section:appendix_tail} (see also
Figure~\ref{fig:appendix_candidates}). One of the two outer candidates, at $\phi_1 \approx +4^\circ$ on the leading-tail side, has an \SSSSS\ heliocentric velocity that sits \emph{between} the $N$-body predictions for the two MW potentials (equivalently, between the fiducial spray track and the lighter $f = 0.5$ variant), qualitatively favoring a MW potential lighter than the fiducial \texttt{McMillan17} but somewhat more massive than \texttt{MWPotential2014}. With only a single tentative tail-candidate star, this is far from a quantitative constraint, but it foreshadows the
diagnostic power of even a modest future tail detection, or a confident dynamical association of the Styx stream with Boo~III's debris, to break the halo-mass
degeneracy along the orbit.

The bottom panel shows the predicted heliocentric distance versus $\phi_1$ for the fiducial particle spray and the $10^6\,{\rm M}_\odot$ $N$-body predictions in both MW potentials. We omit the particle spray MW-mass scan in this panel: the predicted distance track is only weakly sensitive to the halo mass over the range we test, as shown both by Figure~\ref{fig:stream_comparison} and by the near-coincidence of the two $N$-body distance tracks here.  Overplotted are the five RRL distances derived in Section~\ref{section:distance} via the period-independent $M_G([\mathrm{Fe/H}])$ calibration of
\citet{Garofalo2022}.  The five RRLs sit at $|\phi_1| \lesssim 0.4^\circ$ and are consistent with the predicted near-progenitor distance of $\sim 48$~kpc within their
$\sim 3.1$~kpc uncertainties.  Crucially, both spray and $N$-body models predict that the heliocentric distance varies appreciably across even the near-progenitor
footprint: within $|\phi_1| \lesssim 2^\circ$ (roughly the $3\,a_h$ semi-major-axis extent) the predicted distance changes from $\sim 46$~kpc at $\phi_1 \approx -2^\circ$
(the trailing-tail side, behind the progenitor on its orbit) to $\sim 52$~kpc at $\phi_1 \approx +2^\circ$ (the leading-tail side), a distance gradient of $\approx 1.6$~kpc~deg$^{-1}$ near the progenitor in the $N$-body model, over the same
spatial footprint that hosts the 21 RGB members. The cause is geometric: Boo~III's orbit has a significant radial component at the present epoch, so stars displaced along the orbit naturally span a range of heliocentric distances. 

The distance gradient grows steeper still in the tail region: over a $\sim 10^\circ$ $\phi_1$ baseline (a plausible angular extent for a future stream detection), the spray model predicts $\gtrsim 15$~kpc of distance variation along the stream.  This has direct implications for any photometric search for Boo~III's tidal tails: a
search relying on a single fixed distance modulus would systematically lose stars at either end of the stream.  Future deep RRL or BHB samples within the $3\,r_h$
ellipse that can resolve distances at the few-percent level would directly test the near-progenitor prediction, while wide-area searches that incorporate the predicted
distance gradient (rather than assuming a constant heliocentric distance) would be needed for robust tail detection.  We caution, however, that such searches are further limited by
Sgr-stream contamination at the same heliocentric distance (Section~\ref{section:tail_prospects}); the line-of-sight velocity therefore remains the cleanest
discriminator for tail detection.

\begin{figure}[!ht]
    \centering
    \includegraphics[width=\columnwidth]{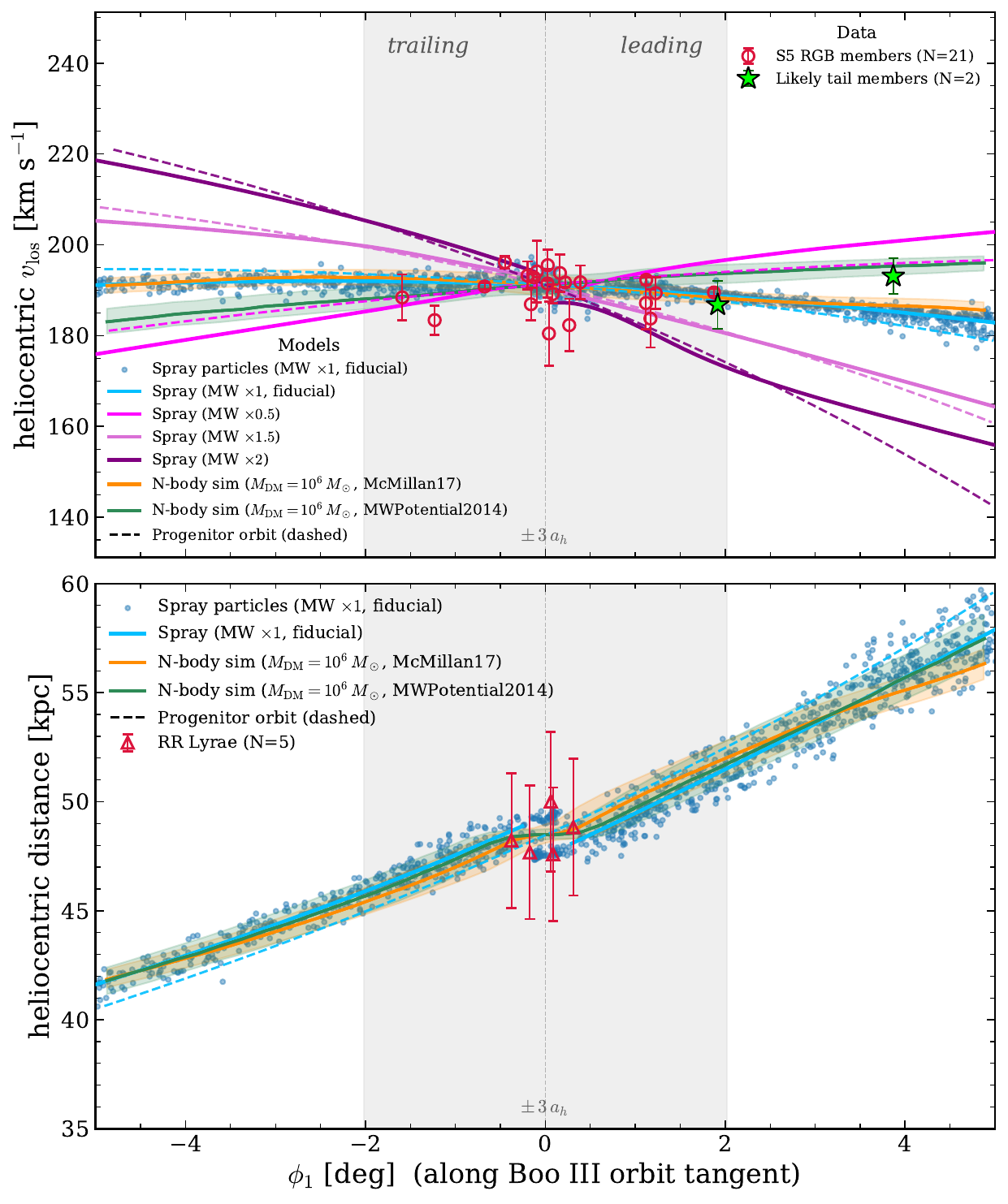}
    \caption{Predicted heliocentric $v_{\rm los}$ (top) and heliocentric distance (bottom) versus $\phi_1$, the great-circle stream coordinate aligned with the Boo~III orbit (Appendix~\ref{section:appendix_phi}). \emph{Top panel:} particle spray model in the fiducial \texttt{McMillan17}~+~LMC potential (blue scatter = raw spray particles; colored solid curves = leading + trailing \texttt{streamTrack}). The solid curves scale the Milky Way halo mass relative to the fiducial \texttt{McMillan17} value -- $\times 1$ (fiducial) in cyan, $\times 0.5$ in magenta, $\times 1.5$ in light purple, and $\times 2$ in purple -- with the LMC held at its fiducial mass throughout (same color scheme as the corresponding column of Figure~\ref{fig:stream_comparison}). Dashed lines in matching colors show the corresponding \emph{progenitor orbits} integrated in each MW-mass potential, demonstrating that the spray particles closely trace the progenitor orbit within the $\pm 3\,a_h$ near-progenitor footprint. The orange and green bands show the median and $16/84$\% range of the $10^6\,{\rm M}_\odot$ $N$-body simulation integrated in the \texttt{McMillan17}~+~LMC and \texttt{MWPotential2014}~+~LMC potentials, respectively. Open red circles are the 21 \SSSSS\ GMM RGB members.  The two green stars in the top panel are the two likely tidal-tail RGB members beyond the $3\,r_h$ ellipse (see Appendix~\ref{section:appendix_tail} for more details).
    \emph{Bottom panel:} same as the top but for heliocentric distance, showing only the fiducial spray (cyan track + blue scatter, with the corresponding progenitor orbit as the cyan dashed line) and the two $N$-body bands. Open red triangles are the five Boo~III RRL distances from Section~\ref{section:distance}. In both panels the gray band marks the $\pm 3\,a_h$ spatial extent of the near-progenitor footprint.}
    \label{fig:phi1_stream}
\end{figure}

\begin{figure*}
    \centering
    \includegraphics[width=\textwidth]{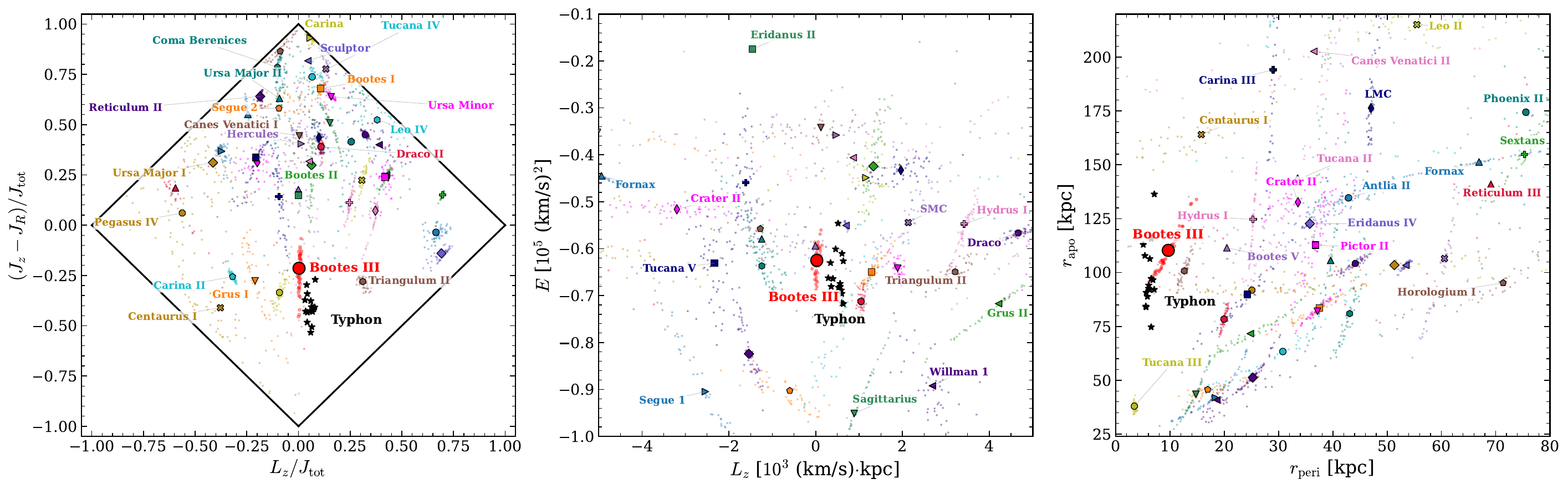}
    \caption{Comparison of the Typhon stream stars and other dwarf galaxies in terms of action, energy, apocenter and pericenter. In these panels, the black star markers show the individual Typhon stars obtained from \gaia DR3. The other data points are systemic values of different dwarf galaxies, computed from their 6D parameters with 50 MC realizations. The small dots clustered around each system's marker (red for Boo~III) are these 50 realizations and illustrate the propagated 6D measurement uncertainty; they are not particle spray stream points. Each dwarf is shown with a unique filled marker, a distinct shape and color held fixed across all three panels, placed at its median value in each parameter space. For Boo~III, the 6D parameters are taken from this analysis (Table \ref{tab:params}), while the 6D parameters of all other dwarfs are taken from the Local Volume Database \citep[LVDB;][]{Pace2025}. Left: the action diamond plot. To reduce clutter, the dwarf labels are distributed across the three panels (each system is generally labeled in only one); the fixed marker shape and color of each system allow it to be matched between panels. Middle: orbital energy vs. angular
    momentum. Right: apocenter vs. pericenter. Across all three panels, of all the dwarf galaxies in the sample, Boo~III (red) lies closest to Typhon in every projection -- the dynamical basis for the possible common-origin association discussed in the text (Figure~\ref{fig:typhon_and_streams} shows the stream-side counterpart).}
    \label{fig:typhon_and_dwarfs}
\end{figure*}

\begin{figure*}
    \centering
    \includegraphics[width=\textwidth]{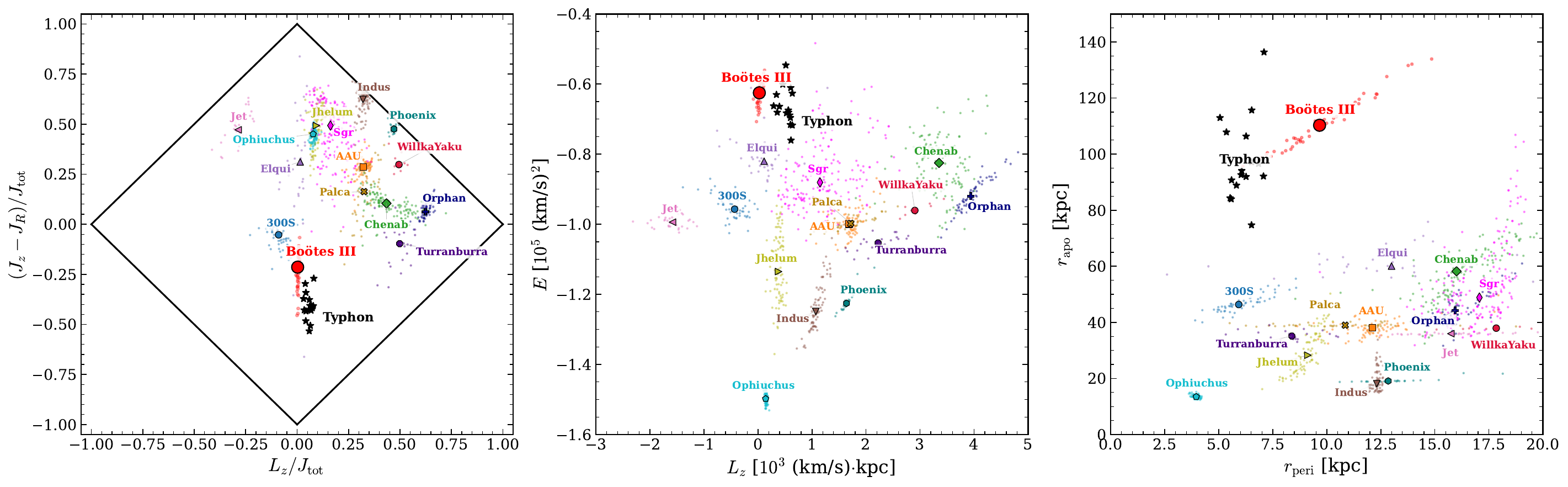}
    \caption{Comparison of Boo~III with the Typhon stream and the stellar streams in \SSSSS, in terms of action, energy, apocenter and pericenter. Refer to Figure~\ref{fig:typhon_and_dwarfs} for a description of the axes. Of all the streams shown, Typhon lies closest to Boo~III in every projection -- the stream-side counterpart of the dwarf-galaxy comparison in Figure~\ref{fig:typhon_and_dwarfs}, where Boo~III is in turn the closest dwarf to Typhon. In these panels, Boo~III and Typhon are shown in the same colors and markers as Figure~\ref{fig:typhon_and_dwarfs}. As in that figure, each \SSSSS\ stream from the stellar-stream catalog of \citet{Li2022} is plotted with a unique filled marker at its median position and is labeled in all three panels; the surrounding small points are its individual measured members. Unlike the dwarf-galaxy medians, the stream medians and members are computed directly from measured 6D members rather than MC realizations. The small red dots around Boo~III are its 50 MC realizations (as in Figure~\ref{fig:typhon_and_dwarfs}), not stream members. Styx, the candidate on-sky tidal stream associated with Boo~III (Section~\ref{section:stream}), is not shown here: only its on-sky track is known, so with no measured 6D phase-space coordinates it cannot be placed in this action and integrals-of-motion comparison.}
    \label{fig:typhon_and_streams}
\end{figure*}

\subsection{Comparison with the Typhon Stream}
The Typhon stream was discovered as a dynamically coherent group of metal-poor halo stars on a highly eccentric, polar orbit by \citet{Tenachi2022}  and characterized further in subsequent spectroscopic analyses \citep{Ji2023Typhon}. Unlike the on-sky Styx stream, Typhon was identified purely as a clustering in integrals-of-motion space, and its members lie much closer to the Sun ($\lesssim 4$~kpc) than Boo~III. Like Boo~III, Typhon's polar orbit and high eccentricity suggest that it is the debris of a tidally disrupted progenitor on a similar dynamical trajectory. The similarities make Typhon a natural comparison case for assessing whether Boo~III shares its dynamical neighborhood and possibly its infall history. To investigate this potential connection, we compute the orbital properties of the Typhon stream members and compare them with those of Boo~III and other dwarf galaxies. Specifically, we use the action diamond, energy vs. angular momentum, and apocenter vs. pericenter parameter spaces. For the 6D parameters of Boo~III we use the values derived in this work (Table~\ref{tab:params}); for all other dwarf galaxies in the comparison sample we adopt the values compiled in the Local Volume Database\footnote{\url{https://github.com/apace7/local_volume_database}} (LVDB; \citealt{Pace2025}).

We again use the \texttt{galpy} Python package for these calculations, in the \texttt{McMillan17} potential. For each input 6D phase-space coordinate we initialize a \texttt{galpy} \texttt{Orbit}, integrate it backward by 10~Gyr, and read off the most recent pericenter and apocenter from the integrated radial distance. The three action components $(J_R, L_z, J_z)$ and the total energy $E$ are obtained directly from the \texttt{Orbit} object's built-in methods (\texttt{Orbit.jr}, \texttt{Orbit.jp}, \texttt{Orbit.jz}, \texttt{Orbit.E}, and \texttt{Orbit.Lz}), which evaluate the actions using the Staeckel-fudge approximation \citep{Binney2012Staeckel} in the same \texttt{McMillan17} potential. For each dwarf galaxy in the comparison sample we propagate its 6D uncertainties via 50 MC realizations (drawn from Gaussian distributions centered on the LVDB median values with widths equal to the quoted $1\sigma$ uncertainties) and take the median of those 50 realizations as the representative value plotted in Figure~\ref{fig:typhon_and_dwarfs}; for Boo~III we use our updated Table~\ref{tab:params} values. For the Typhon stream, every member star has its own measured 6D, so we compute its actions, energy and orbital pericenter/apocenter directly from the per-star measurements (without MC sampling) and plot each member individually as a black star in Figure~\ref{fig:typhon_and_dwarfs}. Figure~\ref{fig:typhon_and_dwarfs} shows the comparison of Boo~III and Typhon against the rest of the LVDB dwarf-galaxy sample. We additionally compute the same set of parameters for the dozen streams in the \SSSSS\ catalog \citep{Li2022}, handled identically to Typhon (per-member values, no MC sampling). Figure~\ref{fig:typhon_and_streams} compares Boo~III and Typhon against this stream sample.

As shown in Figure~\ref{fig:typhon_and_dwarfs}, of all the dwarfs in our comparison sample, Boo~III is the closest neighbor of Typhon in every projection we examined (the action diamond, $E$ vs.~$L_z$, and pericenter vs.~apocenter). Equivalently, as shown in Figure~\ref{fig:typhon_and_streams}, of all the streams, Typhon is the closest neighbor of Boo~III. This dynamical proximity, combined with their shared polar, highly eccentric character, is suggestive of a common-infall origin in which both systems entered the MW potential together. However, Typhon's mean metallicity from high-resolution spectroscopy of seven stars is $\overline{[\mathrm{Fe/H}]} = -1.69 \pm 0.11$ \citep{Ji2023Typhon}, substantially higher than Boo~III's ($\overline{[\mathrm{Fe/H}]} = -2.34 \pm 0.11$). A direct progenitor--debris relationship, e.g.\ Typhon being stripped material from the Boo~III progenitor, is therefore disfavored by the chemistry. The most natural interpretation is that the two systems are kinematically associated as members of the same accreted group but had distinct progenitors with different chemical-evolution histories before infall. This dynamical Boo~III--Typhon association is therefore distinct in kind from the proposed Boo~III--Styx connection: Styx is a candidate on-sky tidal extension of Boo~III itself, whereas Typhon is an independent system that may merely share Boo~III's accretion history.

\begin{figure}[ht]
    \centering
    \includegraphics[width=\columnwidth]{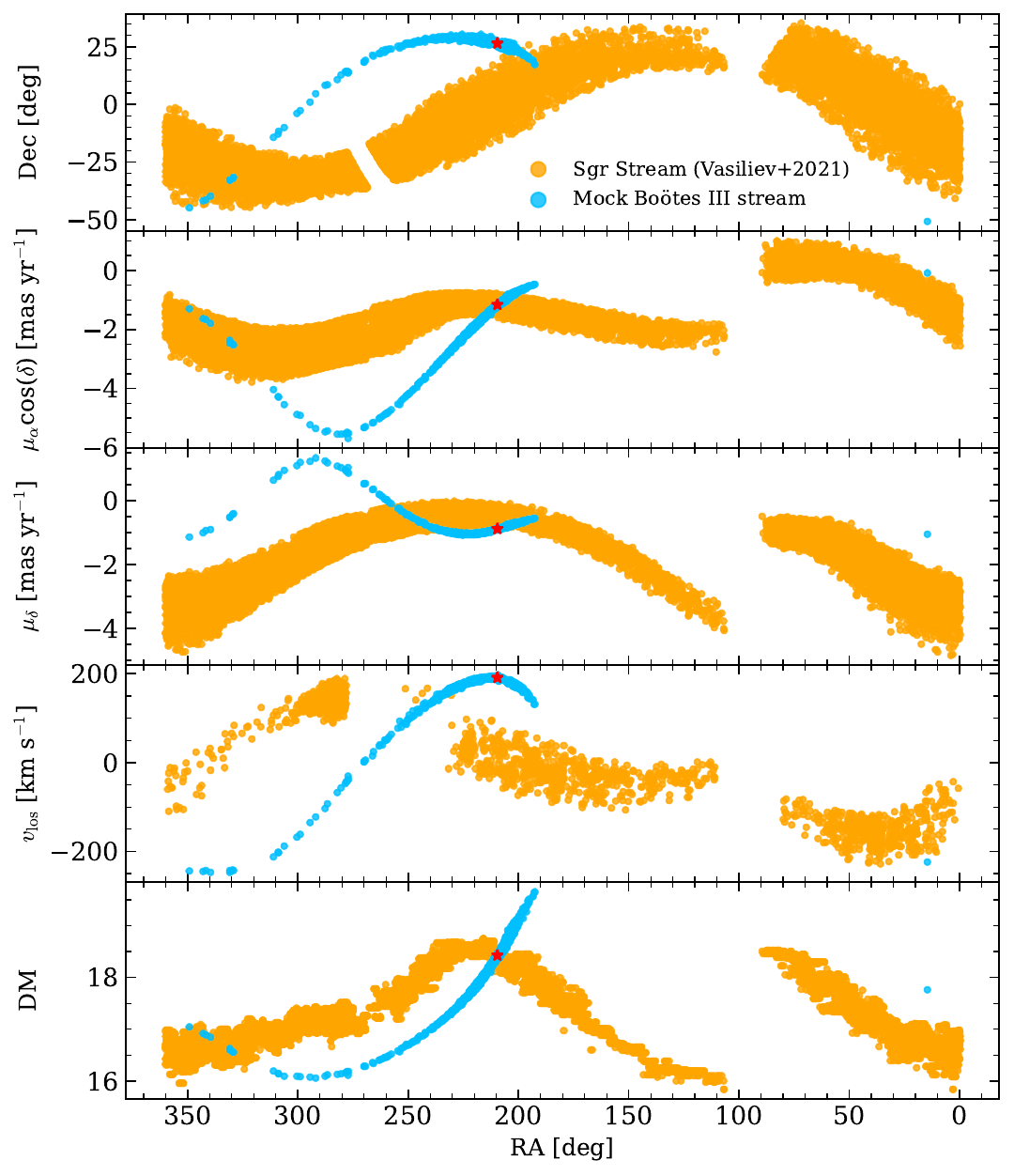}
    \caption{Comparison of the Boo~III mock stream and the Sgr stream. The blue points are from the fiducial particle spray model (Section~\ref{section:stream}), and the orange points are highly probable Sgr stream members from \citet{Vasiliev2021}. The red star in each panel marks the present-day systemic Boo~III value (Table~\ref{tab:params}). From top to bottom the panels show declination, $\mu_\alpha\cos\delta$, $\mu_\delta$, line-of-sight velocity, and distance modulus, all as a function of right ascension. The two structures overlap in sky position, proper motion, \emph{and} heliocentric distance, so photometric tracers alone cannot separate them; only the line-of-sight velocity (fourth panel), where they differ by $\gtrsim 100$~\vlosUnits, provides a clean discriminant, demonstrating that spectroscopic observations are essential for identifying Boo~III tidal-tail members. Although their distance moduli overlap, the two streams' distance gradients run in opposite directions, which can serve as an
    additional consistency check once candidate members are spectroscopically confirmed.}
    \label{fig:sgr_and_boo}
\end{figure}

\subsection{Prospects for Detecting the Boo~III Tidal Tail} \label{section:tail_prospects}
Beyond the 21 GMM members within the 3 $r_h$ ellipse and the two candidate tidal-tail members at $3.7\,r_h$ and $5.8\,r_h$ from the center that we identify in Appendix~\ref{section:appendix_tail}, a direct detection of an extended Boo~III tidal stream would provide an independent confirmation of the disruption picture and a powerful new constraint on the joint MW + LMC potential. We searched for such a stream using two distance-tracer populations from \gaia\ DR3: BHB stars and RRL stars. In each case we selected candidates with sky positions and proper motions consistent with our fiducial particle spray model from Section~\ref{section:stream}. Neither search yielded a convincing extended detection.

A key obstacle is contamination from the Sgr stream, which runs through the same patch of sky as the predicted Boo~III tidal tail. Figure~\ref{fig:sgr_and_boo} compares our fiducial Boo~III stream model with the Sgr-stream members from \citet{Vasiliev2021}. At the on-sky location of the predicted Boo~III tail, the two structures overlap in sky position, proper motion, \emph{and} heliocentric distance (Sgr debris in this longitude range sits at $\sim 40$--$55$~kpc, essentially the same range as the predicted Boo~III tail), so photometric distance tracers alone cannot separate them. Their line-of-sight velocities, however, differ by $\gtrsim 100$~\vlosUnits, providing a clean spectroscopic discriminant. The two streams' distance gradients also run in opposite directions: the Sgr leading arm decreases in distance with increasing RA in this region, while the predicted Boo~III tail increases (Section~\ref{section:phi1_stream}). Once Boo~III members are identified spectroscopically, the opposite-sign distance gradient becomes an additional consistency check.

We therefore conclude that a deep spectroscopic campaign along the predicted stream track, using line-of-sight velocity as the primary discriminant against Sgr, is the most promising path both to confirming that Styx is a genuine stellar stream and to establishing its physical connection to Boo~III. Future wide-field multi-object spectroscopic surveys such as DESI \citep{2016arXiv161100036D}, WEAVE \citep{2024MNRAS.530.2688J}, the Via Project \citep{2026arXiv260618332T}, and Subaru/PFS \citep{2014PASJ...66R...1T}, targeting this region, could definitively establish the physical association, and the spectroscopic 6D of the recovered tail members would simultaneously constrain the MW halo mass and the LMC mass via the gradient diagnostic of Section~\ref{section:phi1_stream} and Appendix~\ref{section:appendix_6dmc}.

\section{Conclusion}\label{section:conclusion}

We have presented updated systemic properties of the UFD Bo\"otes~III based on 24 member stars (21 RGB + 3 RRL) identified from \SSSSS DR2 spectroscopy and \gaia DR3 astrometry. Our main results are as follows.

\begin{enumerate}[leftmargin=*]

\item \textbf{Updated systemic properties, centroid, and heliocentric distance.} From a GMM fit, we measure a velocity dispersion of $\sigma_{v} = 1.69^{+1.03}_{-0.85}$ \vlosUnits, about six times smaller than the $10.7 \pm 3.5$ \vlosUnits reported by \citet{Carlin2018}. This revised value brings Boo~III in line with other known tidally disrupting dwarfs such as Antlia~II ($\sigma_v = 5.98$ \vlosUnits; \citealt{Ji2021}) and Crater~II ($\sigma_v = 2.51$ \vlosUnits; \citealt{Limberg2025}). We also measure a systemic velocity of $v_{\rm los} = 191.22^{+0.64}_{-0.80}$ \vlosUnits, proper motion $(\mu_{\alpha} \cos{\delta}, \mu_\delta) = (-1.163^{+0.017}_{-0.019}, -0.883^{+0.013}_{-0.014})$ \pmUnits, mean metallicity $[\mathrm{Fe/H}] = -2.34 \pm 0.11$~dex, and an updated spatial centroid $(\alpha_0, \delta_0)_{\rm new} = (209.557^\circ, 26.553^\circ) \pm 0.3^\circ$ derived from the iterative median of the GMM members (Section~\ref{section:membership}), offset by $\sim 20'$ ($\sim 0.6\,r_h$) south-east of the \citet{Grillmair2009} discovery position. From five \gaia\ DR3 RRL stars calibrated via \citet{Garofalo2022} we obtain a new heliocentric distance $D_\odot = 48.5 \pm 1.9$~kpc (Section~\ref{section:distance}). Combined with the \citet{Moskowitz2020} angular size, this yields a dynamical mass within the half-light radius of $M_{1/2} \approx 1.24 \times 10^{6}\,{\rm M}_\odot$. All of these updated systemic properties are summarized in Table~\ref{tab:params}.

\item \textbf{Orbit and tidal stream modeling.} Orbit integrations confirm that Boo~III follows a highly eccentric ($e \approx 0.82$ with LMC), polar ($i \approx 89.5^\circ$) orbit with a very recent pericentric passage ($\sim 0.14$~Gyr ago at $r_{\mathrm{peri}} \approx 9.5$~kpc). Using \texttt{galpy}'s particle spray code, we simulate the resulting tidal stream and show that on such an eccentric orbit the debris does not trace the progenitor's orbit, and that the predicted stream track and kinematics are sensitive to the MW halo mass, the LMC mass, and the solar motion (Figure~\ref{fig:stream_comparison}; Section~\ref{section:stream}). The simulated streams are broadly consistent with the Styx stream, supporting a physical association between the two systems, should Styx, which has not yet been independently recovered, prove to be a genuine stellar stream. In particular, the absence of an exact match between Boo~III's \emph{orbit} and the Styx track does not argue against this association, since the appropriate comparison is with the debris track itself.

\item \textbf{Evidence for active tidal disruption.} Three independent diagnostics consistently favor active tidal stripping. (i) The combination of a small pericenter ($r_{\rm peri} \approx 9.5$~kpc) and a very recent pericentric passage ($\sim 0.14$~Gyr ago) means Boo~III has just emerged from the strongest instantaneous tidal field of its orbit and has not had time to relax back to dynamical equilibrium. (ii) Evaluated at pericenter, the tidal radius is $r_t \approx 164$~pc $\approx 0.35\,r_h$, i.e.\ substantially smaller than the system's half-light radius, so material outside the inner $\sim 0.35\,r_h$ is dynamically unbound at closest approach. (iii) Equivalently, Boo~III's mean half-light density is more than an order of magnitude below twice the mean enclosed MW density at its orbital pericenter -- well below the threshold that \citet{Pace2022} identify as marking the onset of tidal disruption. All three diagnostics agree and place Boo~III squarely in the ``actively disrupting'' regime, alongside Antlia~II, Crater~II, Sgr, and Tucana~III. Boo~III is clearly one of the very few UFDs with such direct disruption signatures.

\item \textbf{An unusually low velocity dispersion and its implications for the dark-matter content.} Boo~III's revised dispersion, $\sigma_v = 1.69^{+1.03}_{-0.85}$~\vlosUnits, is among the smallest measured for any UFD. In our $N$-body models with cuspy NFW progenitors, only a progenitor halo mass of order $10^6\,{\rm M}_\odot$ (initialized at the second-most-recent apocenter) reproduces this value, whereas the $\sim 10^8\,{\rm M}_\odot$ halo expected for a UFD of Boo~III's luminosity under $\Lambda$CDM overpredicts $\sigma_v$ (Section~\ref{section:tidaldisrupt_vdisp}). This admits two broad interpretations: either Boo~III formed in a canonical $\sim 10^8\,{\rm M}_\odot$ halo and has since been stripped of most of its dark matter, naturally consistent with the active-disruption picture above, or it retains a more massive halo whose inner profile is cored rather than cuspy (for example, from SIDM; \citealt{Spergel2000, TulinYu2018}), which suppresses $\sigma_v$ at fixed halo mass. This same low-$\sigma_v$, underdense-halo tension is central to the interpretation of Antlia~II and Crater~II \citep{Caldwell2017, Ji2021, Limberg2025}, where it has likewise been attributed to extreme tidal stripping, cored or self-interacting dark-matter halos, or MOND. A systematic comparison of the observed dispersions of disrupting dwarf galaxies against simulations spanning a range of inner-density profiles could therefore turn systems like Boo~III into a distinctive probe of the nature of dark matter. \textbf{Among the known tidally disrupting satellites, Boo~III may be the cleanest laboratory for distinguishing a dark-matter cusp from a core.} Antlia~II and Crater~II are far more luminous, with progenitors likely in the classical-dwarf regime where supernova feedback can itself transform a primordial cusp into a core \citep{Governato2012, DiCintio2014}, so a core in those systems, as in the Sgr dwarf, does not necessarily reflect the microphysics of dark matter. Tucana~III's uncertain nature, globular cluster versus dwarf galaxy, similarly complicates its use for this test \citep{Simon2017}. Boo~III, by contrast, is unambiguously a UFD. Its low mean metallicity alone places it in that regime: on the dwarf-galaxy stellar mass--metallicity relation \citep{Kirby2013}, $\overline{[\mathrm{Fe/H}]} = -2.34$ corresponds to a stellar mass of only $M_* \sim 10^4\,{\rm M}_\odot$, far below the classical dwarfs, while its resolved intrinsic metallicity spread ($\sigma_{[\mathrm{Fe/H}]} = 0.36$~dex) marks a system whose dark-matter halo was deep enough to retain supernova ejecta and self-enrich, rather than a globular cluster \citep{Willman2012, Simon2019}. At such low stellar mass, baryonic feedback is far too weak to core an initially cuspy halo \citep{Penarrubia2012, DiCintio2014}, so an inferred core would instead point to the nature of dark matter itself, for example, self-interactions. This makes Boo~III a uniquely clean, UFD-regime probe of the cusp--core problem.

\item \textbf{Tidal-tail candidates and a path to potential constraints.} Extending the simple box selection of Section~\ref{section:membership} (Eqs.~\eqref{eq:boxcut_rv}--\eqref{eq:boxcut_feh}) beyond the 3 $r_h$ ellipse yields six \SSSSS\ DR2 candidates, of which two are RGB stars at $v_{\rm los} \approx 190$~\vlosUnits\ with $[\mathrm{Fe/H}] \approx -2.6$ that are kinematic and chemical analogues of the inner GMM-selected members  (Appendix~\ref{section:appendix_tail}). These two likely tail members are overplotted as green stars in the $\phi_1$ stream diagnostic of Figure~\ref{fig:phi1_stream}, and both sit close to the predicted spray track in line-of-sight velocity. Notably, the candidate at $\phi_1 \approx +4^\circ$ on the leading-tail side favors a MW potential \emph{lighter} than the fiducial \texttt{McMillan17} but somewhat more massive than \texttt{MWPotential2014}, though with only this single tentative star it is not yet a quantitative constraint. This is consistent with the broader observation that the orbit- and stream-based velocity-gradient diagnostic of Sections~\ref{section:stream}, \ref{section:nbody}, and~\ref{section:phi1_stream} is sensitive to the MW halo mass, the LMC mass, the solar reflex, and Boo~III's own 6D phase-space coordinates -- so a future confirmation of the Boo~III tail (with even a modest number of spectroscopic members at $|\phi_1| \gtrsim 0.5^\circ$) would constrain the joint MW + LMC potential through the gradient predictions, with the dominant 6D-side uncertainty coming from the heliocentric distance (Appendix~\ref{section:appendix_6dmc}). Our broader photometric and astrometric searches using BHB and RRL stars did not yield an extended detection of the predicted tail itself, the primary obstacle being contamination from the Sgr stream, which overlaps with the predicted Boo~III tail in sky position, proper motion, and distance, but differs by $\gtrsim 100$~\vlosUnits\ in line-of-sight velocity (Section~\ref{section:tail_prospects}; Figure~\ref{fig:sgr_and_boo}). Nevertheless, the Styx track was identified purely photometrically, nearly a decade before any proper-motion measurement of Boo~III existed -- that the debris track predicted from our measured 6D phase-space coordinates lands on it would be a remarkable coincidence if the two systems were unrelated, or if Styx were a spurious detection altogether. Since only spectroscopy can supply the line-of-sight velocity that cleanly separates Boo~III debris from the Sgr foreground, deep follow-up along the predicted stream track offers the most promising path both to establishing whether Styx is a genuine stream physically connected to Boo~III and to using any recovered tail members as a potential probe.

\item \textbf{Comparison with the Typhon stream.} Boo~III and the Typhon stream occupy similar regions in action and integrals-of-motion space, consistent with their shared polar orbit and high eccentricity. However, the two systems have distinct metallicities, ruling out a direct progenitor--debris relationship. Instead, their orbital similarity suggests they may share a common infall origin.

\end{enumerate}

\section*{Data Availability} \label{section:data_availability}
All code, analysis notebooks, and figures for this paper are publicly available at \url{https://github.com/s5collab/boo3}.  The repository contains the notebooks and Python scripts that regenerate the main intermediate products (GMM chains, per-component orbital trajectories, particle spray caches) and auto-generate Tables~\ref{tab:boo3_S5_sample}, \ref{tab:rrl_distances}, \ref{tab:appendix_candidates}, and \ref{tab:grad_budget} and Figures~\ref{fig:main_members}--\ref{fig:orbit_params_mwp2014} from the inputs in \texttt{data/}. A few derived quantities quoted in the text (the pericentric timescale, tidal radius, and J-factors) are computed inline from these products; Figure~\ref{fig:pericenter_density} is adapted from \citet{Pace2022}; and larger files, such as the $N$-body simulations described in Section~\ref{section:nbody_setup}, will be made available on Zenodo upon publication of this paper.  

\section*{Acknowledgments}
\begin{acknowledgments}

TSL thanks Jeff Carlin for helpful comments on the manuscript. TSL also thanks Yerkes Observatory for its warm hospitality as a resident astronomer in June 2026, during which much of this manuscript was prepared.

TSL acknowledges financial support from Natural Sciences and Engineering Research Council of Canada (NSERC) through grant RGPIN-2022-04794. SK acknowledges support from the Science \& Technology Facilities Council (STFC) grant ST/Y001001/1. DBZ, GFL, SLM, YY and AMG acknowledge funding from the Australian Research Council through Discovery grant DP220102254.
SLM acknowledges support from the UNSW Scientia Fellowship program. 

This paper includes data obtained with the Anglo-Australian Telescope in Australia and hosted on AAO Data Central (\url{datacentral.org.au}). We acknowledge the traditional owners of the land on which the AAT stands, the Gamilaraay people, and pay our respects to elders past and present.

This work has made use of data from the European Space Agency (ESA) mission \gaia\ (\url{https://www.cosmos.esa.int/gaia}), processed by the \gaia\ Data Processing and Analysis Consortium (DPAC, \url{https://www.cosmos.esa.int/web/gaia/dpac/consortium}). Funding for the DPAC has been provided by national institutions, in particular the institutions participating in the \gaia\ Multilateral Agreement.

This work has made use of data in the Local Volume Database \citep{Pace2025}.
This paper made use of the Whole Sky Database (wsdb) created by Sergey Koposov and maintained at the Institute of Astronomy, Cambridge by Sergey Koposov, Vasily Belokurov and Wyn Evans with financial support from the Science \& Technology Facilities Council (STFC) and the European Research Council (ERC).

This project used public archival data from the Dark Energy Survey and other surveys using the Dark Energy Camera.

This research has made use of NASA's Astrophysics Data System Bibliographic Services.

The scientific results, analysis design, and conclusions of this work were developed by the authors.  Anthropic's \texttt{Claude Code} (Claude Opus 4.8) was used as an interactive coding assistant during preparation of the public release: organizing the analysis into the documented and reproducible notebook + script structure delivered in the companion repository, regenerating the auto-generated tables from the analysis notebooks, cross-checking literature values cited in the text against the primary sources, and running internal-consistency audits on the manuscript (cross-references, defined acronyms, numerical agreement between text, tables, and figure captions).  All scientific interpretation, methodology, and final wording are the responsibility of the authors.

For the purpose of open access, the authors have applied a Creative Commons Attribution (CC BY) license to any Author Accepted Manuscript version arising from this submission.
\end{acknowledgments}

\onecolumngrid
\appendix
\renewcommand{\theHsection}{\Alph{section}}
\renewcommand{\theHfigure}{\thesection.\arabic{figure}}
\renewcommand{\theHtable}{\thesection.\arabic{table}}
\renewcommand{\theHequation}{\thesection.\arabic{equation}}
\makeatletter
\def\@sect#1#2#3#4#5#6[#7]#8{%
  \ifnum #2>\c@secnumdepth
    \def\@svsec{}%
  \else
    \refstepcounter{#1}%
    \ifnum #2=\@ne
      \edef\@svsec{APPENDIX \csname the#1\endcsname:\hskip 0.6em}%
    \else
      \edef\@svsec{\csname the#1\endcsname\hskip 1em}%
    \fi
  \fi
  \@tempskipa #5\relax
  \ifdim \@tempskipa>\z@
    \begingroup #6\relax
    \ifnum #2=\@ne
      \@hangfrom{\hskip #3\relax\@svsec}{\interlinepenalty \@M
        \uppercase{#8}\par}%
    \else
      \@hangfrom{\hskip #3\relax\@svsec}{\interlinepenalty \@M #8\par}%
    \fi
  \endgroup
   \csname #1mark\endcsname{#7}%
   \addcontentsline{toc}{#1}{\ifnum #2>\c@secnumdepth \else
   \protect\numberline{\csname the#1\endcsname}\fi #7}%
 \else
  \def\@svsechd{#6\hskip #3\@svsec
    \ifnum #2=\@ne
        \uppercase{#8}
    \else
        #8
    \fi
    \csname #1mark\endcsname{#7}%
    \addcontentsline{toc}{#1}{\ifnum #2>\c@secnumdepth \else
      \protect\numberline{\csname the#1\endcsname}\fi #7}}%
 \fi
 \@xsect{#5}}
\makeatother

\newpage

\section{Tidal-tail candidates beyond 3 $r_h$} \label{section:appendix_tail}

\begin{figure*}[!htbp]
    \centering
    \includegraphics[width=\textwidth]{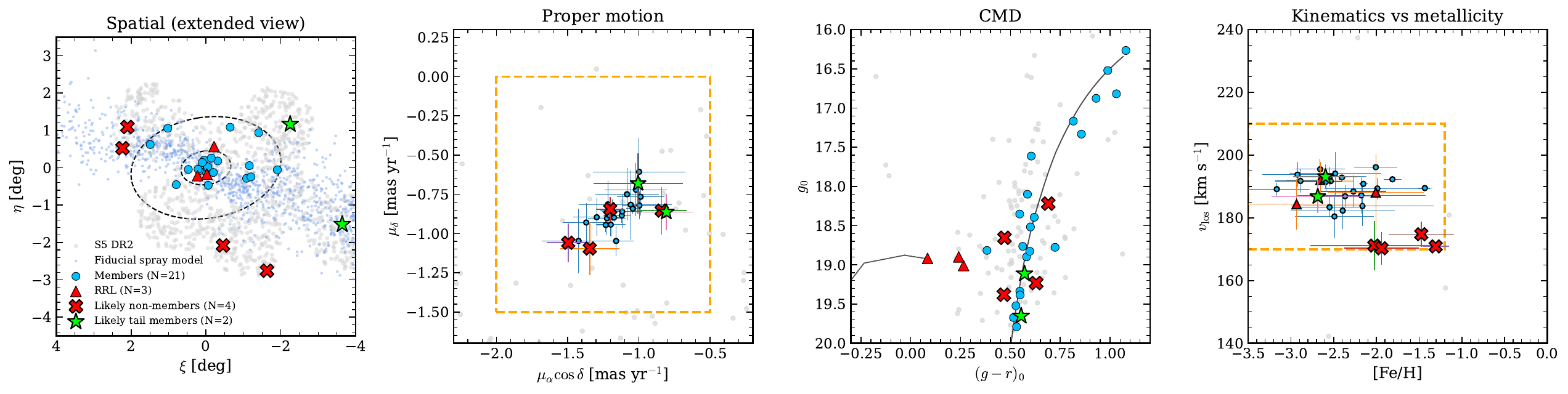}
    \caption{Four-panel diagnostic for the six box-selection candidates beyond the 3 $r_h$ ellipse, in the same observables as Figure~\ref{fig:main_members}. The 21 GMM members within 3 $r_h$ are shown as blue circles for reference; the six candidates outside the 3 $r_h$ ellipse are highlighted with two symbols: green stars for the two \emph{likely tidal-tail members} (low $[\mathrm{Fe/H}]$, $v_{\rm los} \sim 190$~\vlosUnits) and red crosses for the four \emph{likely non-members} clustered at $v_{\rm los} \approx 170$--$175$~\vlosUnits. The orange dashed rectangles in the proper-motion and $v_{\rm los}$-vs-$[\mathrm{Fe/H}]$ panels mark the selection box of Eqs.~\eqref{eq:boxcut_rv}--\eqref{eq:boxcut_feh}. Panel 1 (spatial) shows a square extended view to expose the candidates' positions outside the 3 $r_h$ ellipse. In the same spatial panel, the light-blue points show the fiducial particle spray model (Section~\ref{section:stream}), indicating the expected on-sky location of tidal debris stripped from Boo~III. Interestingly, the more distant of the two likely tidal-tail members (green stars; at $\sim 5.8\,r_h$) lies along this predicted debris track. As in Figure~\ref{fig:main_members}, red triangles mark the three RRL members and small gray points show the \SSSSS\ DR2 field sample.}
    \label{fig:appendix_candidates}
\end{figure*}

The simple box selection of Eqs.~\eqref{eq:boxcut_rv}--\eqref{eq:boxcut_feh} provides an opportunity to extend the Boo~III member search beyond the 3 $r_h$ ellipse used for the systemic-property fit in Section~\ref{section:membership}.
Applied to the parent \SSSSS\ DR2 catalog with the same quality-control cuts, \gaia\ DR3 RRL cross-matches removed, and the spatial cut now \emph{relaxed} to allow stars beyond the 3 $r_h$ ellipse, the box returns six candidates between $3.1\,r_h$ and $5.8\,r_h$ from the center; the exact $r/r_h$ values are listed in the final column of Table~\ref{tab:appendix_candidates}.
Their full \SSSSS\ DR2 properties are listed in Appendix Table~\ref{tab:appendix_candidates}; the four-panel diagnostic plot of Figure~\ref{fig:appendix_candidates} highlights them in the same observables as Figure~\ref{fig:main_members}.

The six candidates fall into two cleanly separated subgroups in line-of-sight velocity.
\emph{Two stars} (\gaia\ DR3 1451743686723428352 and 1445105484053956352) are kinematic and chemical analogues of the 21 GMM members: $v_{\rm los} = 186.8$ and $193.1$~\vlosUnits, $[\mathrm{Fe/H}] = -2.69$ and $-2.59$~dex, with $\log g \approx 3.5$ and $3.2$ consistent with RGB stars at the distance of Boo~III.
Their locations at $r = 3.7\,r_h$ and $5.8\,r_h$ from the center, combined with proper motions and CMD positions consistent with the other high-probability RGB members, make them strong candidates for tidal-tail members.
The remaining \emph{four stars} cluster tightly at $v_{\rm los} \approx 170$--$175$~\vlosUnits, well below the systemic 191~\vlosUnits, and are most plausibly explained as a chance halo subpopulation. Two strong, independent pieces of evidence support this interpretation. First, their metallicities $[\mathrm{Fe/H}] = -1.31$, $-1.47$, $-1.94$, and $-2.02$ (Table~\ref{tab:appendix_candidates}) lie systematically above Boo~III's mean of $\overline{[\mathrm{Fe/H}]} = -2.34 \pm 0.11$~dex, so they are chemically discrepant with the bound population by more than its intrinsic metallicity dispersion ($\sigma_{[\mathrm{Fe/H}]} = 0.36$~dex; Section~\ref{section:membership}). Second, their kinematics are inconsistent with the GMM-fit systemic by $\gtrsim 5\sigma$ given $\sigma_v = 1.7$~\vlosUnits, and one of them (\gaia\ DR3 1251933905373373184) has $\log g = 4.1$, essentially certifying it as a foreground main-sequence dwarf.

We consider the two RGB-like, low-$[\mathrm{Fe/H}]$ candidates at $v_{\rm los} \approx 190$~\vlosUnits\ to be plausible tidal-tail members but do not formally include them in the systemic-property fit of Section~\ref{section:membership}: the 3 $r_h$ ellipse there was chosen as a sensible compromise -- it is large enough to enclose the bulk of the centrally-concentrated GMM members but small enough that the MW field background does not overwhelm the dwarf component, ensuring that the GMM converges on the dwarf parameters cleanly. Extending the spatial cut further introduces enough additional MW contamination to bias the fit and, in extreme cases, prevents convergence on the dwarf component altogether.
Their existence motivates the deeper-survey strategy outlined in Section~\ref{section:tail_prospects}:
targeted spectroscopy outside the 3 $r_h$ ellipse along the predicted stream track from Section~\ref{section:orbit}, with $v_{\rm los}$ as the primary discriminant against halo and Sgr-stream contamination.

\begin{table*}[ht]
\centering
\caption{Six \SSSSS\ DR2 stars beyond 3$r_h$ that pass the box selection of Section~\ref{section:membership} ($v_{\rm los} \in [170, 210]$ \vlosUnits, $\mu_\alpha\cos\delta \in [-2, -0.5]$ \pmUnits, $\mu_\delta \in [-1.5, 0]$ \pmUnits, $[\mathrm{Fe/H}] < -1.2$). Two are kinematic and chemical analogues of the 21 GMM members and are likely tidal-tail members; four cluster at $v_{\rm los} \approx 170$ \vlosUnits\ and are likely halo contaminants (one with $\log g \approx 4$ is a foreground main-sequence star). \label{tab:appendix_candidates}}
\scriptsize
\setlength{\tabcolsep}{3pt}
\begin{tabular}{rrrrrrrrrrrl}
\hline\hline
Gaia Source ID & R.A. & Decl. & $g_0$ & $r_0$ & $v_{\rm los}$ & $[\mathrm{Fe/H}]$ & $\mu_\alpha\cos\delta$ & $\mu_\delta$ & $\log g$ & $r/r_h$ & Class \\
& (deg) & (deg) & (mag) & (mag) & (\vlosUnits) & (dex) & (\pmUnits) & (\pmUnits) & & & \\
\hline
1451743686723428352 & 207.0375 & 27.7010 & 19.65 & 19.10 & 186.77$\pm$5.32 & -2.69$\pm$0.54 & -1.005 & -0.681 & 3.51 & 3.74 & Likely member \\
1445105484053956352 & 205.5562 & 24.9950 & 19.11 & 18.54 & 193.09$\pm$3.92 & -2.59$\pm$0.37 & -0.807 & -0.863 & 3.22 & 5.83 & Likely member \\
1251933905373373184 & 207.7949 & 23.8009 & 19.38 & 18.91 & 171.16$\pm$7.93 & -2.02$\pm$0.74 & -1.344 & -1.096 & 4.08 & 4.73 & Likely non-member \\
1257953177484247040 & 209.0668 & 24.4855 & 19.23 & 18.60 & 170.30$\pm$5.26 & -1.94$\pm$0.44 & -0.840 & -0.854 & 3.24 & 3.14 & Likely non-member \\
1260280877960110464 & 211.9385 & 27.6242 & 18.22 & 17.53 & 170.89$\pm$1.27 & -1.31$\pm$0.16 & -1.201 & -0.845 & 2.28 & 3.53 & Likely non-member \\
1260200957212406528 & 212.0760 & 27.0595 & 18.65 & 18.18 & 174.79$\pm$4.07 & -1.47$\pm$0.38 & -1.495 & -1.059 & 3.44 & 3.43 & Likely non-member \\
\hline
\end{tabular}

\end{table*}

\section{6D Phase-Space Uncertainty Study on the Orbit} \label{section:appendix_6dmc}

To assess how each component of Boo~III's 6D phase-space uncertainty propagates into the inferred orbit, we sample the observed 6D parameters using the \texttt{McMillan17}~+~LMC potential (the same potential as Section~\ref{section:orbit}). For each of the six 6D parameters $(\alpha, \delta, D_\odot, \mu_\alpha\cos\delta, \mu_\delta, v_{\rm los})$ we draw 100 perturbed values from a Gaussian centered on our fiducial measurement and with width equal to the corresponding $1\sigma$ uncertainty from Table~\ref{tab:params} ($\pm 0.3^\circ$ for the centroid, $\pm 1.9$~kpc for the distance, and the GMM posterior $1\sigma$ for the systemic velocity and proper motion), holding the other five parameters fixed at their fiducial values. We integrate each perturbed orbit forward and backward by $\pm 0.5$~Gyr in the \texttt{McMillan17}~+~LMC potential and record the orbital trajectory in the spectroscopic and astrometric observables.

\begin{figure*}
    \centering
    \includegraphics[width=\textwidth]{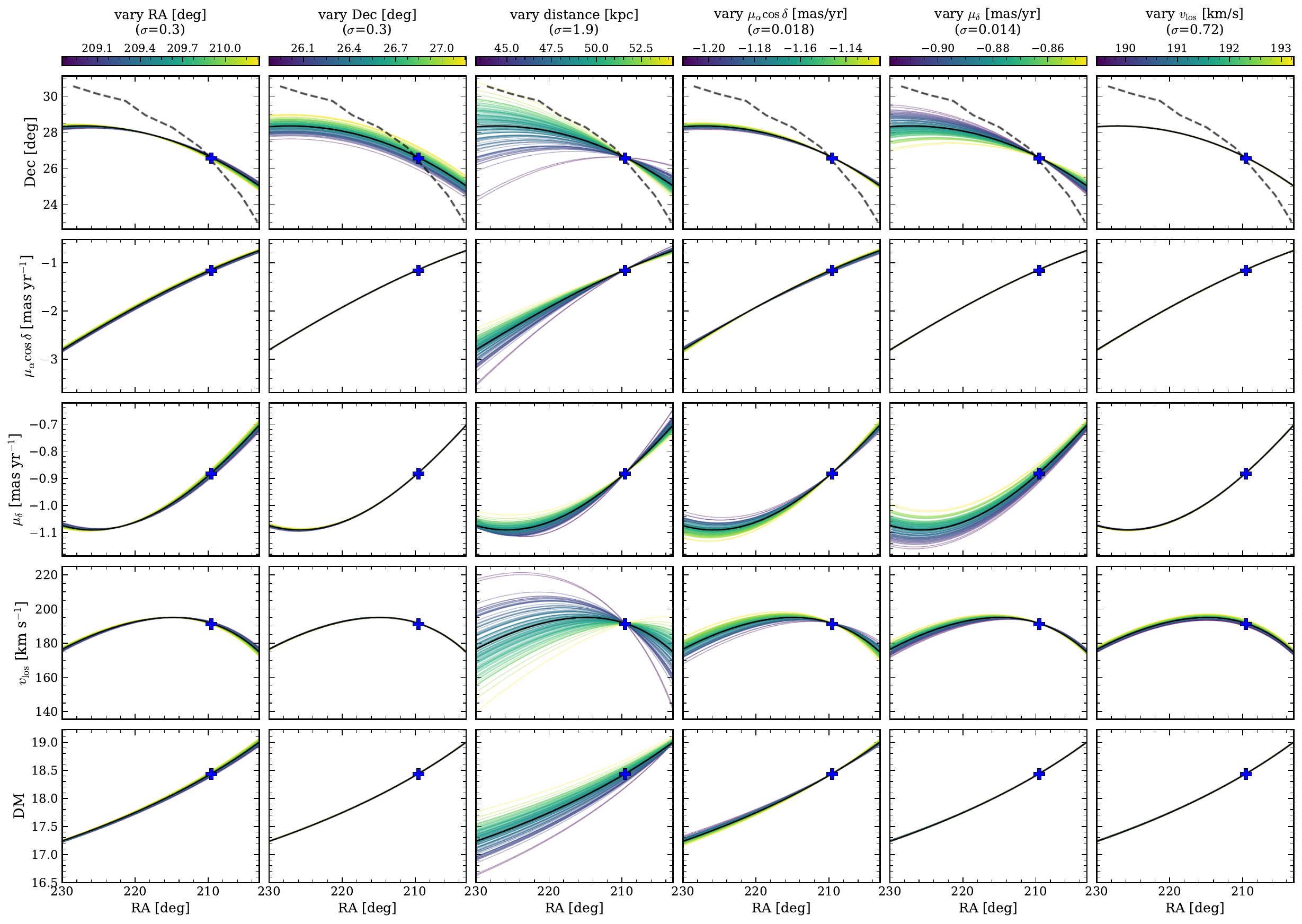}
    \caption{Per-6D-component orbit MC for Boo~III in the \texttt{McMillan17} + LMC potential. Each column varies a single 6D parameter (held at fiducial $\pm$ Table~\ref{tab:params} $1\sigma$, color-coded by the perturbed value); each row is an orbital observable. The 100 perturbed-orbit trajectories per panel sit around the fiducial (black) and the present-day measurement (blue ``+'' marker). The distance column produces by far the widest spread in every observable, identifying $D_\odot$ as the dominant orbit-uncertainty driver. In the Dec row, the gray dashed curve shows the Styx track of \citet{Grillmair2009}.}
    \label{fig:appendix_per_component}
\end{figure*}

Figure~\ref{fig:appendix_per_component} shows the resulting five-by-six panel grid: each column varies a single 6D parameter, and each row shows one orbital observable as
  a function of right ascension along the orbit.  The fiducial trajectory is overplotted as a black curve in each panel, the present-day position is marked with a blue
  filled ``$+$'' symbol, and the 100 perturbed orbits per panel are color-coded by the perturbation value.  The visual
  width of each bundle is a direct proxy for that component's contribution to the orbit uncertainty; the heliocentric distance is clearly the dominant contribution to the orbit-prediction uncertainty.

  We then quantify the same per-component contributions at the level of the predicted Boo~III line-of-sight velocity gradient components $\partial v_{\rm los}/\partial\xi$
  and $\partial v_{\rm los}/\partial\eta$ used in Section~\ref{section:nbody}.  For each 100-realization per-component sub-MC, we evaluate the gradient by linearizing
  the orbit's $v_{\rm los}(\phi)$ track at $\pm 1^\circ$ arc-length around the present-day position and projecting onto $(\hat\xi, \hat\eta)$; the resulting $1\sigma$
  spreads are listed in Table~\ref{tab:grad_budget}.  The budget is dominated by the heliocentric distance: a $\pm 1.9$~kpc spread in $D_\odot$ alone contributes
  $\sigma[\partial v_{\rm los}/\partial\xi] \approx 1.24$ and $\sigma[\partial v_{\rm los}/\partial\eta] \approx 0.17$~\vlosUnits~deg$^{-1}$.  The proper-motion components
  contribute $\sim 5\times$ less, while the centroid (RA, Dec) and systemic-velocity components are negligible at this level.  Adding the six per-component sub-MC
  contributions in quadrature reproduces the full-6D $1\sigma$ error bar shown for the $f = 1$ orbit-prediction diamond in Figure~\ref{fig:fig7_gradient} (top-right panel)
  -- so to a very good approximation, that error bar is just the heliocentric-distance contribution.  Any future improvement of the gradient prediction therefore depends
  primarily on a tighter RRL-based distance, not on tighter centroid or proper-motion measurements.

\begin{table*}[h]
\centering
\caption{Per-6D-component contribution to the orbit-velocity-gradient $1\sigma$ uncertainty, for the \texttt{McMillan17} + LMC fiducial potential, with 100 MC realizations per component and the Table~\ref{tab:params} 1$\sigma$ on each axis.}
\label{tab:grad_budget}
\begin{tabular}{l c c c}
\hline\hline
6D component & $1\sigma$ input & $\sigma[\partial v_{\rm los}/\partial\xi]$ & $\sigma[\partial v_{\rm los}/\partial\eta]$ \\
             &                 & [\vlosUnits~deg$^{-1}$] & [\vlosUnits~deg$^{-1}$] \\
\hline
RA & $0.3^\circ$ & $\sim 0.0$ & $\sim 0.0$ \\
Dec & $0.3^\circ$ & $0.02$ & $\sim 0.0$ \\
$D_\odot$ & $1.9$~kpc & $\mathbf{1.24}$ & $\mathbf{0.17}$ \\
$\mu_\alpha\cos\delta$ & $0.018$ \pmUnits & $0.27$ & $0.07$ \\
$\mu_\delta$ & $0.014$ \pmUnits & $0.14$ & $0.02$ \\
$v_{\rm los}$ & $0.72$ \vlosUnits & $\sim 0.0$ & $\sim 0.0$ \\
\hline
\end{tabular}

\end{table*}

\section{Orbit Integration using MWPotential2014 potential} \label{section:appendix_mwpot}

\begin{figure*}[!b]
    \centering
    \includegraphics[width=\textwidth]{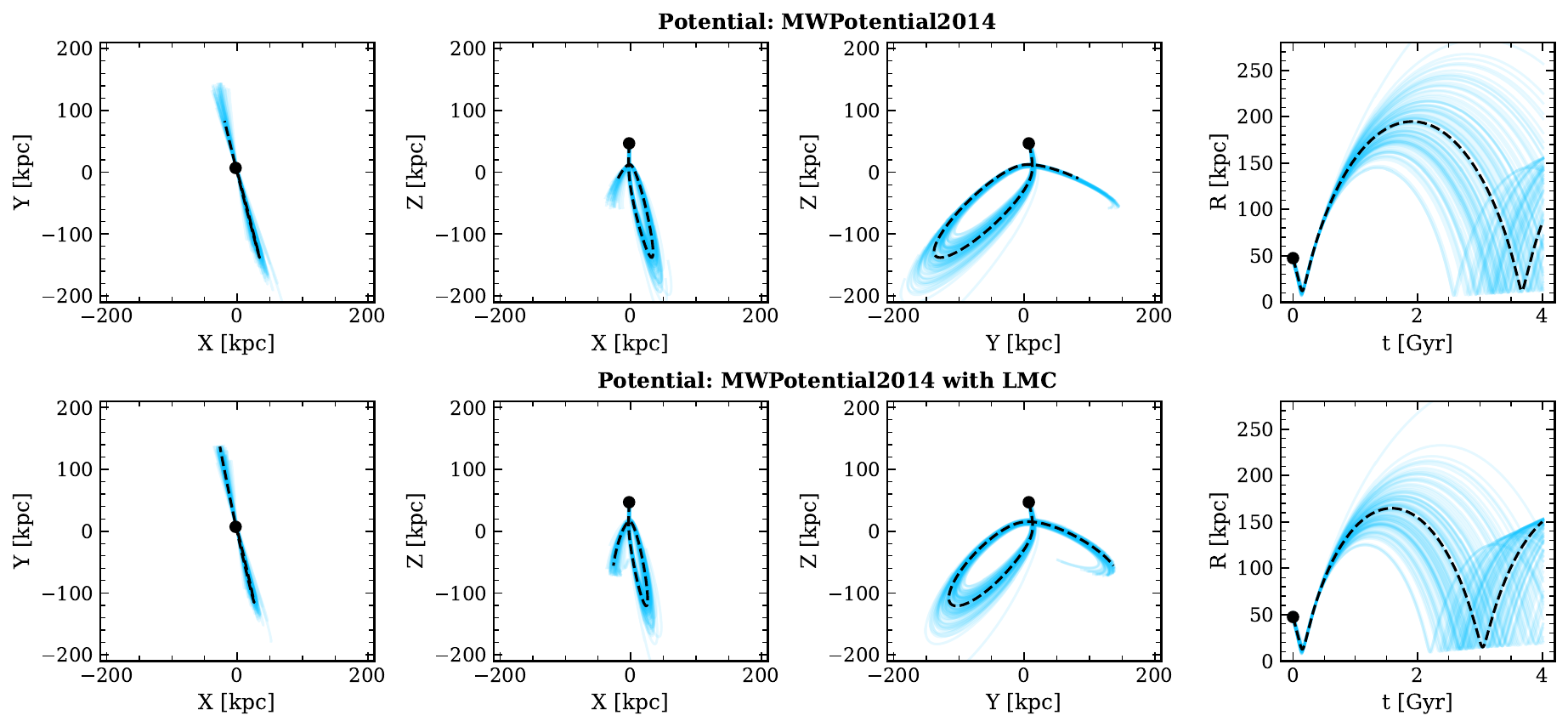}
    \caption{Boo~III orbit integrated with and without LMC, using the \texttt{MWPotential2014} model as base potential. The content of this figure is similar to Figure~\ref{fig:orbits}.}
    \label{fig:orbits_mwp2014}
\end{figure*}

\begin{figure*}[!b]
    \centering
    \includegraphics[width=\textwidth]{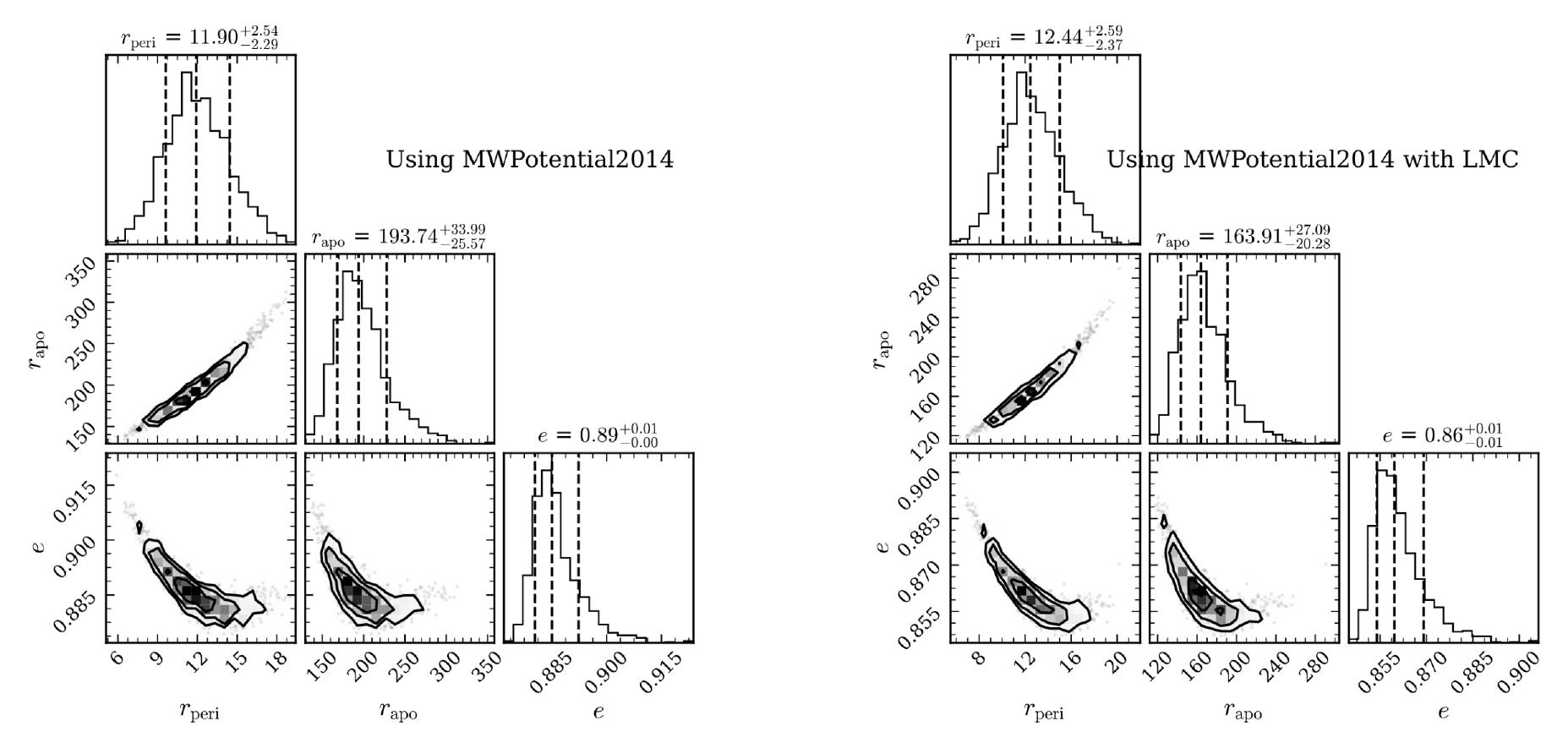}
    \caption{Posterior distributions of the Boo~III orbital parameters from using \texttt{MWPotential2014} as the base potential. The content of this figure is similar to Figure~\ref{fig:orbit_params}.}
    \label{fig:orbit_params_mwp2014}
\end{figure*}

To complement the \texttt{McMillan17}-based orbit integration of Section~\ref{section:orbit}, we repeat the orbit integration with the lighter \texttt{MWPotential2014} model from \citet{Bovy2015} as the base MW potential, with and without the LMC perturbation, propagating the same 6D phase-space uncertainties through 1000 MC realizations. Figure~\ref{fig:orbits_mwp2014} shows the resulting orbital trajectories in Galactocentric Cartesian coordinates and Galactocentric distance vs.\ lookback time, in the same format as Figure~\ref{fig:orbits}; Figure~\ref{fig:orbit_params_mwp2014} shows the corresponding posterior distributions of pericenter, apocenter, and eccentricity, in the same format as Figure~\ref{fig:orbit_params}. The lighter \texttt{MWPotential2014} halo yields a substantially larger apocenter, with more modest increases in pericenter and eccentricity, than the \texttt{McMillan17} model; with the LMC included, the medians shift from $(r_{\rm peri}, r_{\rm apo}, e) = (9.5~\mathrm{kpc}, 98~\mathrm{kpc}, 0.82)$ in \texttt{McMillan17} to $(12.4~\mathrm{kpc}, 164~\mathrm{kpc}, 0.86)$ in \texttt{MWPotential2014}, so the apocenter nearly doubles and the corresponding orbital period stretches from $\sim 1.3$~Gyr to $\sim 3$~Gyr. Despite this quantitative shift, the qualitative orbital character (a polar, highly eccentric trajectory with a recent pericentric passage) is unchanged. As in Section~\ref{section:orbit}, including the LMC primarily reduces the apocenter while leaving the pericenter nearly unchanged. The numerical values quoted in Table~\ref{tab:params} use the \texttt{McMillan17} integration; the \texttt{MWPotential2014} numbers are intended as a sensitivity check on the choice of MW potential.

\section{Boo~III stream coordinates $(\phi_1,\phi_2)$} \label{section:appendix_phi}

The stream diagnostics of Section~\ref{section:phi1_stream} use a great-circle coordinate system $(\phi_1,\phi_2)$ whose origin is the Boo~III centroid $(\alpha_0,\delta_0) = (209.557^\circ, 26.553^\circ)$, matching the updated center in Table~\ref{tab:params}. The $\phi_1$ axis runs along Boo~III's present-day orbit -- the PA of its intrinsic (solar-reflex-corrected) proper motion, $\mathrm{PA} = 257.2^\circ$ east of north, with positive $\phi_1$ toward the leading tail -- so the orbit lies close to $\phi_2 = 0$; the corresponding orbital pole is at $(\alpha_p,\delta_p) = (236.50^\circ, -60.73^\circ)$. The transformation from $(\alpha,\delta)$ to $(\phi_1,\phi_2)$ is the great-circle rotation defined by
\begin{equation}
R = \begin{pmatrix}
-0.778116 & -0.441254 & \phantom{-}0.447024 \\
-0.567164 & \phantom{-}0.799415 & -0.198142 \\
-0.269927 & -0.407714 & -0.872301
\end{pmatrix},
\end{equation}
constructed from this center, PA, and pole.

\section*{Affiliations}
\noindent{\footnotesize
$^{1}$~Department of Astronomy and Astrophysics, University of Toronto, 50 St. George Street, Toronto ON, M5S 3H4, Canada \\
$^{2}$~Dunlap Institute for Astronomy \& Astrophysics, University of Toronto, 50 St George Street, Toronto, ON M5S 3H4, Canada \\
$^{3}$~Data Sciences Institute, University of Toronto, 17th Floor, Ontario Power Building, 700 University Ave, Toronto, ON M5G 1Z5, Canada \\
$^{4}$~Department of Physics, University of Surrey, Guildford GU2 7XH, UK \\
$^{5}$~Department of Astronomy, University of Virginia, 530 McCormick Road, Charlottesville, VA 22904, USA \\
$^{6}$~Institute for Astronomy, University of Edinburgh, Royal Observatory, Blackford Hill, Edinburgh EH9 3HJ, UK \\
$^{7}$~Institute of Astronomy, University of Cambridge, Madingley Road, Cambridge CB3 0HA, UK \\
$^{8}$~Department of Physics and Astronomy, University of Victoria, Victoria, BC, V8W 3P2, Canada \\
$^{9}$~Leibniz-Institut f{\"u}r Astrophysik Potsdam (AIP), An der Sternwarte 16, D-14482 Potsdam, Germany \\
$^{10}$~Research School of Astronomy and Astrophysics, Australian National University, Canberra, ACT 2611, Australia \\
$^{11}$~Department of Astronomy \& Astrophysics, University of Chicago, 5640 S Ellis Avenue, Chicago, IL 60637, USA \\
$^{12}$~NSF-Simons AI Institute for the Sky (SkAI), 172 E. Chestnut St., Chicago, IL 60611, USA \\
$^{13}$~Kavli Institute for Cosmological Physics, University of Chicago, 5640 S Ellis Avenue, Chicago, IL 60637, USA \\
$^{14}$~Lowell Observatory, 1400 W Mars Hill Rd, Flagstaff,  AZ 86001, USA \\
$^{15}$~Sydney Institute for Astronomy, School of Physics, A28, The University of Sydney, NSW 2006, Australia \\
$^{16}$~School of Physics, University of New South Wales, Sydney, NSW 2052, Australia \\
$^{17}$~School of Mathematical and Physical Sciences, Macquarie University, Sydney, NSW 2109, Australia \\
$^{18}$~Macquarie University Research Centre for Astrophysics and Space Technologies, Sydney, NSW 2109, Australia \\
$^{19}$~Department of Astronomy, University of Washington, Seattle, WA 98195, USA \\
$^{20}$~Department of Physics and Astronomy, Texas A\&M University, College Station, TX 77843, USA \\
$^{21}$~Mitchell Institute for Fundamental Physics and Astronomy, Texas A\&M University, College Station, TX 77843, USA \\
\par}

\bibliography{main}{}
\bibliographystyle{aasjournal}

\end{document}